\documentclass[a4paper,fleqn,usenatbib]{mnras}

\usepackage[T1]{fontenc}
\usepackage{ae,aecompl}

\usepackage{subfigure}
\usepackage{graphicx}	
\usepackage{amsmath}	
\usepackage{amssymb}	
\usepackage{color}
\usepackage{mathptmx}
\usepackage{epsfig}
\usepackage{wasysym}
\usepackage{xfrac}
\usepackage{pifont}
\usepackage[table,xcdraw]{xcolor}
\usepackage{booktabs}
\usepackage{multirow}
\usepackage{comment}
\usepackage{soul}
\usepackage{tikz}
\usetikzlibrary{positioning,decorations.pathreplacing,shapes}

\title[Low surface-brightness galaxies]{The formation and evolution of low-surface-brightness galaxies}

\author[G. Martin et al.]{G. Martin,$^{1}$\thanks{E-mail: g.martin4@herts.ac.uk} S. Kaviraj,$^{1}$ C. Laigle,$^{2}$ J. E. G. Devriendt,$^{2}$ R. A. Jackson,$^{1}$ S. Peirani$^{3,4}$ \newauthor Y. Dubois,$^{3}$ C. Pichon,$^{3,5,6}$ and A. Slyz$^{2}$ 
\\
$^{1}$Centre for Astrophysics Research, School of Physics, Astronomy and Mathematics, University of Hertfordshire, College Lane, Hatfield AL10 9AB, UK\\
$^{2}$Dept of Physics, University of Oxford, Keble Road, Oxford OX1 3RH UK\\
$^{3}$Institut d'Astrophysique de Paris, Sorbonne Universit\'{e}s, UMPC Univ Paris 06 et CNRS, UMP 7095, 98 bis bd Arago, 75014 Paris, France\\
$^{4}$ Universit\'e C\^ote d'Azur, Observatoire de la C\^ote d'Azur, CNRS, Laboratoire Lagrange, Bd de l'Observatoire, CS 34229, 06304 Nice Cedex 4, France\\
$^{5}$Korea Institute of Advanced Studies (KIAS) 85 Hoegiro, Dongdaemun-gu, Seoul, 02455, Republic of Korea\\
$^{6}$Institute for Astronomy, University of Edinburgh, Royal Observatory, Blackford Hill, Edinburgh, EH9 3HJ, United Kingdom\\
}

\pubyear{2019}
\begin{document}
\label{firstpage}
\pagerange{\pageref{firstpage}--\pageref{lastpage}}
\maketitle

\begin{abstract}
Our statistical understanding of galaxy evolution is fundamentally driven by objects that lie above the surface-brightness limits of current wide-area surveys ($\mu \sim 23$ mag arcsec$^{-2}$). While both theory and small, deep surveys have hinted at a rich population of low-surface-brightness galaxies (LSBGs) fainter than these limits, their formation remains poorly understood. We use Horizon-AGN, a cosmological hydrodynamical simulation to study how LSBGs, and in particular the population of ultra-diffuse galaxies (UDGs; $\mu > 24.5$~mag~arcsec$^{-2}$), form and evolve over time. For M$_*$>$10^{8}~$M$_{\odot}$, LSBGs contribute 47, 7 and 6 per cent of the local number, mass and luminosity densities respectively ($\sim$85/11/10 per cent for M$_*$>$10^{7}~$M$_{\odot}$). Today's LSBGs have similar dark-matter fractions and angular momenta to high-surface-brightness galaxies (HSBGs; $\mu < 23$ mag arcsec$^{-2}$), but larger effective radii ($\times$2.5 for UDGs) and lower fractions of dense, star-forming gas (more than $\times$6 less in UDGs than HSBGs). LSBGs originate from the same progenitors as HSBGs at $z>2$. However, LSBG progenitors form stars more rapidly at early epochs. The higher resultant rate of supernova-energy injection flattens their gas-density profiles, which, in turn, creates shallower stellar profiles that are more susceptible to tidal processes. After $z\sim1$, tidal perturbations broaden LSBG stellar distributions and heat their cold gas, creating the diffuse, largely gas-poor LSBGs seen today. In clusters, ram-pressure stripping provides an additional mechanism that assists in gas removal in LSBG progenitors. Our results offer insights into the formation of a galaxy population that is central to a complete understanding of galaxy evolution, and which will be a key topic of research using new and forthcoming deep-wide surveys.  
\end{abstract}




\section{Introduction}
Our understanding of galaxy evolution is intimately linked to the part of the galaxy population that is visible at the surface-brightness limits of past and current wide-area surveys. Not only do these thresholds determine the extent of our empirical knowledge, but the calibration of our theoretical models (and therefore our understanding of the physics of galaxy evolution) is strongly influenced by these limits. In recent decades, a convergence of wide-area surveys like the SDSS \citep{Abazajian2009} and large-scale numerical simulations \citep[e.g.][]{Croton2006,Dubois2014,Vogelsberger2014} has had a transformational impact on our understanding of galaxy evolution. While these surveys have mapped the statistical properties of galaxies, comparison to cosmological simulations -- first via semi-analytical models \citep[e.g.][]{Somerville1999,Cole2000,Benson2003,Bower2006,Croton2006} and more recently via their hydrodynamical counterparts \citep[e.g.][]{Dubois2014,Vogelsberger2014,Schaye2015,Kaviraj2017} -- has enabled us to understand the physical drivers of galaxy formation over much of cosmic time.

The SDSS, which has provided much of the discovery space at low and intermediate redshift, starts becoming incomplete at an $r$-band effective surface-brightness, $\langle\mu\rangle_{e}$\footnote{The effective surface-brightness, $\langle\mu\rangle_{e}$, is defined as the mean surface-brightness within an effective radius.}, of $\sim$23~mag~arcsec$^{-2}$ \citep[e.g.][]{Driver2005,Blanton2005,Zhong2008,Bakos2012}. This is primarily due to the lack of depth of the survey but also due, in part, to the standard SDSS pipeline not being optimised for structures that are close to the sky background. Indeed, while bespoke sky subtraction on SDSS images is able to mitigate some of these issues and reveal low-surface-brightness galaxies (LSBGs), these objects do not form the bulk of the population that are visible in such surveys \citep[e.g.][]{Kniazev2004,Williams2016}. Thus, while it is clear that a (largely) hidden Universe exists just below the surface-brightness limits of current large-area surveys, the detailed nature of galaxies in this LSB domain remains largely unexplored, both observationally and in our theoretical models of galaxy evolution. Indeed, the existence of large numbers of faint, undiscovered galaxies has deep implications for our understanding of galaxy evolution. Since our current view of how galaxies evolve is largely predicated on high-surface-brightness galaxies (HSBGs; $\langle\mu\rangle_{e} <$23~mag~arcsec$^{-2}$), this almost certainly leads to potentially significant biases in our understanding of the evolution of the baryonic Universe. Mapping the LSB domain empirically, and exploring the mechanisms by which galaxies in this regime form and evolve, is central to a complete understanding of galaxy evolution. 

The existence of a population of faint, diffuse, (typically) low-mass galaxies has been known since the mid-1980s \citep[e.g.][]{Sandage1984}. However, in the decades following their discovery, very few additional examples were identified \citep[e.g.][]{Impey1988,Bothun1991,Turner1993,Dalcanton1997}, largely due to the surface-brightness limits of contemporary observations. Only very recently, thanks to advances in the sensitivity and field of view of modern instruments \citep[e.g.][]{Miyazaki2002,Kuijken2002,Miyazaki2012,Diehl2012,Abraham2014,Torrealba2018} and the introduction of new observational and data-analysis techniques \citep[e.g.][]{Akhlaghi2015,Prole2018}, has the identification of significant samples of LSBGs become possible \citep[e.g.][]{vanDokkum2015,Koda2015,Munoz2015,vanderBurg2016,Janssens2017,Venhola2017,Greco2018}. 

While modern instruments are enabling the study of systems at significantly fainter surface-brightnesses than was previously possible, deep-wide surveys and spectroscopic follow-up of areas large enough to contain significant populations of LSBGs outside dense, cluster environments remain prohibitively expensive. As a result, the LSB domain remains poorly explored in groups \citep[e.g][]{Castelli2016,Merritt2016,Roman2017b,Roman2017} and the field \citep[e.g][]{MartinezDelgado2016,Papastergis2017,Leisman2017}. This is particularly true for the extremely faint, diffuse end of the LSB population, often referred to, in the contemporary literature, as `ultra-diffuse' galaxies (UDGs; \citet{vanDokkum2015}).  

Recent work suggests that, while LSBGs may be ubiquitous in clusters \citep[e.g][]{Koda2015}, they occur across all environments \citep{Roman2017b,Merritt2016,Papastergis2017}. However, the contribution of the LSB population to the number, mass and luminosity density of the Universe remains unclear. A number of studies \citep[e.g.][]{Davies1990,Dalcanton1997,ONeil2000,Minchin2004,Haberzettl2007} have argued that LSBGs represent a significant fraction of objects at the faint end of the luminosity function and dominate the number density of galaxies at the present day. They may also account for a significant fraction of the dynamical mass budget ($\sim 15$ per cent) \citep[e.g.][]{Driver1999,ONeil2000,Minchin2004} and the neutral hydrogen density \citep{Minchin2004} in today's Universe, although they are thought to contribute a minority (a few per cent) of the local luminosity and stellar mass density \citep{Bernstein1995,Driver1999,Hayward2005}.

While new observations are opening up the LSB domain, the formation mechanisms of LSBGs and their relationship to the HSBG population, on which our understanding of galaxy evolution is predicated, remains poorly understood. Compared to the HSBG population, LSBGs, and UDGs in particular, appear to be relatively quenched, dispersion-dominated systems which largely occupy the red sequence \citep{vanDokkum2015,vanDokkum2016,Ferremateu2018,Ruiz2018}. In lower-density environments, however, they are typically bluer (i.e. unquenched) possibly reflecting a wide range of formation scenarios across different environments \citep[e.g.][]{Roman2017,Zaritsky2019}. LSBGs are typically extremely extended systems for their stellar mass, with low ($n\lesssim 1$) S\'ersic indices  \citep{Koda2015}. While there does not appear to be a single evolutionary path that is able to explain the formation of these objects, a number of mechanisms capable of producing such extended, relatively quenched systems have been proposed. 

For example, \citet{vanDokkum2015} have proposed that UDGs may be failed Milky Way-like ($L^{\star}$) galaxies, which were quenched at high redshift as a result of gas stripping. However, observational evidence using globular cluster abundances \citep{Beasley2016,Peng2016,Amorisco2018}, velocity dispersions \citep[e.g][]{Toloba2018}, weak lensing measurements \citep[e.g][]{Sifon2018}, stellar populations \citep[e.g][]{Ferremateu2018,Ruiz2018}, and the spatial distributions and abundances of the galaxies themselves \citep[e.g][]{Roman2017b}, largely supports the idea that the vast majority of LSBGs are low-mass (i.e. dwarf) galaxies that are hosted by correspondingly low mass dark-matter haloes, except perhaps in a small number of extreme cases \citep[e.g][]{vanDokkum2016,Beasley2016b}. 

UDGs, for example, have been suggested to form as the result of various channels, including anomalously high spin \citep[e.g][]{Amorisco2016,Amorisco2016b,Rong2017,Leisman2017}, gas outflows due to supernova (SN) feedback \citep[e.g][]{DiCintio2017,Chan2018} and strong tidal fields or mergers (e.g. \citealt{Carleton2018, Conselice2018, Abraham2018, Baushev2018}, but see \citealt{Mowla2017}). Thus, while the exact mechanisms responsible for producing UDGs are still debated, there is broad consensus that the progenitors of the majority of UDGs are galaxies in low mass haloes, rather than `failed' high mass haloes where galaxies were prevented from forming in the first place. 

In this paper, we use Horizon-AGN, a cosmological hydrodynamical simulation \citep{Dubois2014,Kaviraj2017}, to perform a comprehensive study of galaxies in the LSB domain. The use of a cosmological simulation is essential for this exercise, since it enables us to study baryonic processes that are likely to drive LSBG formation (e.g. SN feedback, ram-pressure stripping and tidal perturbations) within fully resolved cosmological structure. We explore the predicted properties of a complete sample of LSBGs in today's Universe across all environments, investigate the evolution of their progenitors over cosmic time and study the role of key processes (e.g. SN feedback, tidal perturbations and ram-pressure stripping) in creating these systems. 

This paper is structured as follows. In Section \ref{sec:simulation}, we present an overview of the Horizon-AGN simulation, including the treatment of baryonic physics, the definition of galaxies and their merger trees, and the identification of LSBGs. In Section \ref{sec:properties}, we compare the present-day properties of LSBGs to a sample of their HSB counterparts that have the same distribution of stellar masses. In Section \ref{sec:evolution}, we explore the evolution of key properties in which LSBGs and HSBGs diverge the most (gas fractions, effective radii and density profiles) and which are, therefore, central to the formation of LSB systems. In Section \ref{sec:processes}, we quantify the processes (SN feedback, ram pressure stripping and tidal perturbations) that are responsible for creating LSBGs over cosmic time. We summarise our results in Section \ref{sec:conclusion}.


\section{The Horizon-AGN simulation}

\label{sec:simulation}
In this study we employ Horizon-AGN, a cosmological-volume hydrodynamical simulation \citep{Dubois2014}, that is based on \textsc{ramses} \citep{Teyssier2002}, an adaptive mesh refinement (AMR) Eulerian hydrodynamics code. Horizon-AGN simulates a box with a length of 100~$h^{-1}\, \rm coMpc$. Initial conditions are taken from a \emph{WMAP7} $\Lambda$CDM cosmology \citep{Komatsu2011}, using $1024^3$ dark matter (DM) particles, with a mass resolution of $8\times10^7$~M$_{\odot}$. An initially uniform $1024^3$ cell grid is refined, according to a quasi Lagrangian criterion (when 8 times the initial total matter resolution is reached in a cell), with the refinement continuing until a minimum cell size of $1 \, \rm kpc$ in proper units is achieved. Additional refinement is allowed at each doubling of the scale factor, in order to keep the resolution constant in physical units. Note that, in addition to the hydrodynamics, the AMR cells also define the force softening for the dark matter and baryons. We direct readers to Appendix \ref{sec:resolution} for a discussion of the effect of the resolution of Horizon-AGN on the sizes of galaxies. 

Horizon-AGN produces good agreement with key observables that trace the cumulative evolution of galaxies across at least 95\% of cosmic time: stellar mass/luminosity functions, the star formation main sequence, rest-frame UV-optical-near infrared colours and the merger and star formation histories of galaxies \citep{Kaviraj2015a,Kaviraj2017}. The simulation also reproduces black-hole (BH) demographics, such as the luminosity and mass functions of BHs, the evolution of BH mass density over cosmic time and correlations between BH and galaxy mass from $z=3$ to $z=0$ \citep{Volonteri2016,Martin2018_BH}. Finally, Horizon-AGN produces good agreement with the morphological mix of the local Universe, with the predicted galaxy morphologies reproducing the observed fractions of early and late-type galaxies that have intermediate and high stellar masses \citep{Dubois2016,Martin2018_progenitors}.

In the following sections, we describe aspects of the simulation that are particularly relevant to this study: the treatment of baryonic matter (gas and stars), the identification of galaxies, construction of their merger trees and the selection of LSBGs. 


\subsection{Baryons}
Gas cooling is assumed to take place via H, He and metals \citep{Sutherland1993}, down to a temperature of 10$^4$ K. A uniform UV background is switched on at $z = 10$, following \citet{Haardt1996}. Star formation proceeds via a standard 2 per cent efficiency \citep[e.g.][]{Kennicutt1998}, when the hydrogen gas density reaches $0.1$~H~cm$^{-3}$. The stellar-mass resolution in Horizon-AGN is $4\times10^6$~M$_{\odot}$. 

The simulation employs continuous stellar feedback that includes momentum, mechanical energy and metals from stellar winds and both Type II and Type Ia supernovae (SNe). Feedback from stellar winds and Type II SNe is implemented using \textsc{Starburst99} \citep{Leitherer1999,Leitherer2010}, via the Padova model \citep{Girardi2000} with thermally pulsating asymptotic giant branch stars \citep{Vassiliadis1993}. The `Evolution' model of \citet{Leitherer1992} is used to calculate the kinetic energy of stellar winds. \citet{Matteucci1986} is used to determine the implementation of Type Ia SNe, assuming a binary fraction of 5\% \citep{Matteucci2001}, with chemical yields taken from the W7 model of \citet{Nomoto2007}. Stellar feedback is assumed to be a heat source after 50 Myrs, because after this timescale the bulk of the energy is liberated via Type Ia SNe that have time delays of several hundred Myrs to a few Gyrs \citep[e.g.][]{Maoz2012}. These systems are not susceptible to large radiative losses, since stars will disrupt or migrate away from their birth clouds after a few tens of Myrs \citep[see e.g.][]{Blitz1980,Hartmann2001}.

We note that using an AMR refinement scheme based on total matter density allows us to resolve the gas content of galaxies out to larger radii, since the resolution in the outskirts of the galaxy is principally set by the DM mass, where it dominates rather than the gas mass, which is generally small (as would be the case in smoothed particle hydrodynamics schemes, for example). This is important for the study of diffuse galaxies, particularly those with small gas fractions.


\subsection{Identifying galaxies and merger trees}
\label{sec:merger_trees}
To identify galaxies we use the \textsc{AdaptaHOP} structure finder \citep{Aubert2004,Tweed2009}, applied to the distribution of star particles. Structures are identified if the local density exceeds 178 times the average matter density, with the local density being calculated using the 20 nearest particles. A minimum number of 50 particles is required to identify a structure. This imposes a minimum galaxy stellar mass of $2\times 10^{8}$~M$_{\odot}$. We then produce merger trees for each galaxy in the final snapshot ($z\sim 0.06$), with an average timestep of $\sim$130~Myr, which enables us to track the main progenitors (and thus the assembly histories) of individual galaxies. 

We note that, due to the minimum mass limit described above ($2\times 10^{8}$~M$_{\odot}$), the LSBGs we study in this paper have masses in excess of this threshold. These systems are, therefore, typically at the higher mass end of the LSBG populations that have been studied in recent observational work. 


\subsection{Surface-brightness maps and selection of LSBGs}
\label{sec:SB_maps}
We use the \citet[][BC03 hereafter]{Bruzual2003} stellar population synthesis models, with a \citet{Chabrier2003} initial mass function, to calculate the intrinsic spectral energy distribution (SED) for each star particle within a galaxy, given its metallicity. We assume that each star particle represents a simple stellar population, where all stars are formed at the same redshift and have the same metallicity. The SEDs are then multiplied by the initial mass of each particle to obtain their intrinsic flux. 

We use the \textsc{SUNSET} code to measure dust attenuation, as described in \citet{Kaviraj2017}. Briefly, we first extract the density and metallicity of the gas cells in the galaxy and convert the gas mass within each cell to a dust mass, assuming a dust-to-metal ratio of 0.4 \citep[e.g.][]{Draine2007}. The column density of dust is used to compute the line-of-sight optical depth for each star particle, and dust-attenuated SEDs are then calculated assuming a dust screen in front of each star particle. As shown in \citet{Kaviraj2017}, for optical filters, this produces comparable results to a full radiative transfer approach. The attenuated SEDs are then convolved with the SDSS $r$ band filter response curve and binned to a spatial resolution of 1~kpc.

Following the convention in the observational literature, we identify LSBGs using their effective surface-brightness, $\langle \mu \rangle _{e}$, defined as the average surface-brightness within the effective radius ($R_{\mathrm{eff}}$). We calculate $R_{\mathrm{eff}}$ by performing photometry using isophotal ellipses as apertures, with $R_{\mathrm{eff}}$ defined as the semi-major axis of an isophote containing half of the total galaxy flux. The effective surface-brightness is then calculated using the total flux contained within this ellipse divided by the area of the aperture. We note that the $r$ band surface-brightness is largely insensitive to the specific dust attenuation recipe, especially for LSBGs, which are largely dust poor. 

It is worth noting that the labelling of galaxies as `LSB' systems is strongly determined by the surface-brightness limits of surveys that were available, when the term was coined \citep[e.g.][]{Disney1976}. Galaxies we define as LSBGs in this study are those that are largely invisible at the depth of \textit{current} wide-area surveys, like the SDSS. Indeed, if contemporary large surveys were deeper (e.g. like the forthcoming LSST survey, which will be 5 magnitudes deeper than the SDSS) then our definition of an LSB galaxy would be very different. Surveys like the SDSS start becoming incomplete around $\langle \mu \rangle _{e}$ < 23~mag~arcsec$^{-2}$ \citep[e.g][]{Kniazev2004,Bakos2012,Williams2016} in the $r$ band. The nominal completeness of the survey is $\sim$70 per cent at $\sim$23~mag~arcsec$^{-2}$ \citep[e.g.][]{Zhong2008,Driver2005}, falling rapidly to $\sim$10 per cent for galaxies that are fainter than $\sim$24~mag~arcsec$^{-2}$ \citep[e.g.][]{Kniazev2004}. In our analysis below, we split our galaxies into three categories, defined using effective surface-brightness:

\begin{enumerate}
\item \textit{`High-surface-brightness galaxies' (HSBGs):} These are defined as galaxies with $\langle \mu \rangle _{e}$ < 23~mag~arcsec$^{-2}$ in the $r$ band. They represent the overwhelming majority of galaxies that are detectable in past surveys like the SDSS, and which underpin our current understanding of galaxy evolution.\\
\item \textit{`Classical low-surface-brightness galaxies' (Cl. LSBGs)}: These are defined as galaxies with 24.5 > $\langle \mu \rangle _{e}$ > 23~mag~arcsec$^{-2}$ in the $r$ band. They represent the brighter end of the LSBG population and are the `classical' LSB galaxy populations that have been studied in the past literature, particularly that which preceded the SDSS.\\
\item \textit{`Ultra-diffuse galaxies' (UDGs):} These are defined as galaxies with $\langle \mu \rangle _{e}$ > 24.5~mag~arcsec$^{-2}$  in the $r$ band \citep[e.g.][]{Laporte2018}. They represent the fainter end of the LSB galaxy population. 

We note that there is no standard definition in the literature of what constitutes a UDG, owing to the often specialised nature of the instruments and techniques involved in their detection. However, most definitions are roughly equivalent. For example, \citet{vanDokkum2015} and \citet{Roman2017} both use a $g$ band central surface-brightness ($\mu_{0}$) of 24~mag~arcsec$^{-2}$, \citet{Koda2015} use an $R$ band effective surface-brightness of 24~mag~arcsec$^{-2}$ and \citet{vanderBurg2016} use an $r$ band effective surface-brightness of 24~mag~arcsec$^{-2}$. Often, UDGs are also selected using an effective radius threshold of $R_{\mathrm{eff}} \gtrsim 1.5$ in order to differentiate them from more compact, lower mass objects with equivalent surface-brightnesses \citep[e.g.][]{vanDokkum2015,Koda2015,vanderBurg2016,Roman2017}. While this is an important consideration over the mass ranges that these observational studies examine ($M_{\star} < 10^{8}$ M$_{\odot}$), the range of masses that we consider in Section \ref{sec:present_day_props} onwards ($10^{8.5}$--$10^{10}$~M$_{\odot}$) precludes such objects.
\label{fig:radii_compare}
\end{enumerate}

Note that, in the following sections, we use `low surface-brightness galaxy' (LSBG) to refer to any galaxy in Horizon-AGN with $\langle \mu \rangle _{e}$ > 23~mag~arcsec$^{-2}$ (i.e. any galaxy that falls in either the Cl. LSBG or UDG categories). As we describe below, the threshold $\langle \mu \rangle _{e}$ $\sim$ 24.5~mag~arcsec$^{-2}$ between our two LSBG categories (Cl. LSBGs and UDGs), appears to demarcate two galaxy populations that are reasonably distinct, both in terms of the redshift evolution of their properties and their formation mechanisms. The  Cl. LSBGs are much closer to the HSBGs in terms of their formation histories, with the real distinctions emerging between HSBGs and UDGs. The differences between the evolution of HSBGs and UDGs is therefore the principal focus of this study. 

\begin{figure*}
	\centering
    \includegraphics[width=0.95\textwidth]{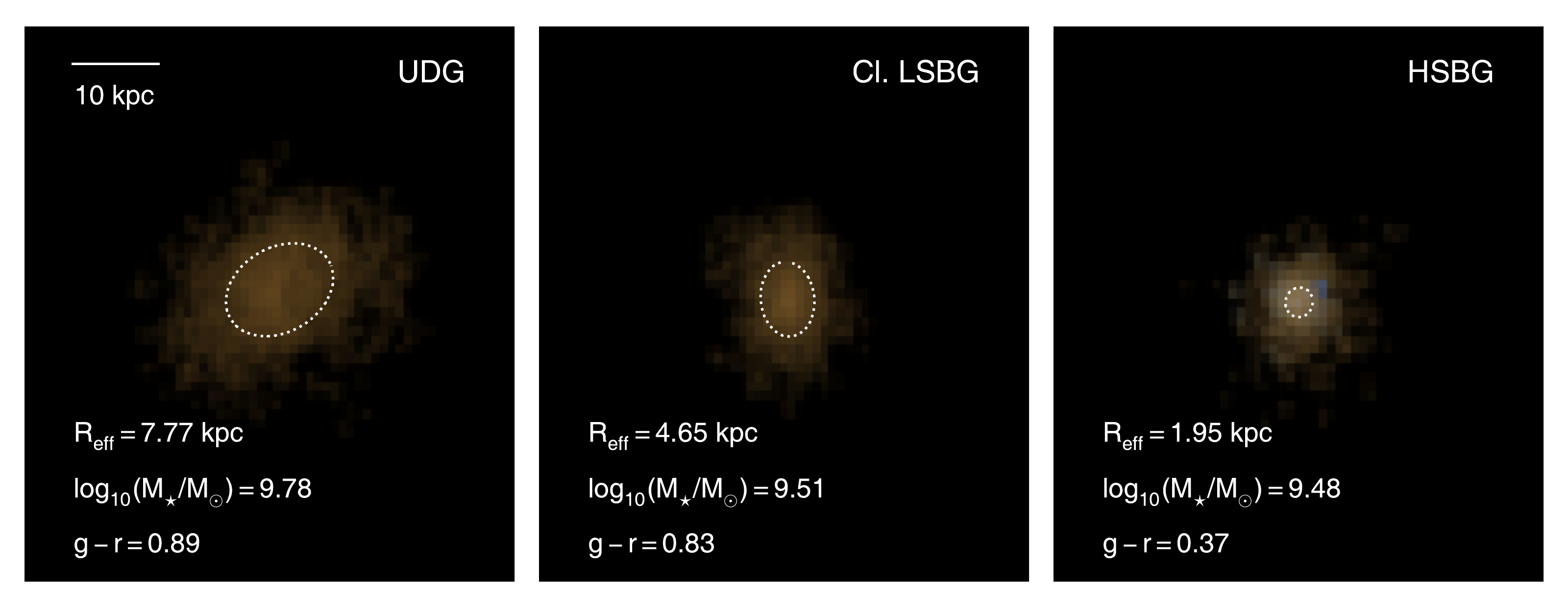}
    \caption{Example $g$ $r$ $i$ band false colour images of low-mass Horizon-AGN galaxies. The left, middle and right hand panels show a typical example of galaxies identified as UDGs, Cl. LSBGs and HSBGs respectively. The dotted white ellipses are isophotes which contain half of the galaxy's $r$ band flux. A common spatial scale (indicated in the top-left corner of the left-hand panel) is used for all three images.}
    \label{fig:mock_image}
\end{figure*}

\autoref{fig:mock_image} shows an example of a galaxy from our three populations, with the dashed ellipses indicating the apertures used to calculate the effective surface-brightness. \autoref{fig:radii_compare} shows the effective radii and stellar masses of a random selection of Horizon-AGN galaxies that fall into each of the three categories described above. For comparison, we show observed galaxy populations in the nearby Universe. We note that, even for relatively low stellar masses ($M_{\star}\sim 10^{8.5}$M$_{\odot}$), the LSBGs in Horizon-AGN are well-resolved enough to recover accurate effective radii. However, depending on the implementation of sub-grid physics (e.g. prescriptions for feedback), effects other than resolution can produce some systematic offset in galaxy sizes (see Appendix \ref{sec:resolution} for a full discussion).

Our simulated HSBGs fall along the same locus as observed HSBGs and dwarf ellipticals from \citet{Cappellari2011} and \citet{Dabringhausen2013}. Although the mass resolution of Horizon-AGN ($2\times$10$^{8}$~M$_{\odot}$) does not allow us to probe the stellar mass regime where the majority of UDGs have been discovered observationally, many observed UDGs from e.g. \citet{vanDokkum2015}, \citet{Mihos2015} and \citet{Yagi2016} that are massive enough do occupy the same region in parameter space as their model counterparts. Furthermore, as we describe in Appendix \ref{sec:UDG_MF}, while past observational studies are dominated by low-mass LSBGs, this is largely due to the small volumes probed in these works. These small volumes do not preclude the existence of massive LSBGs in new and forthcoming deep-wide surveys. Indeed, some massive LSBGs, such as Malin 1 and UGC 1382, are already known (see Figure \ref{fig:radii_compare} below), although the small observational volumes probed so far mean that such objects are rare in current (and past) datasets.

\begin{figure}
	\centering
    \includegraphics[width=0.4\textwidth]{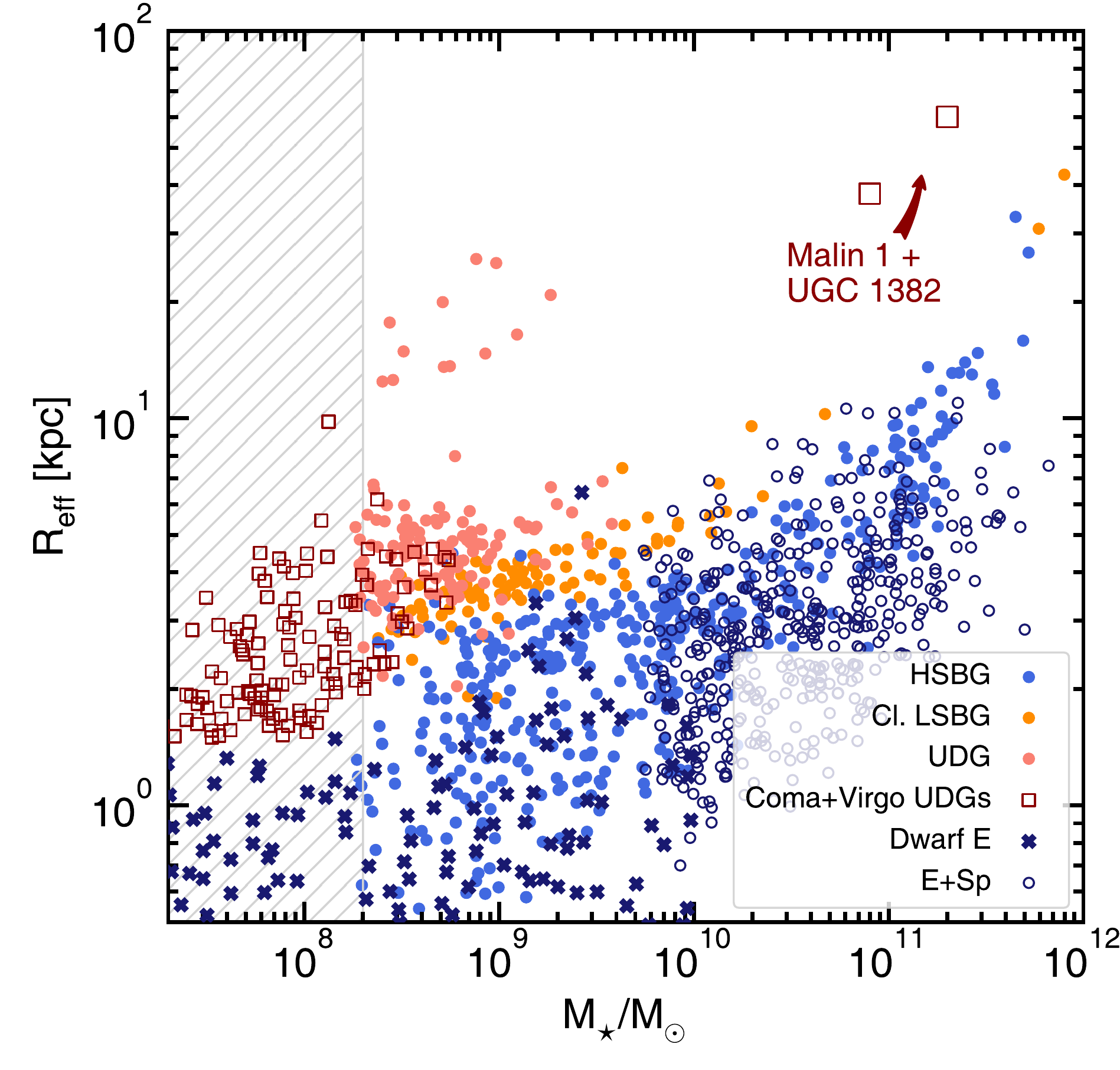}
    \caption{Effective radius ($R_{\rm eff}$) vs. stellar mass ($M_{\star}$) for a random selection of galaxies from Horizon-AGN, compared to observed galaxies in the local Universe. Blue, orange and red filled circles show simulated galaxies identified as HSBGs, Cl. LSBGs and UDGs respectively. Open red squares show UDGs from the Coma and Virgo clusters \citep{vanDokkum2015,Mihos2015,Yagi2016,Gu2018}. Dark blue crosses indicate dwarf ellipticals, and open dark blue circles indicate high mass ellipticals and spirals, from \citet{Dabringhausen2013} and \citet{Cappellari2011}. Large open red squares show the giant LSBGs Malin 1 and UGC 1382 \citep{Bothun1987,Hagen2016}. The grey hatched region falls below the mass resolution limit of the simulation. 
    }
    \label{fig:radii_compare}
\end{figure}


\section{The low-surface-brightness Universe at the present-day}
\label{sec:properties}
We begin by studying the contributions of LSBGs to the number, mass and luminosity densities at low redshift (Section \ref{sec:contribution}).
We then compare key properties of LSBGs (effective radii, local environments, dark matter fractions, stellar ages and star-formation histories) to their HSB counterparts at $z\sim0$ (Section \ref{sec:present_day_props}). 

\subsection{Contribution of LSBGs to the local number, stellar mass and luminosity densities}
\label{sec:contribution}

\begin{figure}
	\centering
    \includegraphics[width=0.45\textwidth]{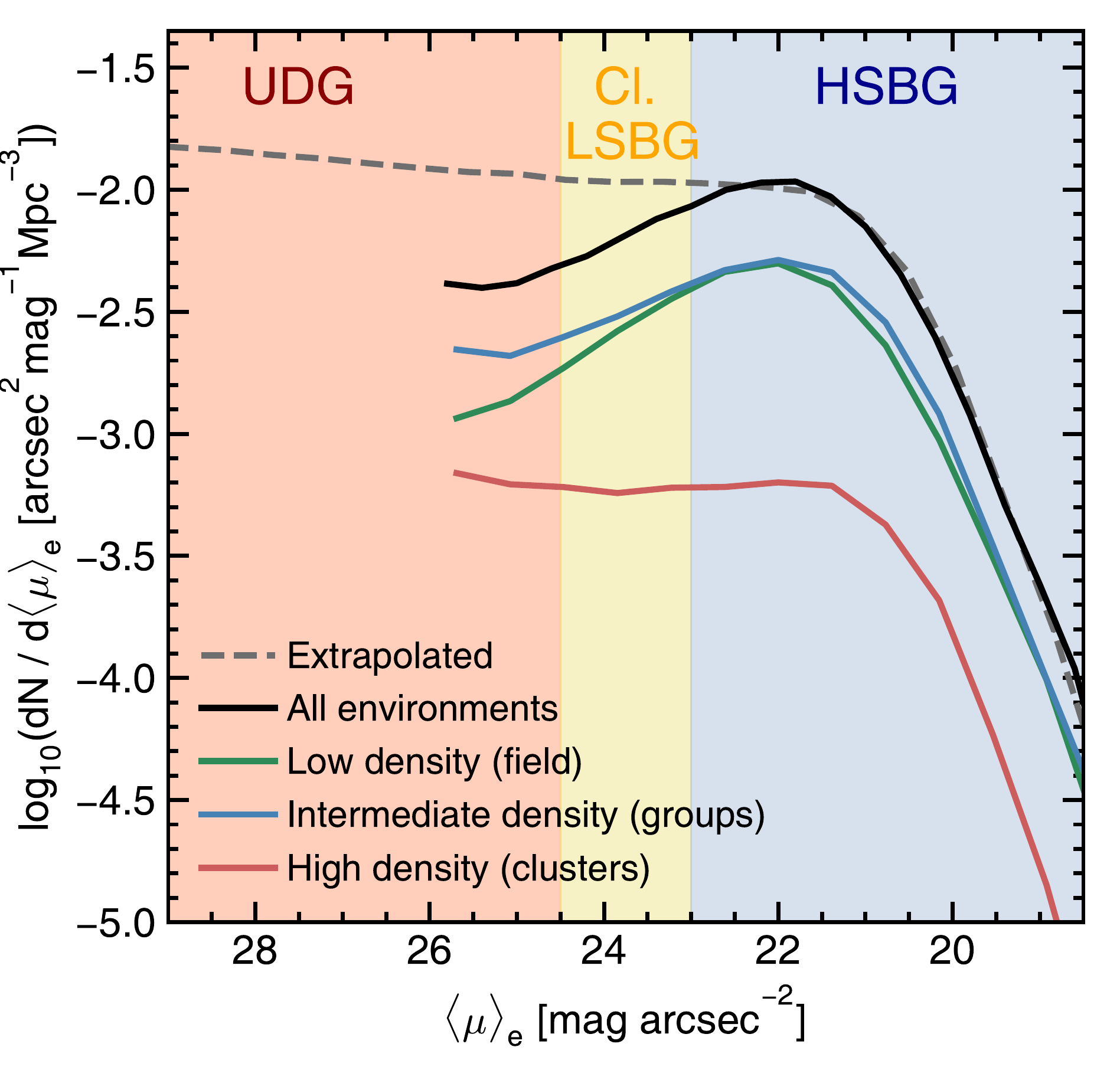}
    \caption{The surface-brightness function, showing the number density of galaxies as a function of their $r$ band effective surface-brightness at $z=0$. We show separate curves for low (green), intermediate (blue) and high (red) density environments and all environments (black). Low, intermediate and high density environments roughly correspond to the field, groups and clusters respectively. The dashed line shows the surface-brightness function that is produced by extrapolating the stellar mass function down to $10^{7}~$M$_{\odot}$, as described in Appendix \ref{sec:appendix_SB}.}
    \label{fig:sbfunction}
\end{figure}

\autoref{fig:sbfunction} shows the surface-brightness function in Horizon-AGN i.e. the number density of galaxies as a function of $\langle \mu \rangle _{e}$ in the $r$ band (solid line). The coloured lines indicate galaxies in different environments. Following \citet{Martin2018_progenitors}, environment is defined according to the 3-D local number density of objects around each galaxy. Local density is calculated using an adaptive kernel density estimation method\footnote{The sharpness of the kernel used for multivariate density estimation is responsive to the local density of the region, such that the error between the density estimate and the true density is minimised.} \citep{Breiman1977,Ferdosi2011,Martin2018_progenitors}. The density estimate takes into account all galaxies above $2\times10^{8}$M$_{\odot}$.

Galaxies are then split into three bins in local density: `low density' corresponds to galaxies in the 0th -- 40th density percentiles, `intermediate density' correspond to the 40th -- 90th percentiles and `high density' corresponds to galaxies in the 90th -- 100th percentiles. The low, intermediate and high density bins roughly correspond to the field, groups and clusters (see \citet{Martin2018_progenitors} for more details). Typically, galaxies in the intermediate and high density bins are found in halos with masses $10^{12.5}<M_{halo}<10^{13.5}$~M$_{\odot}$ and $M_{halo}>13.5$~M$_{\odot}$ respectively. In the low density bin, most galaxies ($\sim 70$ per cent) are isolated (i.e. they are not a sub-halo of a larger halo). Of the galaxies in the low-density bin that are satellites, typical halo masses are $\sim 10^{12}$~M$_{\odot}$. We note that there is no perfect correspondence between number density and halo mass - for example, at fixed density, UDGs are typically hosted by haloes that are $\sim$0.5 dex more massive than HSBGs.

Since we do not consider objects with stellar masses below $2\times10^{8}$~M$_{\odot}$, the predicted surface-brightness function starts becoming incomplete as we approach this limit. In order to account for this when estimating the LSBG contribution to the local number, mass and luminosity densities, we extrapolate the galaxy stellar-mass function down to $10^{7}$M$_{\odot}$ (as described in Appendix \ref{sec:appendix_SB}). The dashed black line indicates the corresponding extrapolated surface-brightness function, using a combination of surface-brightnesses drawn from the extrapolated fits (between $10^{7}$~M$_{\odot}$ and $10^{9}$~M$_{\odot}$) and the raw simulation data ($10^{8}$~M$_{\odot}$ to $10^{12}$~M$_{\odot}$) - see Appendix \ref{sec:appendix_SB} for more details. 

\begin{table*}
\centering
\caption{The frequency (col 2, 3) and number (col 4, 5) of LSBGs of different surface-brightnesses (in $r$ band mag~arcsec$^{-2}$) as a function of environment in the present-day universe in the Horizon-AGN simulation, for stellar masses greater than $2 \times 10^{8}$~M$_{\odot}$. The numbers in brackets indicate the corresponding fractions produced by extrapolating the stellar mass function down to $10^{7}~$M$_{\odot}$. The `low' (local number density in the 0th -- 40th density percentile), `intermediate' (40th -- 90th density percentile) and `high' (90th -- 100th density percentile) density bins correspond roughly to field, group and cluster environments respectively \citep[see][]{Martin2018_progenitors}.}
\label{table_LSB}
\begin{tabular}{lllll}
\toprule
      & $f$($24.5 > \langle \mu \rangle _{e} > 23$) & $f$($\langle \mu \rangle _{e} > 24.5$) & $N$($24.5 > \langle \mu \rangle _{e} > 23$) & $N$($\langle \mu \rangle _{e} > 24.5$) \\ \midrule
Low density (Field) & 0.23 (0.09)    & 0.18 (0.77) & 10760  & 5634  \\
Intermediate density (Groups) & 0.21 (0.09)   & 0.27 (0.74) & 12691 & 12119 \\
High density (Clusters) & 0.19 (0.07) & 0.46 (0.83) & 2310  & 4572 \\ \bottomrule
\end{tabular}
\end{table*}

\begin{table}
\centering
\caption{The fraction of the local stellar mass, luminosity and number density budget contributed by galaxies of different $r$ band surface-brightnesses (in units of mag~arcsec$^{-2}$) in the Horizon-AGN simulation, for stellar masses greater than $2 \times 10^{8}$~M$_{\odot}$. The numbers in brackets indicate the corresponding fractions produced by extrapolating the stellar mass function down to $10^{7}~$M$_{\odot}$.}
\label{table_n}
\begin{tabular}{llll}
\toprule
      & $\langle \mu \rangle _{e} < 23$ & $24.5 > \langle \mu \rangle _{e} > 23$ & $\langle \mu \rangle _{e} > 24.5$ \\ \midrule
$f_{M_{\star}}$ & 0.924 (0.902)    & 0.059 (0.067) & 0.014 (0.030)   \\
$f_{L}$ & 0.939 (0.892)   & 0.049 (0.071) & 0.012 (0.037)  \\
$f_{N}$ & 0.534 (0.145) & 0.214 (0.093) & 0.252 (0.762)   \\ \bottomrule
\end{tabular}
\end{table}

Table \ref{table_LSB} summarises the absolute numbers and number fractions of HSBGs and LSBGs in the present-day Universe, as a function of local environment. The numbers in brackets indicate the corresponding values using the extrapolated mass function. For galaxies with stellar masses above the resolution limit of the simulation ($2 \times$10$^{8}$~M$_{\odot}$), LSBGs account for a significant fraction (over half) of the galaxy population in clusters and a significant minority (40-50 per cent) of objects in low-density environments (groups and the field). 

However, for stellar masses down to $10^{7}$~M$_{\odot}$, LSBGs are expected to overwhelmingly dominate the number density of the Universe, accounting for more than 70 per cent of galaxies, irrespective of the local environment being considered. It is worth noting that the \textit{absolute} numbers of LSBGs across different environments (see col 4, 5 in \autoref{table_LSB}) are similar. For example, the absolute numbers of UDGs in the Horizon-AGN volume that inhabit the field and those that inhabit clusters are predicted to be almost the same (col 5 in \autoref{table_LSB}). This is because, although the LSBG fraction is higher in clusters, the total number of galaxies that inhabit low-density environments (e.g. the field) is much larger.

\autoref{table_n} summarises the contribution of HSBGs and LSBGs to the mass, luminosity and number density budgets of the local Universe. For galaxies with stellar masses greater than $2 \times 10^{8}$~M$_{\odot}$, LSBGs contribute around 47 per cent of the total number density and make a small but non-negligible contribution to the stellar mass (7.5 per cent) and luminosity (6 per cent) budgets. These numbers change to 85 (number density), 10 (mass density) and 11 (luminosity density) per cent respectively, when we extrapolate down to a stellar mass of $10^{7}$~M$_{\odot}$. Although they account for the majority of the number density budget (76 per cent with extrapolation to $10^{7}$~M$_{\odot}$) at low redshift, the extreme end of the LSBG population, i.e. UDGs ($\langle \mu \rangle _{e} > 24.5$), account for only a small fraction of the mass or luminosity budget (less than 4 per cent in both cases).

We note that the extrapolated quantities above are used only to estimate the overall contribution of LSBGs to the number, stellar mass and luminosity density down to a stellar mass of $10^{7}$~M$_{\odot}$. For the rest of the analysis that follows, we use galaxies that are actually resolved in the simulation and for which the minimum stellar mass is $2\times 10^{8}$~M$_{\odot}$. 


\subsection{Properties of LSB galaxies at the present day}
\label{sec:present_day_props}

\begin{figure}
	\centering
    \includegraphics[width=0.45\textwidth]{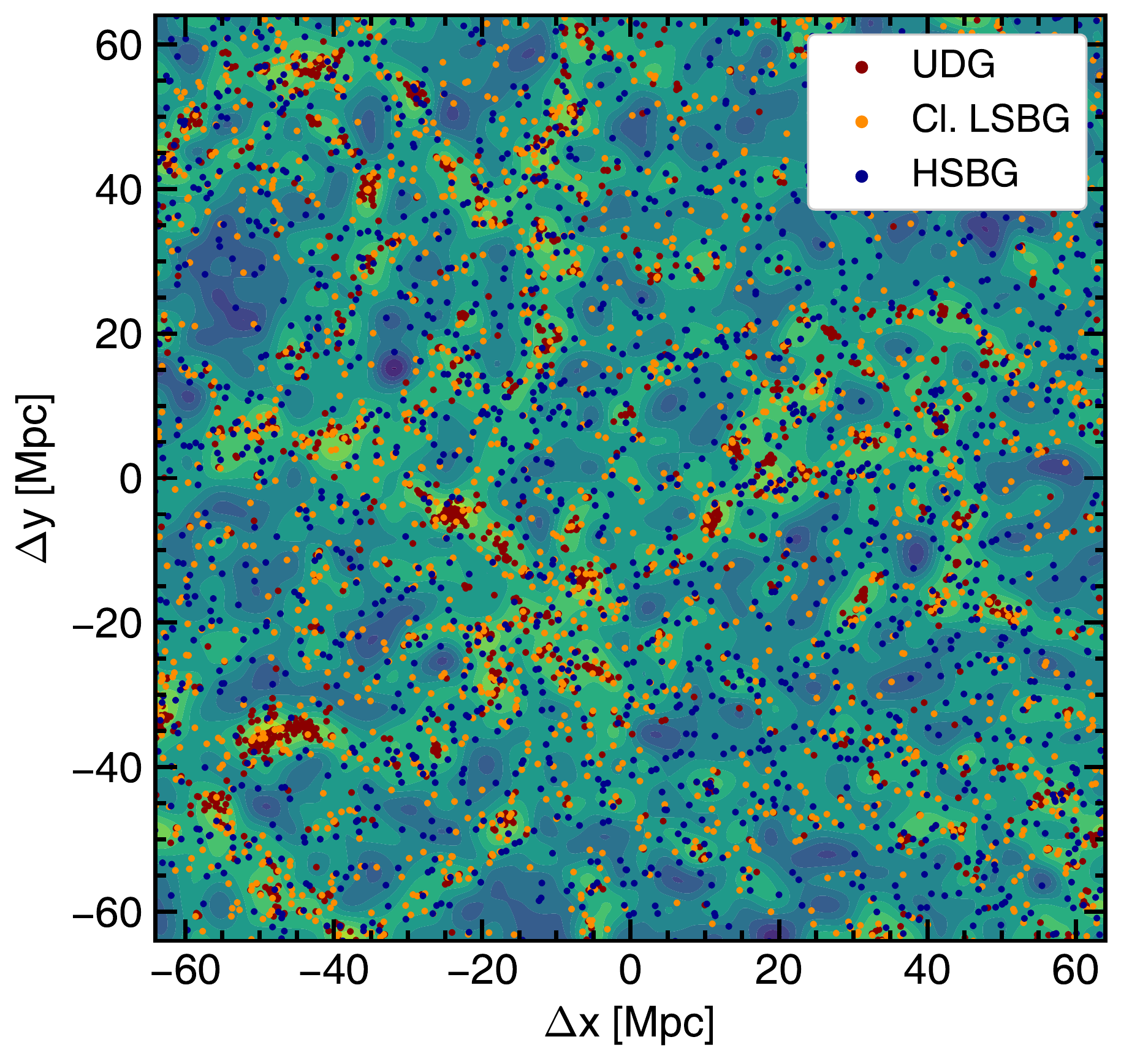}
    \caption{The spatial distribution, within the cosmic web, of the UDG, Cl. LSBG and HSBG populations. Red, orange and blue coloured points show the positions of individual UDGs, Cl. LSBGs and HSBGs. Contours indicate the surface density, calculated using all objects in the simulation, with lighter colours indicating higher densities.}
    \label{fig:cosmic_web}
\end{figure}

\begin{figure*}
	\centering
    \includegraphics[width=0.95\textwidth]{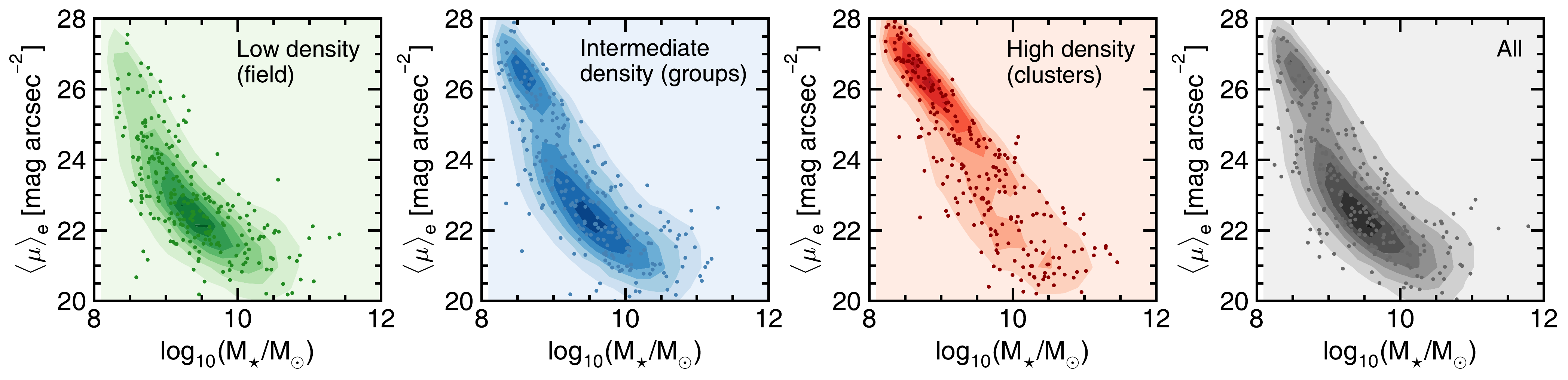}
    \caption{Contour plots showing the number density of galaxies as a function of effective surface-brightness ($\langle \mu \rangle _{e}$) and stellar mass ($M_{\star}$) at $z=0.06$, split by local environment. Low, intermediate and high density environments correspond roughly to field, group and cluster environments respectively. A random sample of galaxies is plotted using points in each panel. The right-hand panel shows the same for all galaxies in the simulation.}
    \label{fig:mu_vs_mass}
\end{figure*}

\begin{figure*}
	\centering
	\subfigure[]{\includegraphics[width=0.33\textwidth]{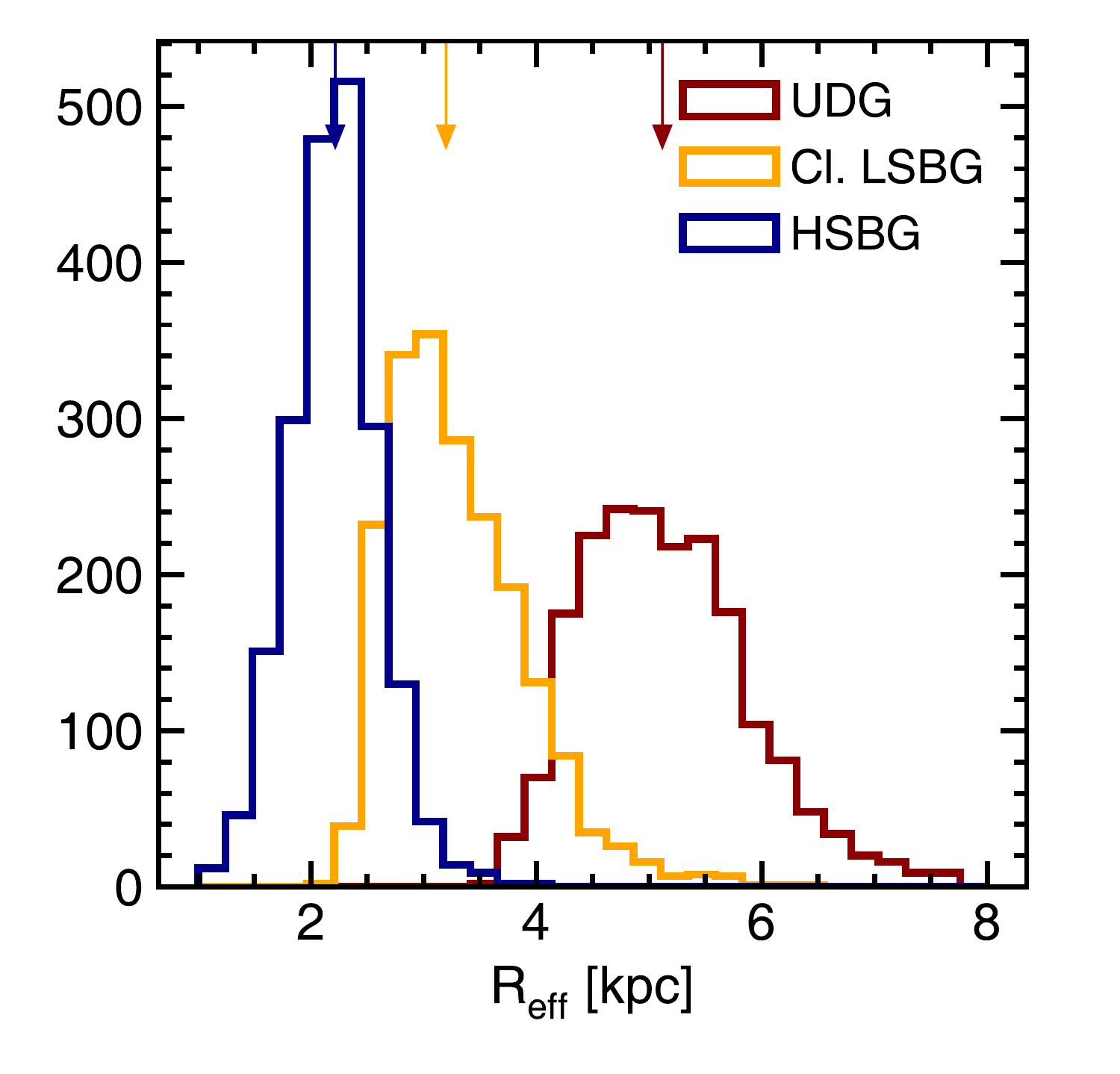}\label{fig:a}}
   	\subfigure[]{\includegraphics[width=0.33\textwidth]{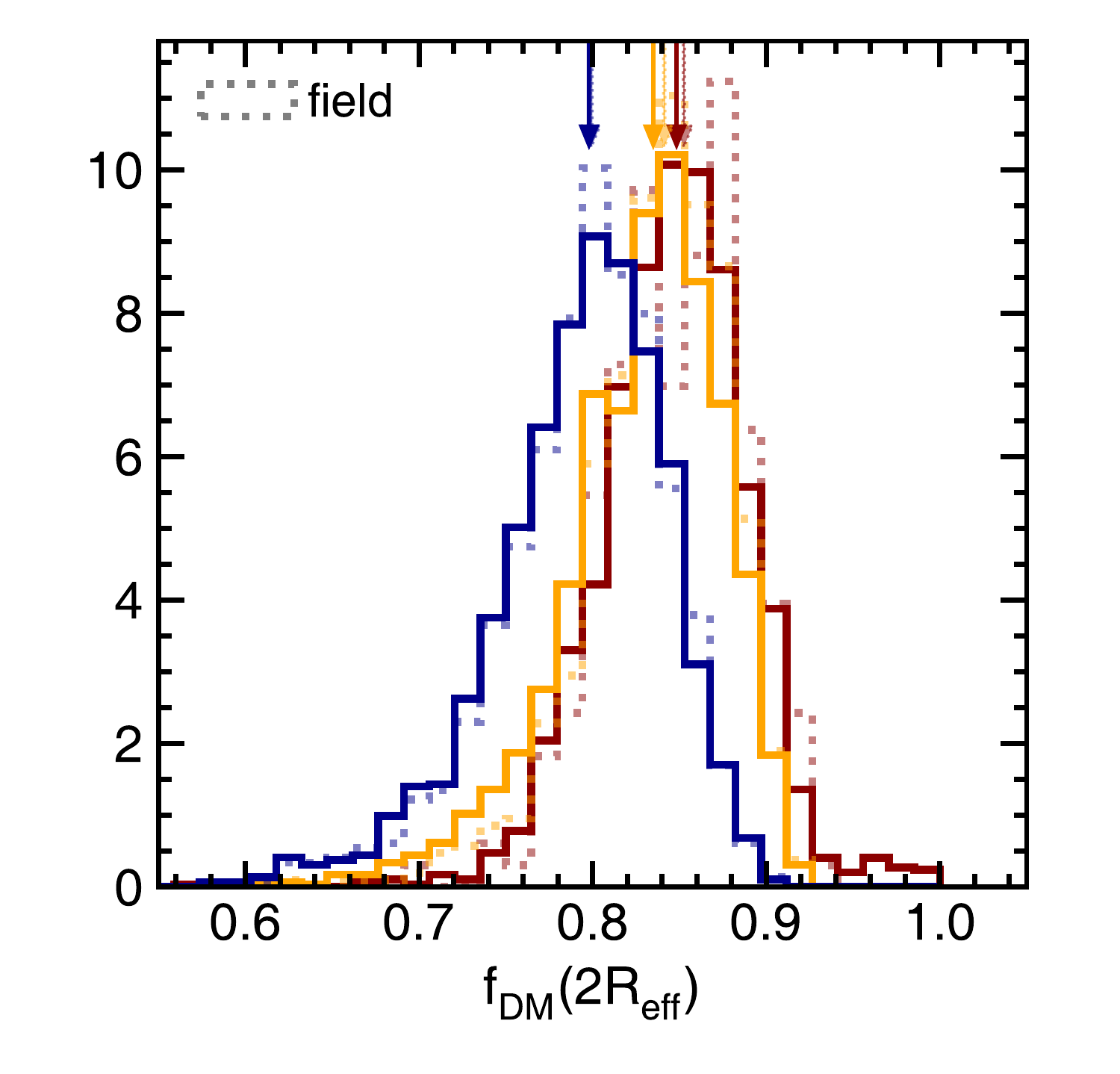}\label{fig:b}}
    \subfigure[]{\includegraphics[width=0.33\textwidth]{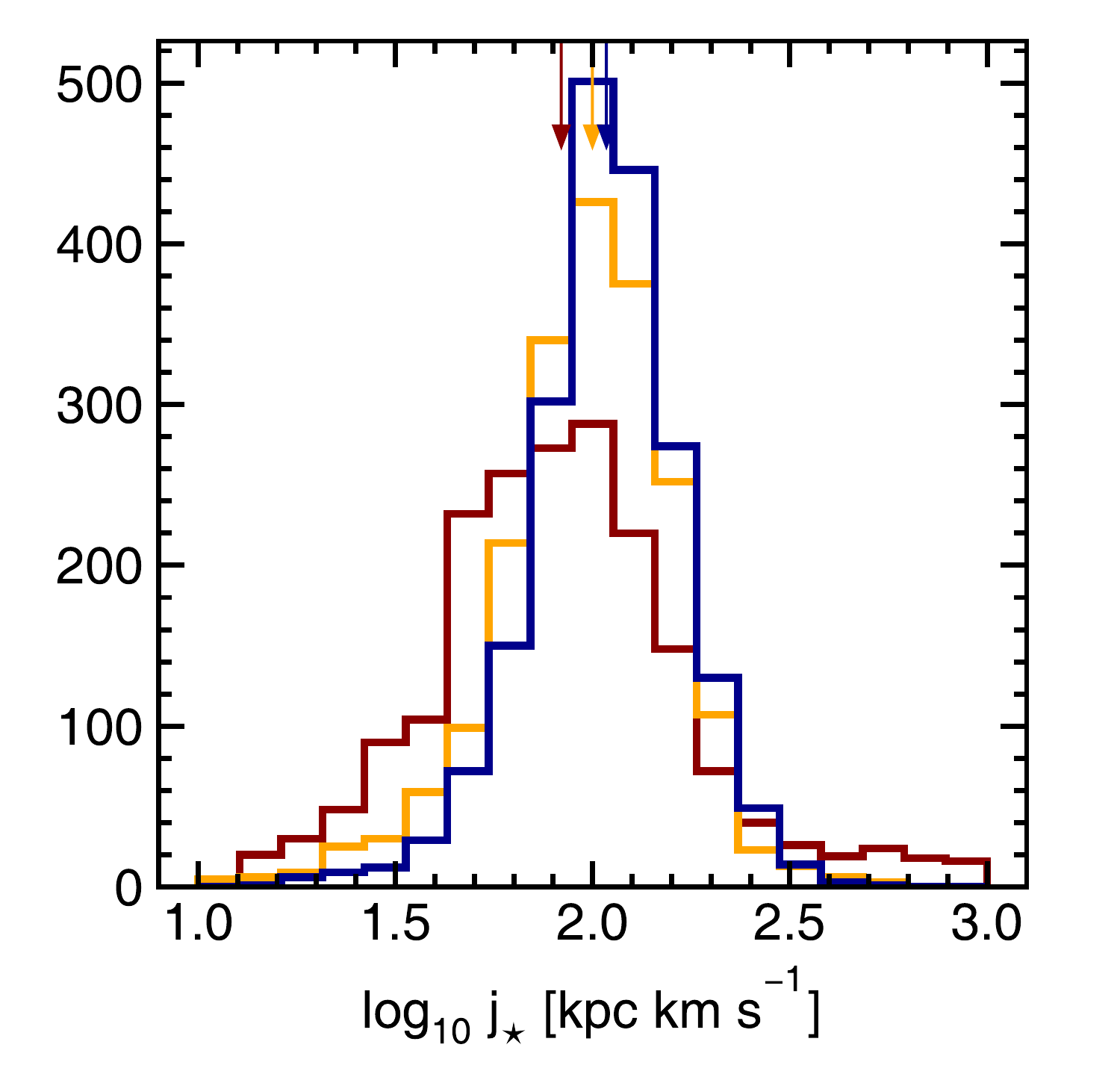}\label{fig:c}}
    \subfigure[]{\includegraphics[width=0.33\textwidth]{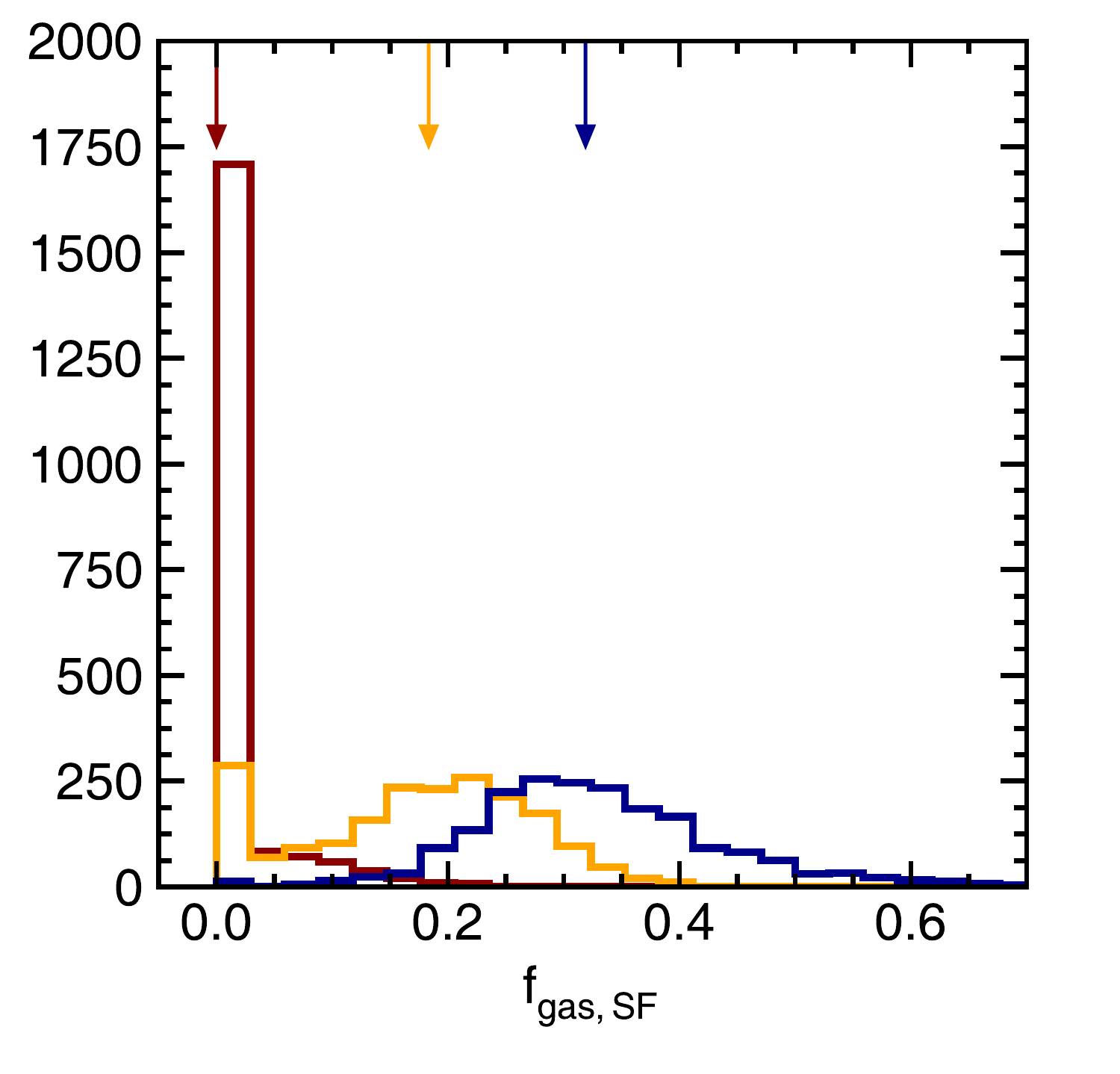}\label{fig:d}}
   	\subfigure[]{\includegraphics[width=0.33\textwidth]{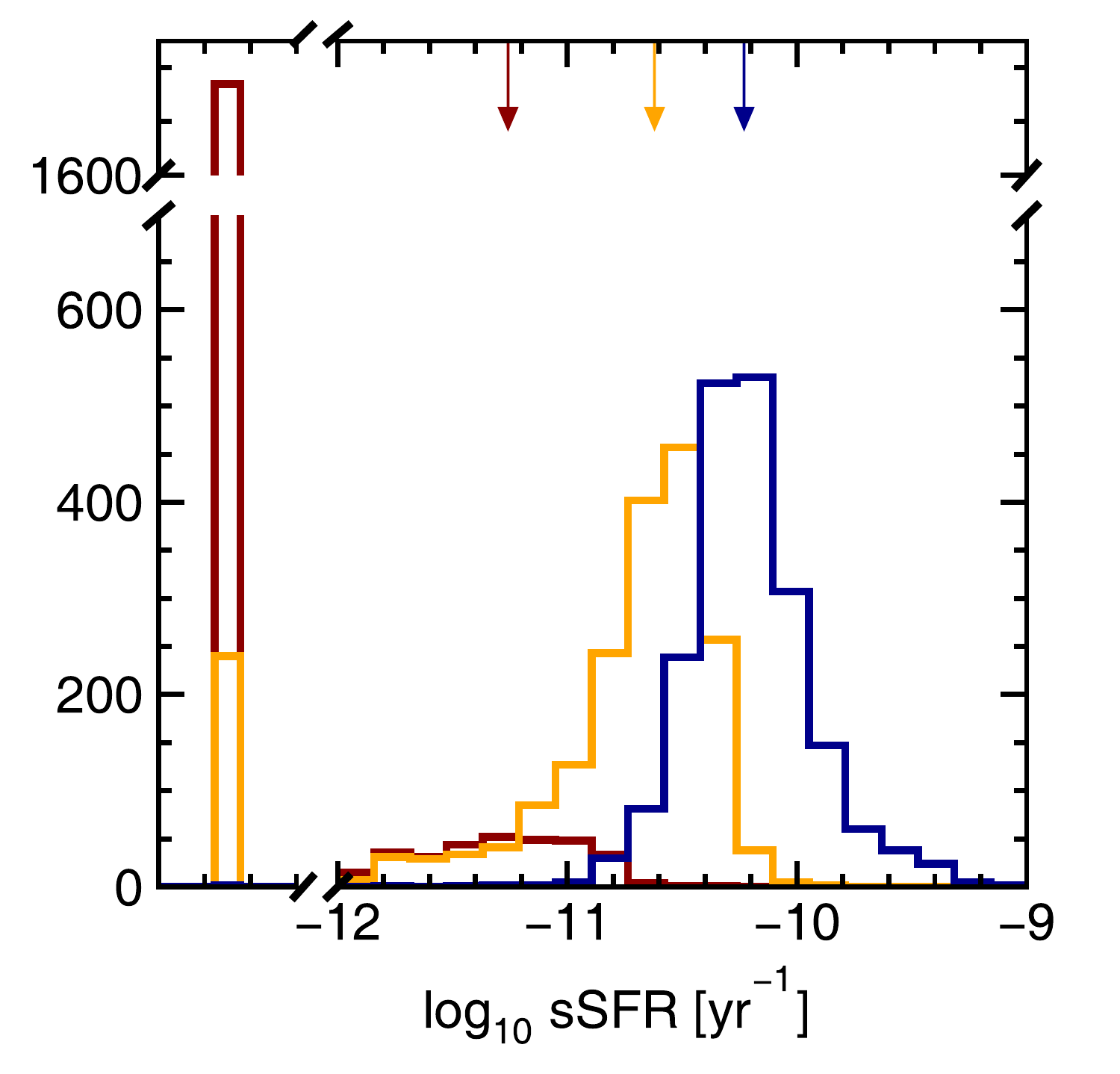}\label{fig:e}}
  	\subfigure[]{\includegraphics[width=0.33\textwidth]{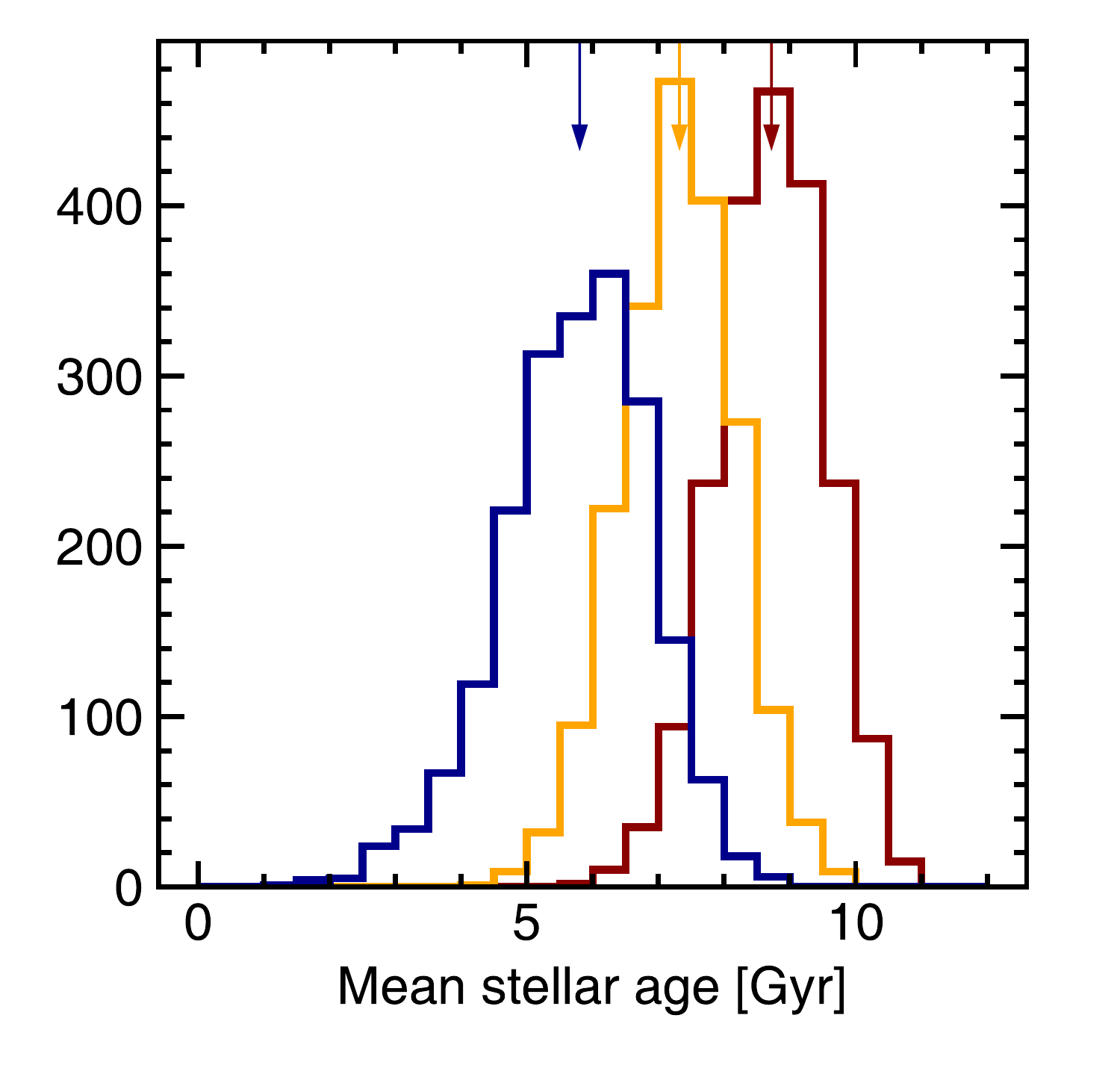}\label{fig:f}}
    \caption{Properties of present-day UDGs and Cl. LSBGs compared to those of HSBGs at $z\sim0$ (red, orange and blue histograms respectively). Coloured arrows indicate the median values of each population. Fainter dashed arrows indicate the median values for field populations only. Panels are as follows: \textbf{(a)} effective radius measured using the stellar distribution of each galaxy \textbf{(b)} the dark-matter fraction ($M_{\rm DM}/(M_{\rm DM}+M_{\star})$) measured within the central $2~R_{\mathrm{eff}}$, for all galaxies (solid line) and galaxies in the field (dotted lines) - note that the histograms are normalised in order to easily compare the two populations \textbf{(c)} stellar specific angular momentum \textbf{(d)} star-forming ($\rho_{gas \ cell}>0.1\mathrm{H}~\mathrm{cm}^{-3}$) gas fraction measured within $2~R_{\mathrm{eff}}$ ($M_{\rm gas,SF}/(M_{\star}+M_{\rm gas,SF})$) \textbf{(e)} specific star formation rate - the bar to the left indicates galaxies with sSFRs of 0 \textbf{(f)} mass-weighted mean stellar age ($(\sum_{i} \  \mathrm{age}_{i} m_{\star,i})/\sum_{i} \ m_{\star,i}$).}
    \label{fig:properties}
\end{figure*}

In this section we compare the properties of LSBGs to their HSB counterparts at the present day. \autoref{fig:cosmic_web} shows the spatial distribution of a random selection of UDGs, Cl. LSBGs and HSBGs within the cosmic web. The contours indicate the surface density of galaxies calculated using all objects in the simulation. Although they appear to exist preferentially in regions of high number density, many UDGs occur in regions of much lower density. On the other hand HSBGs appear to be essentially uniformly distributed.

In \autoref{fig:mu_vs_mass}, we show contour plots of the distribution of galaxies as a function of $r$ band effective surface-brightness, $\langle \mu \rangle _{e}$, and stellar mass at $z=0$, split by local environment. The histogram for all galaxies across all environments is bimodal. However, the bimodality varies strongly with environment. At a given stellar mass, the frequency of LSBGs is higher in denser environments. While in the field most galaxies inhabit the HSB peak, the LSB peak progressively dominates as we move to higher density environments. Indeed, for low-mass galaxies, in clusters, the LSB peak overwhelmingly dominates the population (this is partly the reason why much of the UDG literature has been focussed on clusters to date).

Since the frequency of LSBGs is a strong function of stellar mass (see e.g. \autoref{fig:mu_vs_mass}), we first construct mass-matched samples of 2000 HSBGs, LSBGs and UDGs with stellar masses between $10^{9} $M$_{\odot}$ and $10^{10} $M$_{\odot}$, each of which have the same  distribution in stellar mass. Due to the shape of the UDG mass function (see Appendix \ref{sec:UDG_MF}), the stellar mass distribution of our sample peaks close to $10^{9}$M$_{\odot}$ and declines such that $\sim95$ per cent of galaxies are less massive than $10^{9.5}$M$_{\odot}$. We then use these mass-matched samples to explore key properties of LSB systems -- effective radii, dark-matter fractions, specific angular momenta, gas densities, specific star formation rates and mean stellar ages. Note that the analysis presented in all subsequent sections, which explore how LSBG progenitors evolve with time, is also based on these mass-matched samples. 

\autoref{fig:properties} shows histograms of these properties. LSBGs have larger effective radii (panel (a)), with the mean effective radii of UDGs around 2.5 times larger than HSBGs. The dark-matter (DM) fractions in LSBGs and HSBGs (panel (b)) are similar, with the median value for  LSBGs predicted to be slightly ($\sim$ 5 per cent) higher than in HSBGs. The overwhelming majority of LSBGs are, therefore, not devoid of DM, nor do they have anomalously large DM fractions for their stellar mass. Contamination due to galaxies being embedded in more massive DM haloes does not appear to have a significant impact on the ratios shown -  when we restrict our sample to field galaxies only (dotted histograms), there is no difference in the median DM to stellar mass ratio. This suggests that high-DM-fraction UDGs (i.e. failed $L^{\star}$ galaxies) \citep[e.g.][]{vanDokkum2016,Beasley2016b} are extremely uncommon, at least in the stellar mass range we study here ($10^{9}$ M$_{\odot}$$<M_{\star}<$$10^{10}$ M$_{\odot}$). 

It is worth noting here that, while recent observations have suggested that at least some UDGs may have very low dark matter fractions (e.g. \citealt{vanDokkum2018}, but see \citealt{Laporte2018, Trujillo2018}), a small fraction of low mass DM-free galaxies can form naturally within the LCDM paradigm as tidal dwarf galaxies in galaxy mergers \citep[e.g.][]{Barnes1992,Okazaki2000, Bournaud06,Kaviraj2012}. However, mergers typically produce tidal dwarfs with very low stellar masses \citep{Kaviraj2012}, and the mass range that we consider ($M_{\star}>10^{9}$M$_{\odot}$) precludes significant numbers of these objects in our sample. It may not be surprising, therefore, that we do not find any evidence of UDGs with anomalously low DM fractions in Horizon-AGN, even if this were a significant channel for their production. 

The distribution of the stellar specific angular momenta (panel (c)) of LSBGs and HSBGs is similar, indicating that the formation of LSBGs, and UDGs in particular, is not primarily due to them being the high spin tail of the angular momentum distribution \citep[e.g.][]{Yozin2015,Amorisco2016,Amorisco2016b,Rong2017}. The LSBGs in this study typically have spins that are not significantly different from, or indeed, are slightly below, those seen in HSBGs \citep[see also][]{DiCintio2017,Chan2018}.

Finally, we consider quantities that trace the star formation properties of galaxies. Panel (d) shows the `star-forming' gas fraction, defined as the ratio of the gas mass that is dense enough to form stars ($\rho_{gas \ cell}>0.1\mathrm{H}~\mathrm{cm}^{-3}$) to the stellar mass ($M_{\rm gas, SF}/(M_{\rm gas, SF}+M_{\star})$), measured within the central 2~$R_{\mathrm{eff}}$\footnote{We note that calculating the gas fraction within a fixed radius does not alter our conclusions.}. Gas fractions in LSBGs are lower than those in their HSB counterparts. For example, the gas fractions of UDGs mostly lie around zero, with 4 out of 5 UDGs being completely devoid of star forming gas in their central 2~$R_{\mathrm{eff}}$. HSBGs, on the other hand, still retain fairly significant fractions of star-forming gas ($f_{\rm gas, SF} \sim 0.3$). UDGs that do contain some star-forming gas at the present day have median values that are around one sixth of this value. The lower gas fractions are reflected in lower specific star formation rates (sSFRs; panel (e)) and higher mass-weighted mean stellar ages ($\sum_{i} \  \mathrm{age}_{i} m_{\star,i})/\sum_{i} \ m_{\star,i}$; panel (f)) in LSB systems. For example, the sSFRs in UDGs are an order of magnitude lower than in HSBGs, when galaxies with zero sSFR (again, around 4 out of 5 UDGs) are neglected. The median age of UDG stellar populations is 9~Gyrs, 50 per cent older than their HSB counterparts. The large age differences between LSBGs and HSBGs indicates that the LSB nature of these systems must be partly driven by gas exhaustion at \textit{early} epochs and consequently a more quiescent recent star history. 

We note here that the production of UDGs may be too efficient in clusters leading to quenched HSB galaxies being relatively unrepresented. Additionally, since the quenched fraction (especially at low redshift) is somewhat inconsistent with observations, and produces an offset in the star formation main sequence between the observed and theoretical populations in low-mass galaxies \citep[e.g. see][]{Kaviraj2017}, this may lead to relatively diffuse HSB or LSB galaxies becoming UDGs due to fading stellar populations.


\begin{figure}
	\centering
    \subfigure{\includegraphics[width=0.45\textwidth]{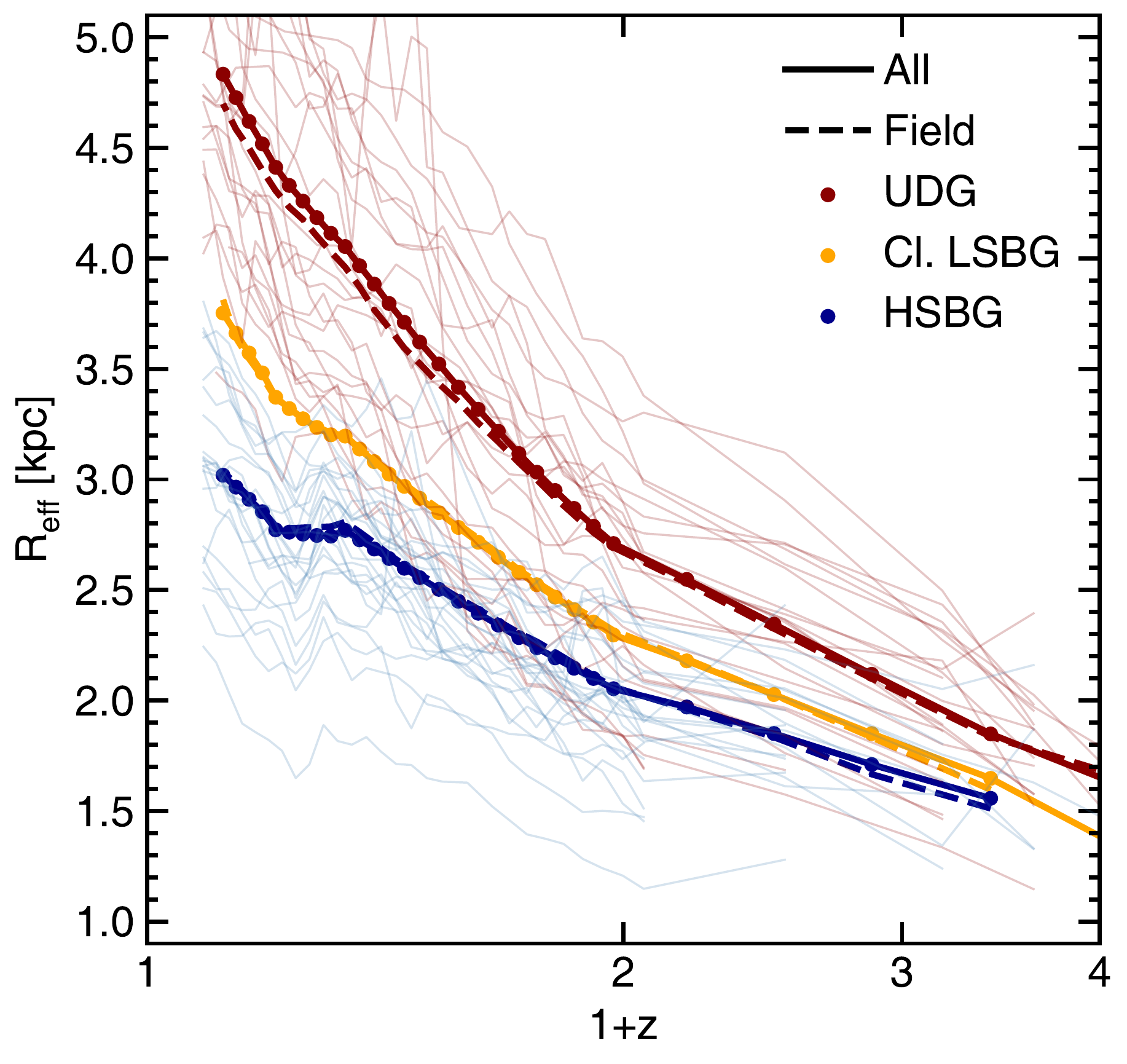}}\label{fig:a}
    \subfigure{\includegraphics[width=0.45\textwidth]{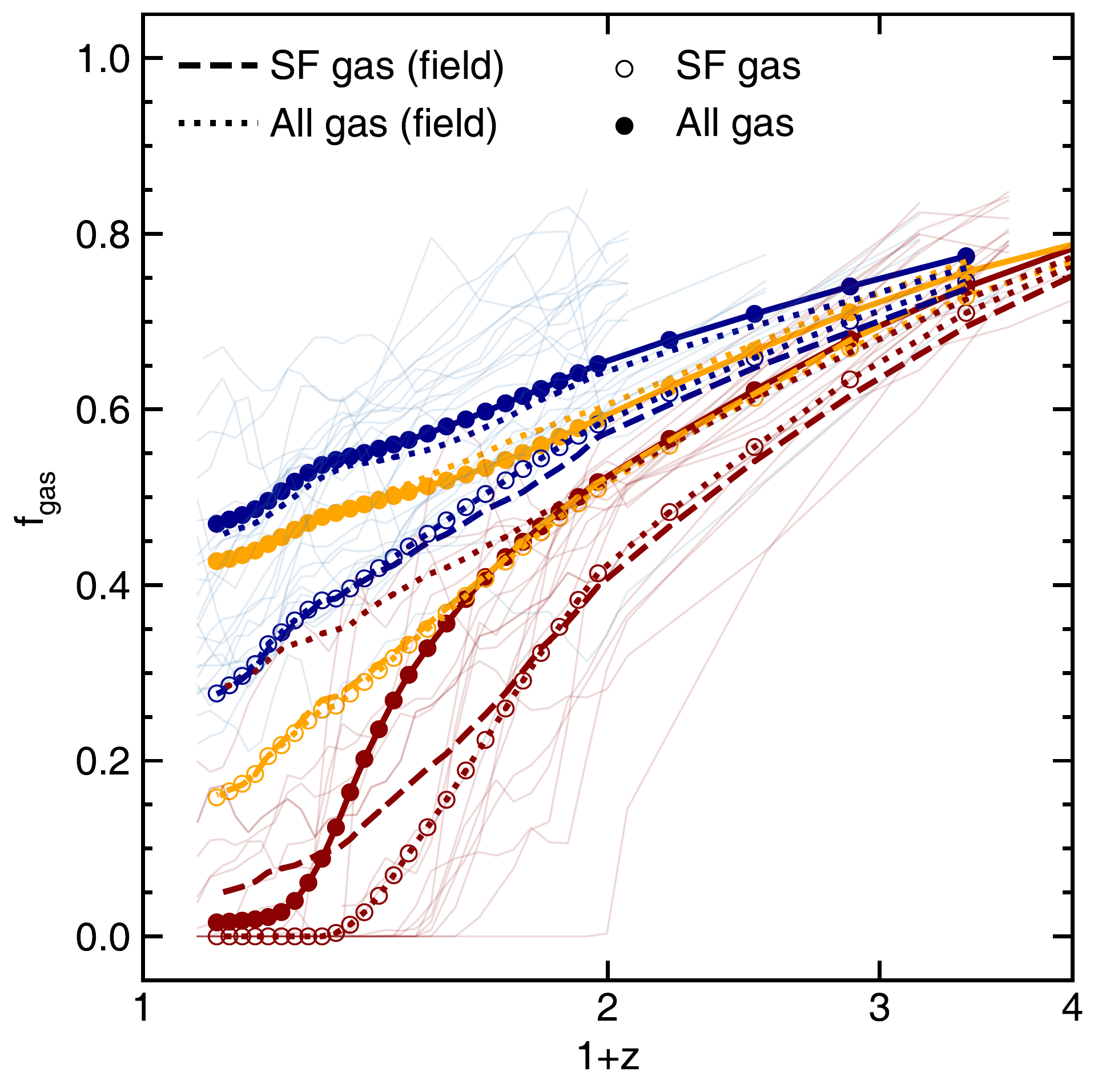}}\label{fig:b}
    \caption{\textbf{Top}: The redshift evolution of effective radii. Solid blue, orange and red coloured points show the redshift evolution of the median effective radii of the HSBG, Cl. LSBG and UDG populations respectively. The dashed lines shows the evolution of $R_{\mathrm{eff}}$ for galaxies in the field only. Note that rather than attempt to emulate observational methods to calculate $R_{\rm eff}$ at all redshifts, we instead use the average projected half mass radius in the $xy$, $xz$ and $yz$ planes here \textbf{Bottom}: The redshift evolution of the median gas fraction, defined as $M_{gas}/(M_{gas}+M_{\star})$, for total gas (solid coloured points) and star-forming gas (open coloured points) for the HSBG, LSBG and UDG populations. Dashed and dotted lines without points show the evolution of $f_{gas}$ for total gas and star-forming gas respectively for field galaxies only. Pale red and blue lines show tracks for the effective radii and star-forming gas fractions of a random sample of individual UDGs and HSBGs.}
    \label{fig:reff_fgas_evo}
\end{figure}

\section{Redshift evolution of LSBG progenitors}
\label{sec:evolution}

We proceed by comparing the redshift evolution of LSBG progenitors to the progenitors of their HSB counterparts. We focus, in particular, on the evolution of the effective radii and gas fractions which, as we showed in Section \ref{sec:properties}, are the quantities in which HSBGs and LSBGs diverge the most at the present day. We note that, since we restrict our study to resolved progenitors, There is some incompleteness in the sample at higher redshifts. This is due to the limit of 50 particles that we impose on the structure finder (see Section \ref{sec:merger_trees}), which renders their merger trees incomplete after galaxies fall below this level. The merger trees of the LSBG and UDG samples are largely complete after $z=2$ (80 and 90 per cent of main progenitors at $z=2$ are accounted for respectively) owing to their rapid assembly histories (see Section \ref{sec:SN} below). For the HSBG sample, around 60 per cent of main progenitors are accounted for at $z=2$ (rising to 100 per cent by $z=1$), which may lead to the exclusion of more slowly evolving HSBGs before $z=1$.

\subsection{Gas fractions and effective radii}
The top panel of \autoref{fig:reff_fgas_evo} describes how the effective radii of the main progenitors of LSBGs and HSBGs evolve as a function of redshift. LSBGs, and UDGs in particular, are consistently larger, on average, than their HSB counterparts. Furthermore, after $z\sim1$, the rate of increase in the effective radii of UDGs is higher  compared to that in HSBGs. \autoref{fig:reff_fgas_evo} shows that the evolution of the effective radii of all galaxy populations is not abrupt but relatively steady and smooth with time, both galaxy by galaxy (pale lines) and as a population. \textit{It is unlikely, therefore, that the large radii of LSBGs today are the result of single, violent events at early epochs.} 

The dashed lines indicate the evolution of galaxy populations in field environments only. As the dashed red line indicates, the evolution of the effective radii of field UDGs proceeds almost identically to the general UDG population, despite the frequency of UDGs being higher in very dense (cluster) environments. This implies that the process(es) that produce the large sizes seen in today's UDGs are the same regardless of environment (although they may occur less frequently in the field). In particular, the principal mechanism for UDG production is not cluster-specific i.e. galaxies do not have to inhabit cluster environments to be the progenitors of UDGs at the present day. 

The bottom panel of \autoref{fig:reff_fgas_evo} describes how the gas fractions of the main progenitors of LSBGs and HSBGs evolve as a function of redshift. While the gas fractions are similar for progenitors of all galaxies at high redshift, they begin to diverge rapidly at $z\sim2$. The \textit{total} gas fractions in HSBGs and Cl. LSBGs evolve similarly to each other and both HSBGs and Cl. LSBGs retain relatively high total gas fractions at $z=0$. In these populations the reduction in the average gas fraction is primarily due to gas being converted into stars, rather than as a result of gas being expelled from the galaxy. As we will also show in Section \ref{sec:processes}, most of this gas in HSBGs that is turned into stars is not replenished, at least after $z=1$, so that the decreasing gas fractions are due the gas masses steadily decreasing rather than the stellar masses simply increasing in these galaxies. There is a more pronounced divergence in terms of the fraction of \textit{star-forming} gas. By $z=0$, Cl. LSBGs have significantly lower fractions of star-forming gas compared to their HSB counterparts.

While Cl. LSBGs and HSBGs retain relatively significant reservoirs of gas as they evolve, the same is not true of UDGs. By $z=0.5$, the majority of UDGs have lost almost all of their star-forming gas, essentially terminating star formation, and by $z=0.25$, the majority of UDGs have been almost completely stripped of all of their gas. In around half of the cases, the gas fractions of the main progenitors of UDGs do not evolve linearly with time. Instead they undergo a phase of rapid gas loss lasting a few Gyrs around $z \sim 0.5$, which significantly reduces their gas content towards the present day.

The evolution of UDGs in field environments (dotted red lines) is slightly different from that of the global UDG population. There is no phase of rapid gas stripping and both the total and star-forming gas fractions in field UDGs evolve with a similar pattern to their HSB counterparts, albeit much more rapidly. Ultimately, the rate of gas heating is intense enough that the  \textit{star-forming} gas fraction is still reduced to similar levels to the wider UDG population ($<5$ per cent by $z=0$) by the present day. Note that the loss of star-forming gas is not due to gas being physically removed (i.e. gas stripping), since field UDGs retain fairly high \textit{total} gas fractions ($\sim 30$ per cent on average, as shown by the dotted red line).

The complete removal of gas is, therefore, not a necessary criterion for the production of UDGs. Gas heating alone produces the low star-forming gas fractions in these objects (regardless of local environment), without requiring that the gas be removed from the galaxy entirely. Whether UDGs have had their gas entirely removed or have just undergone heating makes little difference to their stellar populations at $z=0$. The median stellar ages of UDGs that have been completely stripped of gas, and those in field environments that have only undergone heating, are 8.7~Gyrs and 8.5~Gyrs respectively. In Section \ref{sec:processes}, we explore the processes that lead to the removal or heating of gas in the LSBG population. 

Note that although some galaxies ($\sim 30$ per cent) in the low-density `field' environment are actually satellites of another galaxy, the average properties of field UDGs (or LSBGs and HSBGs) do not change significantly if we select genuinely isolated galaxies only (i.e. those that are not satellites). Isolated UDGs have typical effective radii that are only slightly larger than field UDGs generally (5.15~kpc) and have slightly higher gas fractions (0.11).

\begin{figure*}
	\centering
    \includegraphics[width=0.95\textwidth]{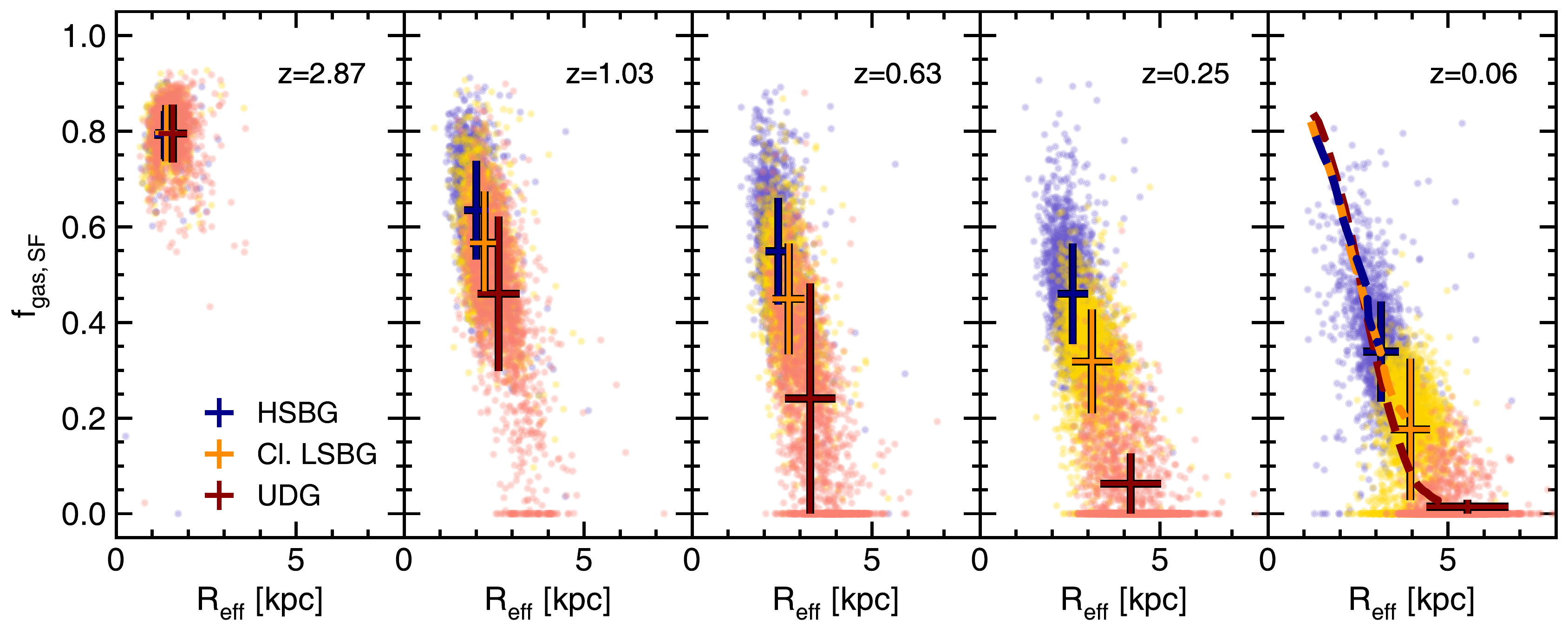}
    \caption{Redshift evolution of the star-forming gas fraction, defined as $M_{\rm gas, SF}/(M_{\rm gas, SF}+M_{\star}$), and effective radii of the progenitors of the HSBG (blue), Cl. LSBG (orange) and UDG (red) populations. The time between each redshift snapshot is $\sim$ 2.5~Gyr. Coloured points indicate the position of individual galaxies in the $f_{gas}$--$R_{\textrm{eff}}$ plane. The error bars in each panel show the median values and $1\sigma$ dispersions for the distributions of the HSBG, Cl. LSBG and UDG populations at each redshift. The dashed lines in the right-hand panel indicate the average locii followed by the main progenitors of HSBGs, Cl. LSBGs and UDGs in the $f_{gas}$--$R_{\textrm{eff}}$ plane over cosmic time.}
    \label{fig:fgas_Reff_evo2}
\end{figure*}

In \autoref{fig:fgas_Reff_evo2}, we show the redshift evolution of LSBGs and HSBGs in the star-forming gas fraction vs effective radius plane. As shown in the left-hand panel, the main progenitors of the different populations are very similar at high redshift ($z\sim 3$). Although they differ somewhat in terms of their other properties (e.g. stellar mass and environment), the progenitors of today's LSBGs and HSBGs share essentially identical effective radii and gas fractions in the early Universe. \textit{This indicates that LSBGs emerged from a common population of progenitors as HSBGs.} The three populations only begin to diverge significantly around $z\sim2$ (see \autoref{fig:reff_fgas_evo}) and then separate rapidly at intermediate redshifts ($z<1$). UDGs, in particular, diverge  quickly from their HSB counterparts, both in terms of rapidly increasing their effective radii and losing significant fractions of their gas reservoirs at these redshifts. 

We note that LSBGs appear to be part of a smooth distribution of properties across the general galaxy population. The dashed blue, orange and red lines in the right-hand panel show the average evolutionary tracks followed by HSBGs, Cl. LSBGs and UDGs respectively over cosmic time. LSBGs  do not take a different route through the $f_{gas}$--$R_{\textrm{eff}}$ plane. Instead, they follow very similar locii, although their evolution (particularly for the UDG population) is more rapid. Together with the fact that their high-redshift progenitors share very similar properties with the progenitors of HSBGs, this suggests that \textit{LSBGs are not a special class of object in terms of the populations from which they originate.} 

\subsection{Density profiles}
\label{sec:dens_slope}
Our mass-matched population of LSBGs exhibit somewhat larger effective radii compared to their HSB counterparts, even at high redshift. This can either be a result of processes that directly influence the distribution of the stellar component of the galaxy, or a result of processes that influence the distribution of the gas from which these stars form. Establishing which of these is the case is important for understanding what triggers the formation of LSBGs at early epochs.

In this section, we consider how the slope of the median gas and stellar density profiles of the different galaxy populations evolve over time. The slope of the stellar density profile determines the measured effective radius of the galaxy, with shallower slopes typically resulting in larger effective radii at a given stellar mass. Shallower density slopes (and therefore shallower gravitational potentials) also reduce the energy required to displace material in the system. In the case of the gas content, the shape of the potential defines the distribution of stellar mass that forms from this gas. Galaxies with shallower slopes are more vulnerable to the effects of encounters with other galaxies or interactions between the galaxy and the intergalactic medium (tidal heating, harassment, gas stripping etc.), which may be important factors in their subsequent evolution.

We calculate the mass-weighted log-log slope of each galaxy's gas and stellar outer density profile between 0.5$R_{\mathrm{eff}}$ and 3$R_{\mathrm{eff}}$. We calculate the density profile using radial bins of 30 particles. The log-log density slope is parametrised by $\gamma^{\prime}$ \citep{Dutton2014}:

\begin{equation}
\gamma^{\prime} = \frac{1}{M(3R_{\mathrm{eff}})-M(0.5R_{\mathrm{eff}})} \int_{0.5R_{\mathrm{eff}}}^{3R_{\mathrm{eff}}} \gamma(r) 4 \pi r^{2} \rho(r) dr,
\end{equation}

\noindent where $\gamma = -d \,\mathrm{log} (\rho) / d\, \mathrm{log} (r)$ is the local log-log slope of the density profile, $M(R)$ is the mass enclosed within a radius $R$, and $\rho(r)$ is the local density at radius $r$. Lower values of $\gamma$ indicate shallower density slopes. The density slopes that we recover are consistent with previous studies using the Horizon-AGN simulation \citep{Peirani2017}.

\begin{figure}
	\centering
    \includegraphics[width=0.45\textwidth]{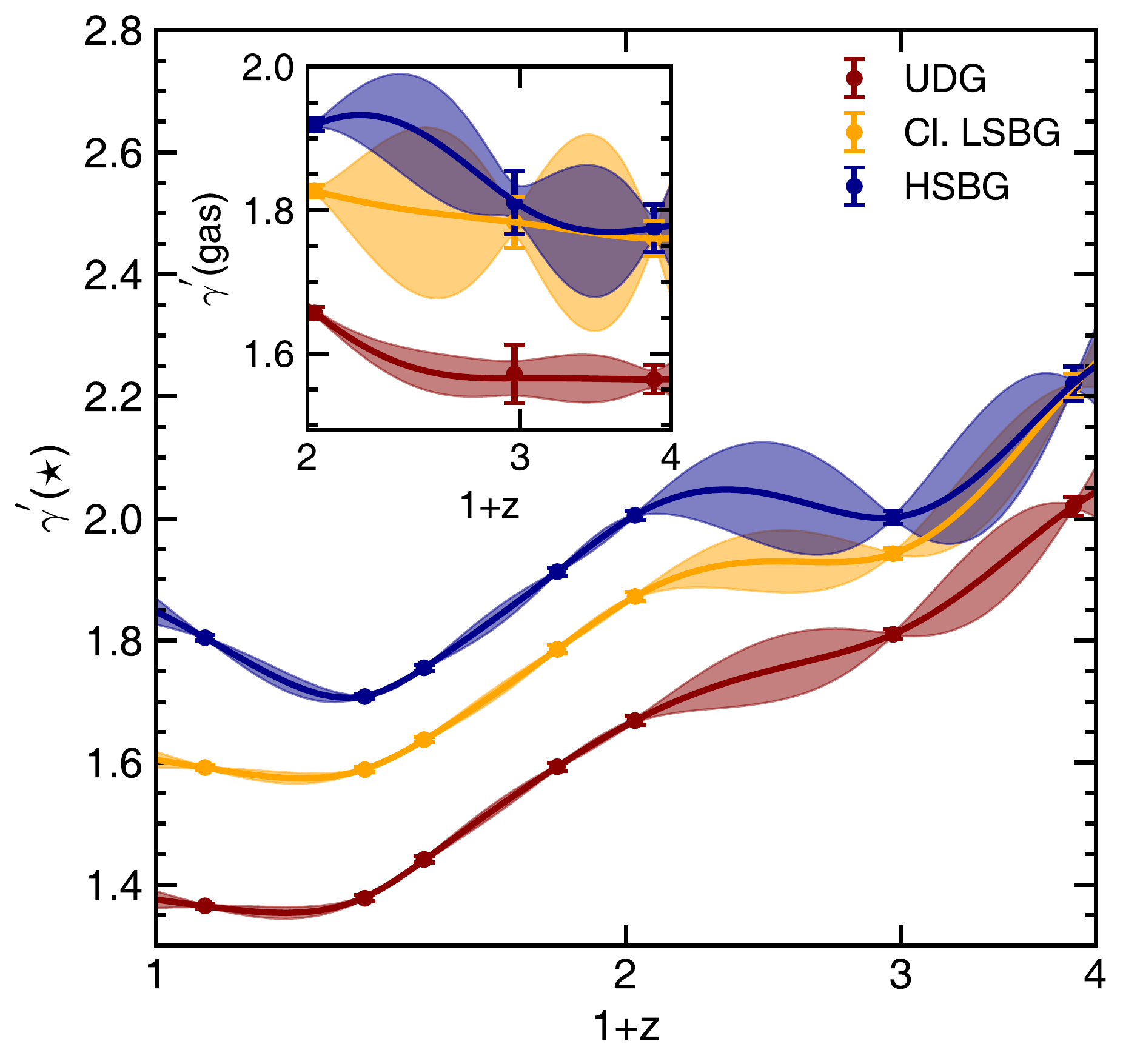}
    \caption{The evolution of the median log-log stellar density slope, $\gamma^{\prime}(\star)$, calculated within $0.5 < R_{\mathrm{eff}}<3$ for UDGs, Cl. LSBGs and HSBGs as a function of redshift. Error bars indicate the error on the median value of $\gamma^{\prime}$ at each redshift and solid filled regions show the $1\sigma$ confidence interval for a Gaussian process regression to these points. \textbf{Inset:} the corresponding plots for the log-log star-forming gas density slope, $\gamma^{\prime}(\rm gas)$. Note that, in the case of the gas density slope, galaxies with very low star-forming gas fractions, i.e. those less than 0.05, are excluded when we calculate the median values of $\gamma^{\prime}$(\rm gas).}
    \label{fig:slopes_evo}
\end{figure}

\begin{figure}
    \subfigure[]{\includegraphics[width=0.235\textwidth]{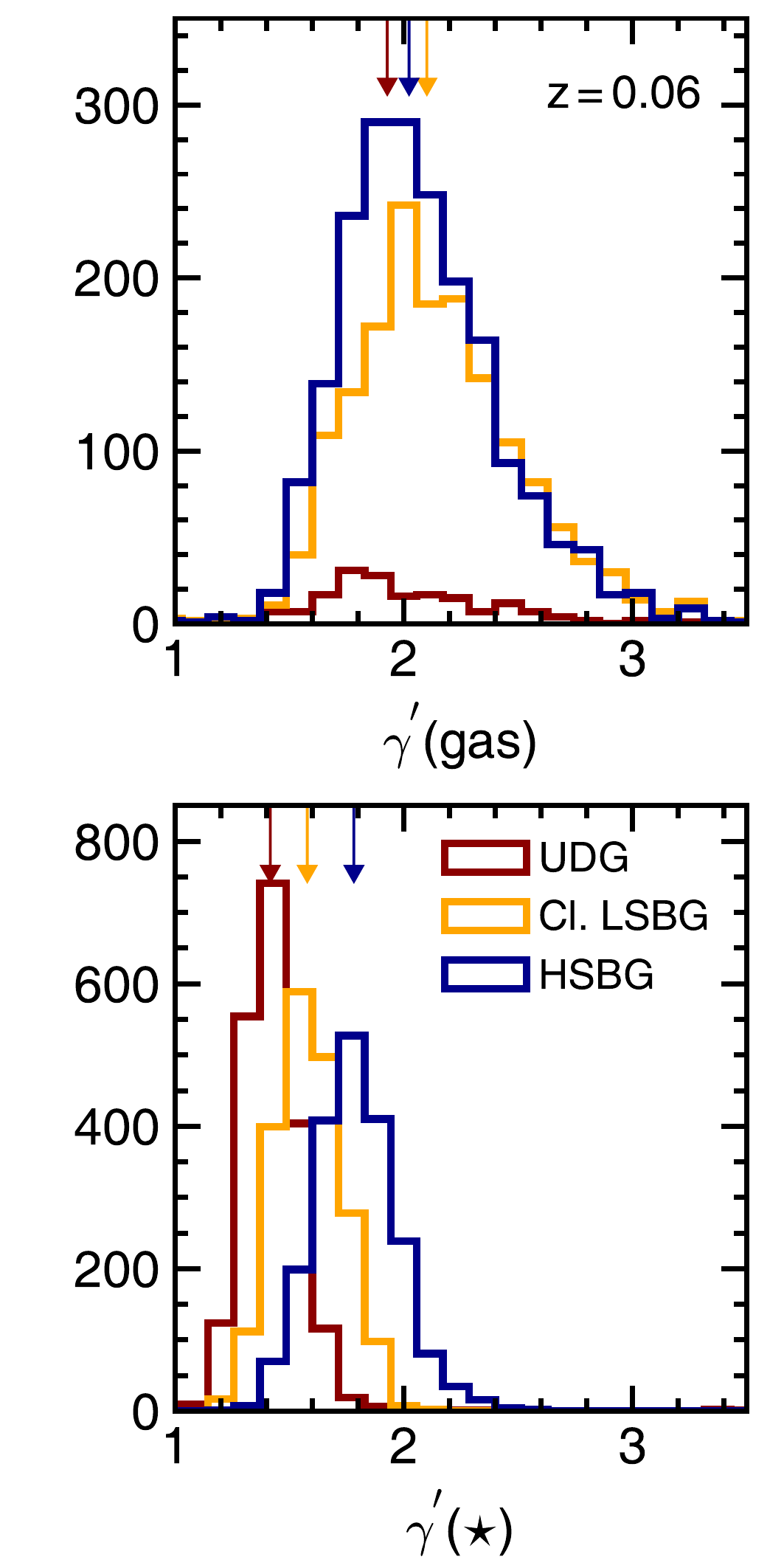}}\label{fig:a}
    \subfigure[]{\includegraphics[width=0.235\textwidth]{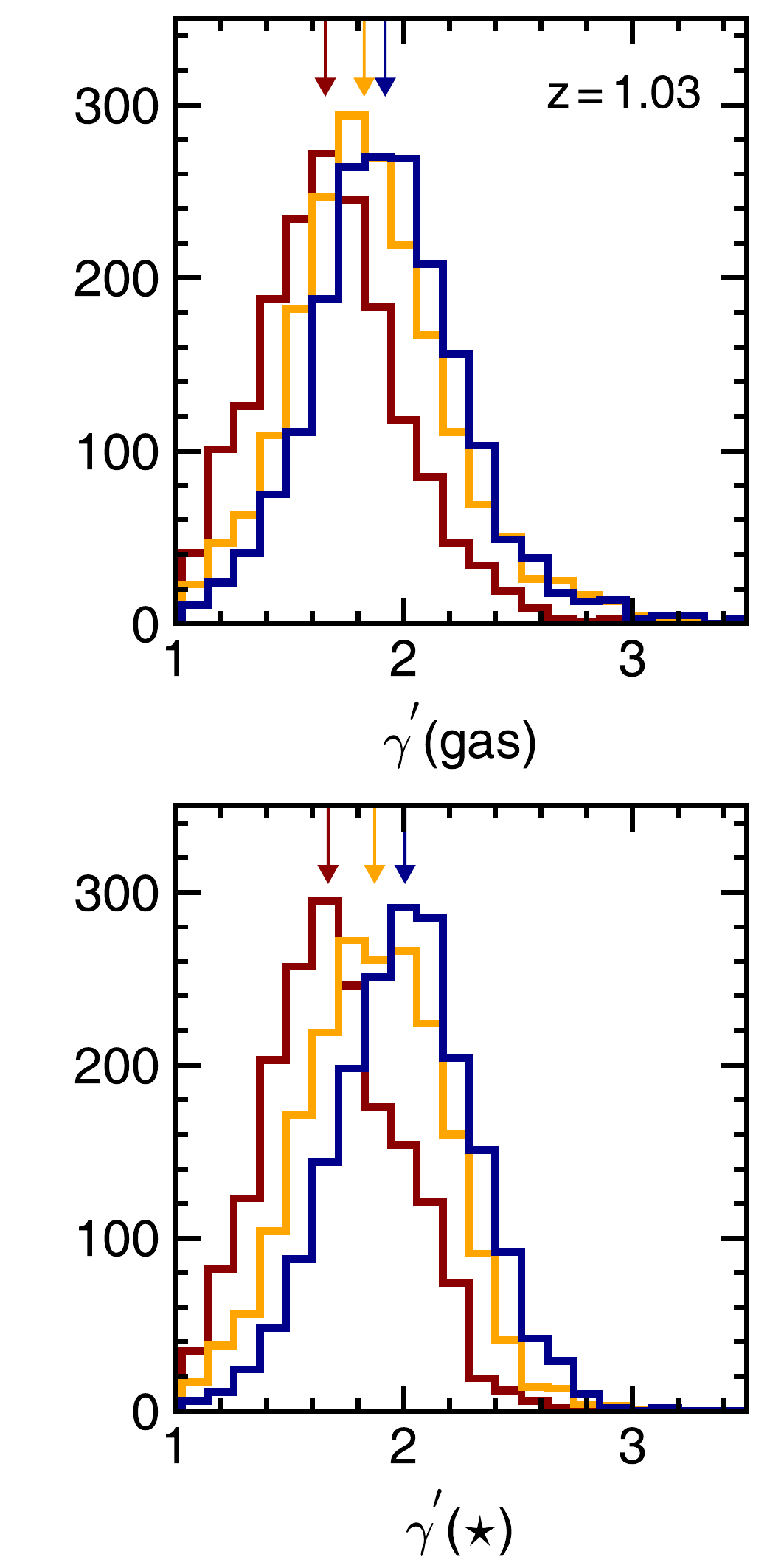}}\label{fig:a}
    \caption{\textbf{(a)} The distribution of the gas (top) and stellar (bottom) log-log density slopes, $\gamma^{\prime}$, calculated within $0.5 < R_{\mathrm{eff}}<3$ for UDGs, Cl. LSBGs and HSBGs at $z=0.06$ . Coloured arrows indicate the median value for each histogram. \textbf{(b)} The same for the distribution of the log-log density slopes at $z=1.03$. As in \autoref{fig:slopes_evo}, we exclude galaxies with star-forming gas fractions smaller than 0.05 when plotting the gas density slopes.}
    \label{fig:slope_hists}
\end{figure}

The main panel of \autoref{fig:slopes_evo} shows the redshift evolution of the median stellar density slopes for HSBGs, Cl. LSBGs and UDGs. The inset shows the evolution of the median gas density slopes for the same populations between $z=3$ and $z=1$. This is an epoch at which galaxies are forming significant fractions of their present-day stellar mass. This is particularly true of UDGs which, as we show in Section \ref{sec:SN} below, form the bulk ($\sim$75 per cent) of their stellar mass by $z=1$. At these early epochs, therefore, the gas distribution is actively driving the creation of the stellar distribution. In calculating the median gas density slope, we exclude any galaxies with star-forming gas fractions ($M_{\rm gas, SF}/(M_{\rm gas, SF}+M_{\star}$)) smaller than 0.05, so as to remove galaxies where the gas is no longer  influencing the stellar distribution 
(since the star-forming gas mass is negligible and star-formation has effectively ceased). 

At high redshift, the median value of $\gamma^{\prime}$(gas) is lower (i.e. the gas density slopes are shallower) in UDGs compared to both Cl. LSBGs and HSBGs ($\sim$1.56 at $z=2$ compared with $\sim$1.8 for Cl. LSBGs and HSBGs).  Between $z=3$ and $z=1$, the gas density slopes in the UDGs remain at a level significantly below the Cl. LSBG and HSBG populations, while their stellar density slopes decline faster than those of the Cl. LSBG and HSBG populations. Thus, at the epochs where UDGs are actively forming the bulk of their stellar mass, their gas density profiles are significantly flatter than that of the HSBGs (and also the Cl. LSBGs).

After $z=1$, the stellar density slopes decline rapidly, even though most LSBGs have assembled the majority of their stellar mass by this time. By $z=0.06$ the median value of $\gamma^{\prime}(\star)$ for UDGs has fallen by $\sim 0.32$, from 1.67 at $z=1$ to 1.35. The median value of $\gamma^{\prime}(\star)$ for HSBGs (most of which have not yet assembled the majority of their stellar mass at $z=1$), falls from 2.0 to 1.8 between $z=1$ and $z=0.06$. 

\autoref{fig:slope_hists} shows the distributions of the gas and stellar density slopes at two epochs: $z=0.06$ and $z=1.03$ (where the divergence in the effective radii, gas fractions and stellar density slopes between the LSB and HSB populations accelerates). At $z=1.03$, the distribution of $\gamma^{\prime}$(gas) strongly resembles that of $\gamma^{\prime}(\star)$ for all three populations. For example, for the UDG gas and stellar density slope distributions, a two-sample Kolmogorov--Smirnov test \citep{Smirnov1939} yields a $D$-statistic of 0.033 and a $p$-value of 0.28, indicating a strong likelihood that the two samples are drawn from the same distribution. This is a natural consequence of the fact that, at early epochs ($z>1$), the gas distribution is the principal factor driving the development of the stellar profile, especially in UDGs, which form the bulk of their stellar mass at these redshifts. The stellar density slope is, therefore, gradually driven towards the gas density slope over this epoch.

After $z\sim 1$ the gas and stellar slopes progressively diverge, with the divergence being fastest in UDG progenitors. By $z=0.06$, the stellar density slopes in UDGs have decoupled completely from the gas density slopes, with the average stellar density slope becoming much shallower than the average gas density slope. Thus, the trigger for the initial divergence of HSBGs and UDGs at high redshift is likely to be processes that act on the gas profiles in UDGs to make them shallower, rather than those that directly affect the stellar components of these galaxies.

In the next section we explore some of the processes that lead to the divergence in the evolution in effective radius, gas fraction and density slopes of LSBGs compared to their HSB counterparts.

\section{How do low-surface-brightness galaxies form?}
\label{sec:processes}
The analysis presented above shows that the formation mechanisms that produce LSBGs act to both increase the effective radii of their progenitors and drive the steady loss of star-forming gas (either by ejection from the galaxy or by heating). This produces diffuse systems with low SFRs and older stellar populations which, together, result in systems that exhibit low surface-brightnesses. In this section, we study the mechanisms which drive these changes over cosmic time: SN feedback, perturbations due to the ambient tidal field and ram-pressure stripping. 

\begin{figure}
	\centering
  	\includegraphics[width=0.45\textwidth]{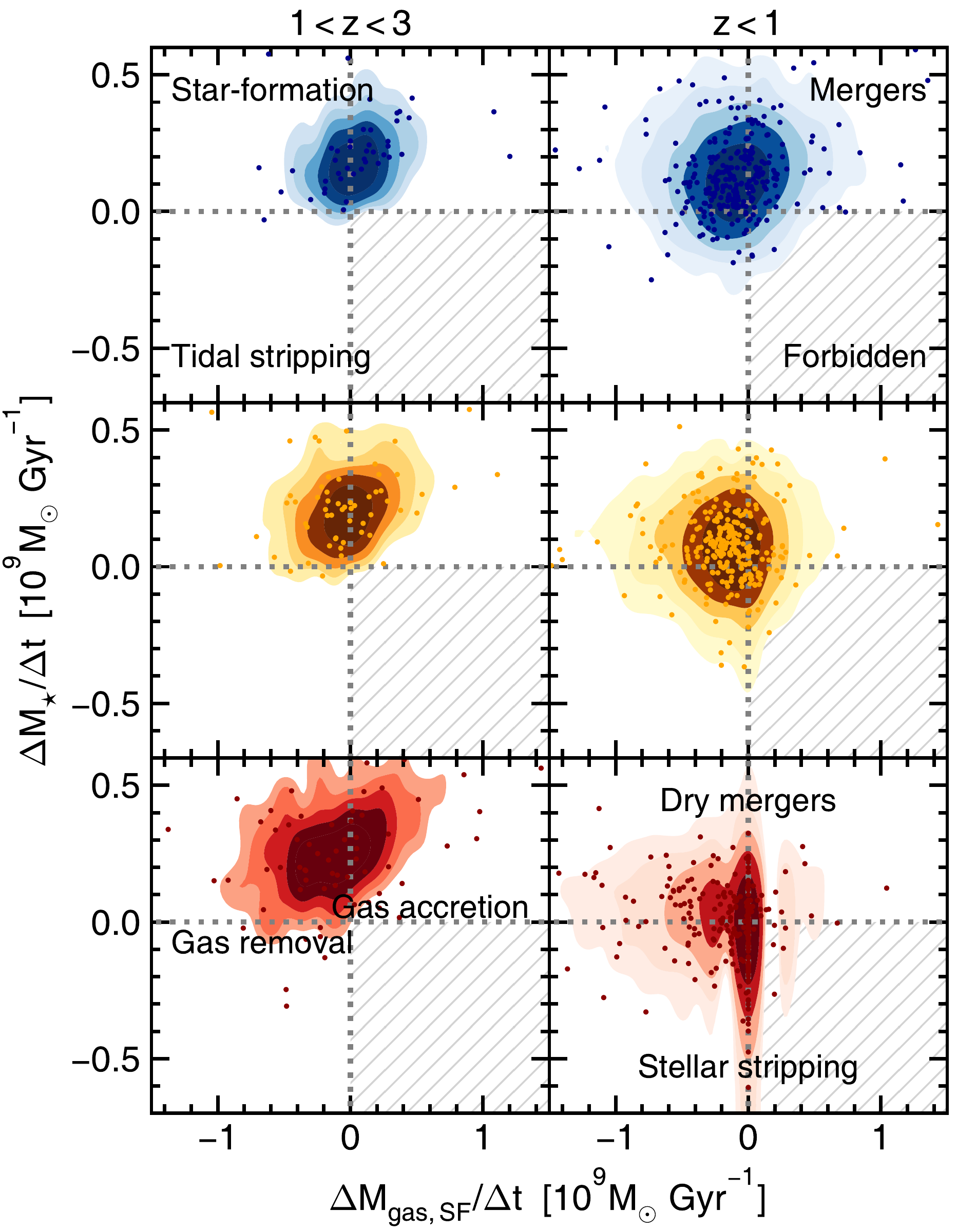}
    \caption{Contour plots showing the distribution of the rate of change in star-forming gas mass vs the rate of change in stellar mass, in units of $10^{9}$M$_{\odot}$~Gyr$^{-1}$. A random selection of the data is plotted using the coloured points in each panel. Each point represents the change between two consecutive timesteps. The left panels show changes in the redshift range $1<z<3$ and the right panels show the same for $z<1$. The top, middle and bottom rows show distributions for HSBGs, Cl. LSBGs and UDGs respectively. Labels in each quadrant of the top panels indicate the main processes operating in that section of the parameter space. Labels in the bottom panels indicate the processes that cause points to fall close to the horizontal and vertical axes.}
    \label{fig:rates}
\end{figure}

We begin our analysis by taking an aggregate view of the role of key processes that could drive LSBG formation. \autoref{fig:rates} shows distributions of the change in star-forming gas and stellar mass (in units of $10^{9}$M$_{\odot}$~Gyr$^{-1}$), for approximately evenly spaced simulation outputs ($\sim$250~Myrs), in the redshift range $1<z<3$ (left) and at $z<1$ (when the HSB and LSB populations diverge most rapidly; right). The top, middle and bottom panels show distributions for the progenitors of HSBGs, Cl. LSBGs and UDGs respectively.

Different regions in this plot indicate different processes that act to produce each of these galaxy populations. For example, star formation will increase the stellar mass while decreasing the gas mass, as it fuels the star formation. Galaxies undergoing star formation will, therefore, populate the upper-left quadrant of this plot. Mergers increase both stellar and gas mass (upper right quadrant), with dry mergers towards the left-hand side of this quadrant. The signature of gas removal (e.g. ram-pressure stripping and/or gas heating) is a decrease in gas mass which is not accompanied by a corresponding change in stellar mass (i.e. the negative half of the $x$-axis), while gas accretion causes points to accumulate close to the positive half of the $x$-axis. Tidal stripping (which is driven by tidal heating) results in stripping of both stellar and gas mass (lower left quadrant), typically from the outskirts of a galaxy. Tidal heating will also cause the entire distribution of stars to expand, although this is not possible to show in this plot. Finally, the lower right quadrant is typically forbidden, because galaxies tend not to increase their gas mass while simultaneously losing stars.

The top panel in \autoref{fig:rates} indicates that HSBG evolution at all epochs is largely driven by gas accretion, star formation and mergers, with little impact from processes such as ram-pressure or tidal stripping/heating. Star formation at high redshift is smooth and the gas mass lost to star formation is typically replenished by accretion. At lower redshifts, star formation remains at similar levels, but accretion is typically no longer fast enough to offset the gas that is transformed into stars. The plots show that the degree of gas stripping and heating experienced by HSBGs must be small as, in the vast majority of cases, any decrease in gas mass is accompanied by an increase in stellar mass of a similar magnitude. 

However, as we transition to populations that have lower surface-brightnesses at the present day, the relative role of these processes changes. Cl. LSBGs and UDGs both show similar evolution to HSBGs at $z>1$, the epoch at which the bulk of their stellar mass forms (see Section \ref{sec:SN} below). However at $z<1$, Cl. LSBGs and, in particular, UDGs, have both experienced large decreases in gas mass that are not the result of star formation. In the case of UDGs, ram pressure stripping and tidal stripping/heating of both stars and gas are clearly important processes in their evolutionary history (particularly at lower redshifts), as shown by the much higher fraction of such systems that inhabit the lower left quadrant compared to HSBGs. In the following sections, we study the mechanisms that drive these processes and explore their relative role in creating LSB systems over cosmic time. 


\subsection{Supernova feedback - a trigger for LSBG formation}
\label{sec:SN}
Theoretical studies by \citet{DiCintio2017} and \citet{Chan2018} show that, at least at low stellar masses, SN feedback may be capable of producing UDGs by fuelling outflows which create flattened total density profiles \citep[e.g.][]{Navarro1996,Governato2010,Pontzen2012,Teyssier2013,Errani2015,Onorbe2015,Carleton2018,Sanders2018}. These outflows may be effective, not only at removing gas from the galaxy, but also at producing shallower gas density profiles \citep{Brook2011,Brook2012,Pontzen2014,DiCintio2014,DiCintio2014b,Dutton2016,DiCintio2017} and through the dynamical heating of stars, increasing their effective radii \citep{Chan2015,El-Badry2016} \citep[although there may be some tension with observations e.g.][]{Patel2018}. It may also be the case, as we show later, that, rather than directly influencing the size and gas content of galaxies, SN feedback instead allows other processes to work more efficiently (e.g. tidal heating and ram-pressure stripping).

It has been shown using the Horizon suite of simulations that the inclusion of baryons and their associated feedback processes results in shallower stellar density slopes, compared to an otherwise identical DM-only simulation \citep{Peirani2017}. As we have already shown in Section \ref{sec:properties}, the DM masses of UDGs and Cl. LSBGs are not dissimilar to that in their HSB counterparts. It is therefore not the case that UDGs are massive haloes that have been quenched before their reservoir of star-forming gas has been used up. Instead, it is worth considering whether differences in the actual stellar assembly history of LSBGs, especially at early epochs where the bulk of the stellar mass was formed, may be a contributing factor to creating their LSB nature over cosmic time.

\begin{figure}
	\centering
    \includegraphics[width=0.45\textwidth]{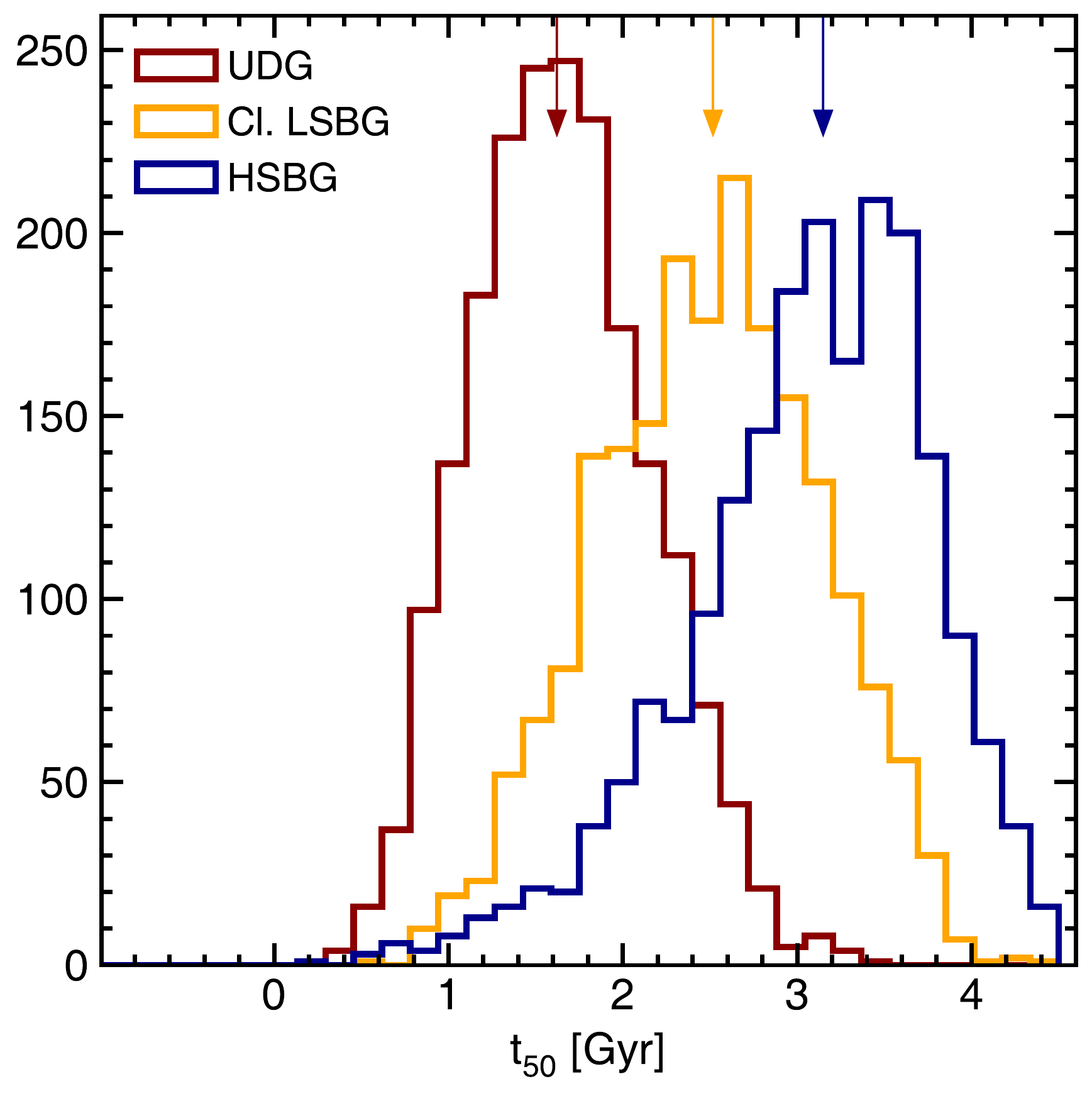}
    \caption{Distribution of $t_{50}$, (i.e. the minimum time required to form 50 per cent of a galaxy's present-day stellar mass) for UDGs, Cl. LSBGs and HSBGs. Coloured arrows indicate the median value for each histogram.}
    \label{fig:t50}
\end{figure}

Since the amount of SN feedback energy deposited in the potential well will be sensitive to the star formation history, we consider the burstiness of star formation in our different galaxy populations. To quantify the burstiness we define $t_{50}$, which measures the minimum amount of time required to form 50 per cent of the stellar mass in a galaxy\footnote{In order to calculate $t_{50}$, we first produce a histogram of the distribution of star-particle ages in each galaxy at $z=0.06$, using bin widths of 100~Myrs. The bins in each histogram are then re-sorted, in order of decreasing frequency, and the cumulative distribution function (CDF) is calculated. $t_{50}$ is the time at which this CDF reaches 50 per cent.}. \autoref{fig:t50} shows the distribution of $t_{50}$ values for the HSBG, Cl. LSBG and UDG populations. The median value of $t_{50}$ for HSBGs is typically $\sim$3~Gyrs, while for UDGs it is $\sim$1.5~Gyr (with the value for Cl. LSBGs falling in between these values). This indicates that the formation of UDGs is much more rapid than HSBGs of similar stellar masses. UDGs typically assemble earlier and, on average, they have already formed 75 per cent of their stellar mass by $z=1$ \citep[i.e. as a result of halo assembly bias,][]{Sheth2004}. 

On the other hand, HSBGs have  formed only 30 per cent of their stellar mass by this time. The median SFR for HSBGs falls only modestly between $z=3$ and the present day. As a result, energy released by supernovae (SNe) and stellar winds is distributed over most of the lifetime of the galaxy, whereas feedback energy is almost entirely concentrated before $z=1$ in the case of UDGs. This, in turn, means that the maximum instantaneous energy imparted into the gas is much larger in UDGs than in their HSB counterparts. 

We proceed by quantifying the impact that SN and stellar feedback may have on the galaxy populations due to their disparate formation histories. We define the total mechanical and thermal energy released by stellar processes between two timesteps, $t_{0}$ and $t_{1}$, by summing the energy released by each star particle within this interval:

\begin{equation}
E_{\mathrm{SN}} = \sum_{i} m_{\star,i} (E(z_{1},Z)_{i}-E(z_{0},Z)_{i})
\end{equation}

\noindent where $m_{\star,i}$ is the mass of a star particle and $E(z,Z)_{i}$ is the cumulative mechanical and thermal energy released by that star particle as a result of Type Ia SNe, Type II SNe and stellar winds per unit stellar mass, for a metallicity $Z$, and between the time of its formation and a redshift of $z$.

\begin{figure}
	\centering
    \includegraphics[width=0.45\textwidth]{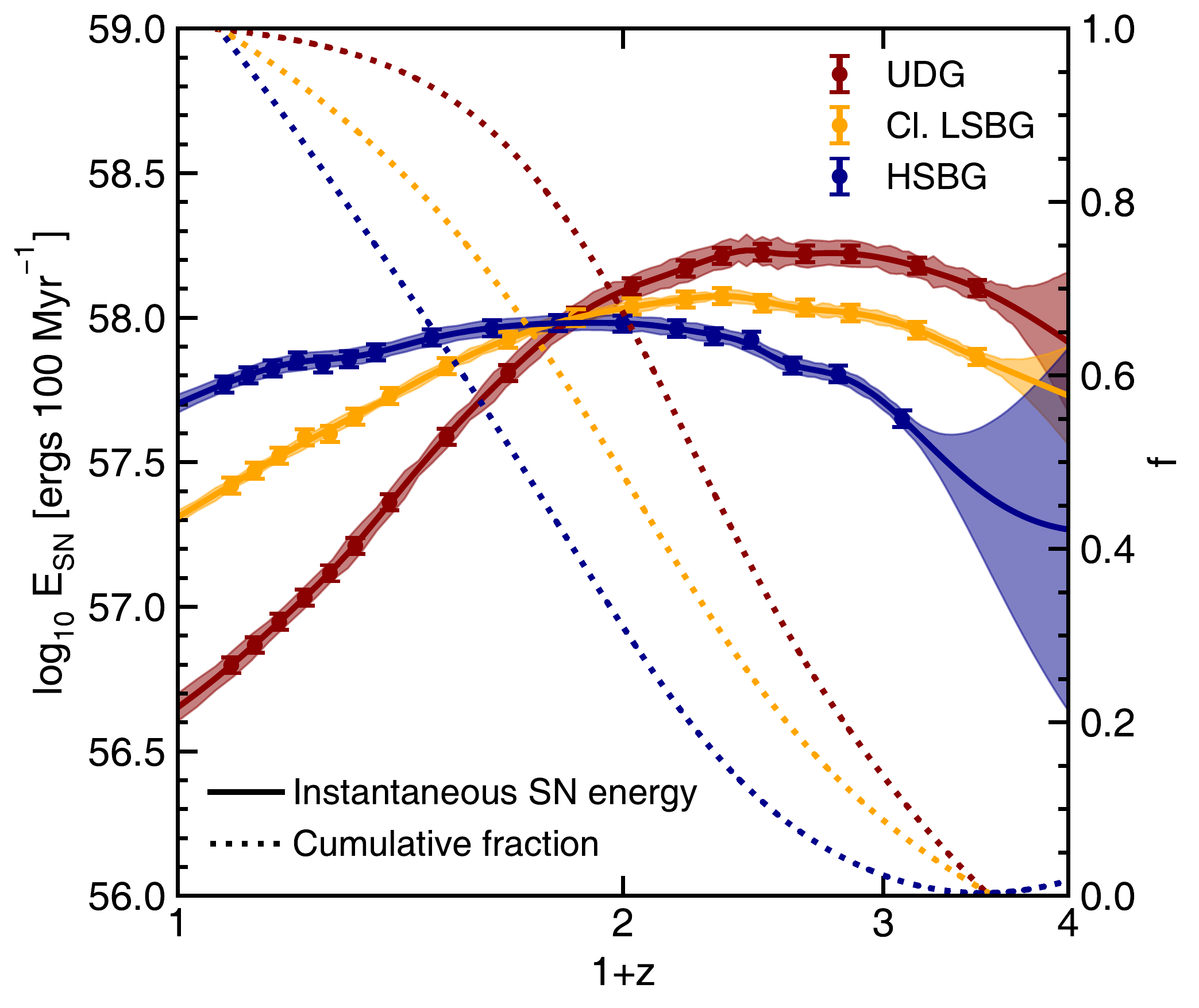}
    \caption{Median mechanical and thermal energy injected as a result of Type Ia and Type II SNe and stellar winds. Solid lines indicate the SN energy released per 100~Myr as a function of redshift. Dotted lines show the average cumulative energy as a fraction of the value at $z=0.06$. The fractions are indicated by the values on the right-hand axis.}
    \label{fig:SNevo}
\end{figure}

\begin{figure}
	\centering
    \label{fig:AGN}
    \includegraphics[width=0.45\textwidth]{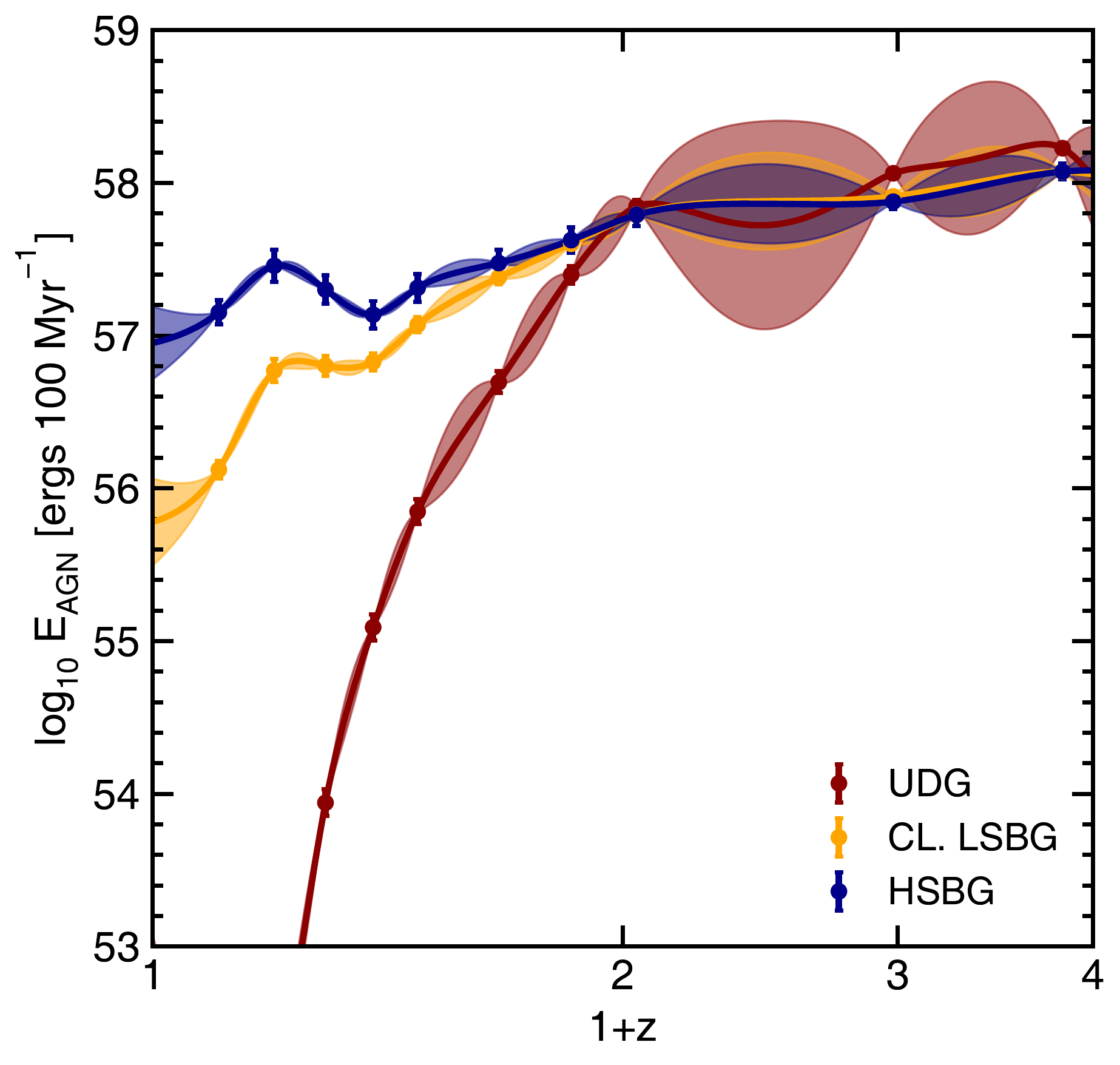}
    \caption{Median mechanical and thermal energy released as a result of AGN feedback. Solid lines indicate the AGN energy released per 100~Myr as a function of redshift.}
    \label{fig:AGNevo}
\end{figure}

\autoref{fig:SNevo} shows the median mechanical and thermal energy released by stars over the last 100~Myr, as a function of redshift. Since our samples are matched in stellar mass, the total cumulative feedback energy for each sample reaches the same value at $z=0.06$ but the pattern of energy injection differs between the populations. Since they form the majority of their stellar mass early on (75 per cent before $z=1$), the progenitors of UDGs release energy over a shorter period of time. 

As a result, UDGs experience high levels of SN feedback at early times. Between $z=3$ and $z=1$, UDGs have already released 75 per cent of their integrated stellar feedback energy, compared with 50 per cent for Cl. LSBGs and only 30 per cent for HSBGs. The SN energy released in HSBGs remains roughly constant as a function of redshift, decreasing by only 0.25 dex between the peak at $z\sim1$ and $z=0$. In comparison, the SN energy released in UDGs peaks between $z=2$ and $z=1$ and then declines rapidly towards $z=0$ to a value 1.5 dex lower. 

We note that the same patterns are \textit{not} observed for AGN feedback. As \autoref{fig:AGNevo} shows, the evolution of AGN feedback energy in UDGs, Cl. LSBGs and HSBGs proceeds similarly at high redshift ($z>1$), falling rapidly for low redshift UDGs as hot gas in cluster environments quenches the Bondi accretion rates of their BHs. Additionally, BH growth in low-mass haloes is regulated by SN feedback \citep[e.g.][]{Volonteri15,Habouzit17,Bower17}, so that SN feedback is the principal feedback process in the UDG population. 

As we have shown in Section \ref{sec:dens_slope} (Figure \ref{fig:slopes_evo}), the gas density slopes of UDG progenitors are significantly and consistently shallower than their HSB counterparts in the early Universe (between $z=1$ and $z=3$). This coincides with the period where instantaneous SN feedback energy is at its peak in the UDG progenitor population. It is worth noting that the profiles of HSBG progenitors behave very differently at these early epochs. Both their gas and stellar density slopes tend to \textit{increase} with time at these early epochs, as baryons accumulate in the centres of their gravitational potential wells. SN feedback therefore has a much greater impact on LSBG progenitors than it does on their HSB counterparts. 

We note that, while large amounts of energy are released into the gas in UDG progenitors at these early epochs, the fraction of star-forming gas (\autoref{fig:reff_fgas_evo}, bottom panel) in UDG progenitors remains significant ($f_{\rm gas,SF}>0.4$) at these times. This indicates that, while the slope of the gas density profile is made shallower due to this SN feedback, the feedback is not so strong that the gas is completely removed and star-formation quenched. 

As was noted in Section \ref{sec:dens_slope}, stars forming from this gas progressively flatten the stellar density slopes, leading to the decrease in $\gamma^{\prime}(\star)$ shown in Figure \ref{fig:slopes_evo}. SN feedback, therefore, appears to be the mechanism that drives the creation of shallower gas and stellar density slopes in UDG progenitors at high redshift, which leaves these systems more vulnerable to tidal processes (e.g. tidal heating and, additionally, ram-pressure stripping in dense environments) over cosmic time. It is worth noting here that the specific angular momenta of LSBG and HSBG progenitors are very similar at $z\sim3$, indicating that the flatter density profiles of LSBG progenitors is not due to them initially forming with higher values of spin. 

Although UDGs clearly increase in size (\autoref{fig:reff_fgas_evo}, top panel) and gain flatter density slopes (\autoref{fig:slopes_evo}, bottom panel) compared to HSBGs and other LSBGs at $z>1$, the difference is fairly modest compared to the much greater divergence in effective radii and gas fractions seen \textit{after} $z=1$ (\autoref{fig:reff_fgas_evo}, top panel). Thus, SN feedback appears to be the initial \textit{trigger} for the divergence of UDGs from the rest of the galaxy population, rather than the principal cause of their large sizes at $z=0$. A combination of a shallower potential and a broader distribution of stars is likely to contribute to the steep rise in the effective radii of UDG progenitors, in contrast to their HSB counterparts seen after $z=1$.

Much of this evolution must be due to external processes that act to increase the effective radii steadily over cosmic time. Since they would be expected to operate more efficiently on systems where galaxies have shallower gravitational potentials (and where the material, at least in the outer regions, is more weakly bound), environmental processes such as perturbations from the ambient tidal field and ram-pressure stripping are likely to amplify the initial divergence produced by SN feedback \citep{Pontzen2012,Errani2015,Carleton2018,Sanders2018}. We explore the effect of these processes in the next two sections.

It is worth noting here that processes other than SN feedback could assist in the initial creation of shallower density slopes in UDG progenitors. For example, an accretion history that is rich in low-mass-ratio (i.e. minor) mergers may also act to broaden the stellar distribution \citep[e.g.][]{Naab2009,Bezanson2009,Hopkins2010,Bedorf2013}. However, while there is some evidence that LSBG progenitors do exhibit some level of enhancement in their merger histories in the \textit{early} Universe ($\sim$ twice the number of major mergers undergone by HSBGs between $z=3$ and $z=1$), it is difficult to draw concrete conclusions, as the merger histories of low mass-ratio mergers are typically highly incomplete in the simulation at high redshift \citep{Martin2018_sph}\footnote{Due to the stellar mass resolution of the simulation, only objects that are more massive than $2\times10^{8}$~M$_{\odot}$ are detectable. As a result, only 50 per cent and 20 per cent of the ($z=0$) progenitors of $10^{9.5}$~M$_{\odot}$ galaxies are massive enough for a 1:10 mass ratio merger to be detectable at $z=2$ and $z=3$ respectively (\citealt{Martin2018_sph}, Figure 1)}. 

In order to quantify the relative (and probably additive) roles of feedback \citep[e.g.][]{Dashyan2018} and minor mergers \citep[e.g.][]{DiCintio2019} in triggering the initial shallower gas density profiles, a higher resolution simulation is required. In a forthcoming paper (Jackson et al. in prep) we will use New-Horizon (Dubois et al. in prep), a 4000~Mpc$^3$ zoom-in of a region of Horizon-AGN, which has 64 times better spatial resolution to probe this `trigger epoch' in more detail. 

\subsection{Perturbations due to the ambient tidal field - a key driver of LSBG evolution}
Recall first from the arguments above that the processes that produce  LSBGs operate steadily over cosmic time (since the effective radii and gas fractions change gradually with redshift) and are not specific to cluster environments (since UDGs are found in all environments). Mergers and tidal interactions with nearby objects offer an attractive mechanism for LSBG formation because they act to dynamically heat galaxies and destroy cold, ordered structures \citep{Moore1996,Moore1998,Gnedin2003,Johansson2009}. These processes are therefore likely contributors to both the observed increase in the effective radii and the decrease in the star-forming gas fractions seen in the LSBG population, regardless of local environment. 

It is worth noting first that, compared to HSBGs and Cl. LSBGs, UDGs in our sample are considerably more `spheroidal' (i.e. a larger fraction of their stars are on random orbits compared to ordered, rotational ones). While the median value of the ratio of rotational to dispersional velocities of the stellar component, $(V/\sigma)_{\star}$, is 0.4 for Cl. LSBGs, it is only 0.15 for UDGs. In comparison, late-type i.e. disc-dominated galaxies typically exhibit $(V/\sigma)_{\star}>0.55$ \citep{Martin2018_progenitors}. Since mergers and interactions are efficient drivers of (disc-to-spheroid) morphological transformation \citep{Martin2018_sph}, this is evidence that the UDGs have indeed undergone a larger number of interactions (but not necessarily actual mergers) that have shaped their structural evolution. 

Recent observational work lends support to the idea that the formation of LSBGs is connected to the tidal effects of nearby galaxies. Some studies have pointed to the idea that UDG progenitors may be more massive star-forming dwarfs that are destroyed as a result of interactions within a cluster environment \citep[e.g][]{Conselice2018}. Alternatively, they may be less massive dwarfs that have undergone considerable expansion \citep[e.g.][]{Carleton2018} due to tidal interactions. It has also been suggested that at least some UDGs may be tidal dwarfs \citep[e.g.][]{vanDokkum2018, Ogiya2018, Greco2018b}, formed when material is stripped from larger galaxies. However, since mergers typically produce tidal dwarfs with low stellar masses (less than 1 per cent of the mass of the merging progenitors, see e.g. \citet{Barnes1992,Okazaki2000,Kaviraj2012}), the mass range that we consider in this study ($M_{\odot}>10^{8}$) precludes significant numbers of these objects in our sample. 

In the context of mergers (i.e. interactions which result in the actual coalescence of the interacting progenitors) it is worth noting that both LSBGs and HSBGs undergo very few actual mergers at low redshift, where the effective radii and star-forming gas fractions change significantly. Indeed, only a few percent of galaxies have undergone mergers of mass ratios larger than 1:4 since $z=1$; see e.g. \citealt{Darg2010,Martin2018_BH,Martin2018_sph}). While UDGs do undergo more mergers than HSBGs at high redshift (as was noted earlier in Section \ref{sec:SN}), they experience a relative dearth of mergers (a factor of 2.5 fewer major mergers) than their HSB counterparts between $z=1$ and the present day, when much of the increase in radii and decrease in gas content takes place. Galaxy mergers, therefore, are unlikely to be the principal driver of LSBG evolution over cosmic time.

However, tidal interactions (or fly bys) between galaxies can produce similar effects to that due to actual mergers \citep[e.g.][]{Martin2018_sph,Choi2018}. To explore the effect of tidal interactions on LSBGs and HSBGs, we employ a perturbation index (PI) which quantifies the environmental tidal field due to objects in the vicinity of the galaxy in question. We define the PI \citep[e.g.][]{Byrd1990,Choi2018} between $z=3$ and the redshift in question, by calculating the \textit{cumulative} contribution of all galaxies within 3~Mpc:

\begin{equation}
\label{eqn:PI}
PI = \int^{z}_{z=3} \sum_{i}\left ( \frac{M_{i}}{M_{gal}}\right ) \left ( \frac{R_{\mathrm{eff}}}{D_{i}} \right )^{3} dt \ / \ {\rm Gyr}
\end{equation}

\noindent where $M_{gal}$ is the stellar mass of the galaxy in question and $M_{i}$ is the stellar mass of the $i$th perturbing galaxy. $R_{\mathrm{eff}}$ is the effective radius as defined in Section \ref{sec:SB_maps}, $D_{i}$ is the distance from the $i$th perturbing galaxy and $dt$ is in units of Gyrs. By this definition, galaxies that are more massive and/or approach more closely will contribute more to the PI, with each galaxy's contribution dropping off steeply with distance. For example, a perturbation index $PI=10^{-1}$ is equivalent to a single 1:10 mass ratio merger or an equal mass galaxy moving within 2 effective radii. We note that our definition of PI is a cumulative one, so that we integrate the perturbations felt by individual galaxies between $z=3$ and the redshift in question ($z$). The PI is calculated at evenly spaced timesteps of $\sim$130~Myr and we do not attempt to integrate galaxy orbits, as the relatively coarse time resolution makes this unreliable.

\begin{figure}
	\centering
    \subfigure{\includegraphics[width=0.45\textwidth]{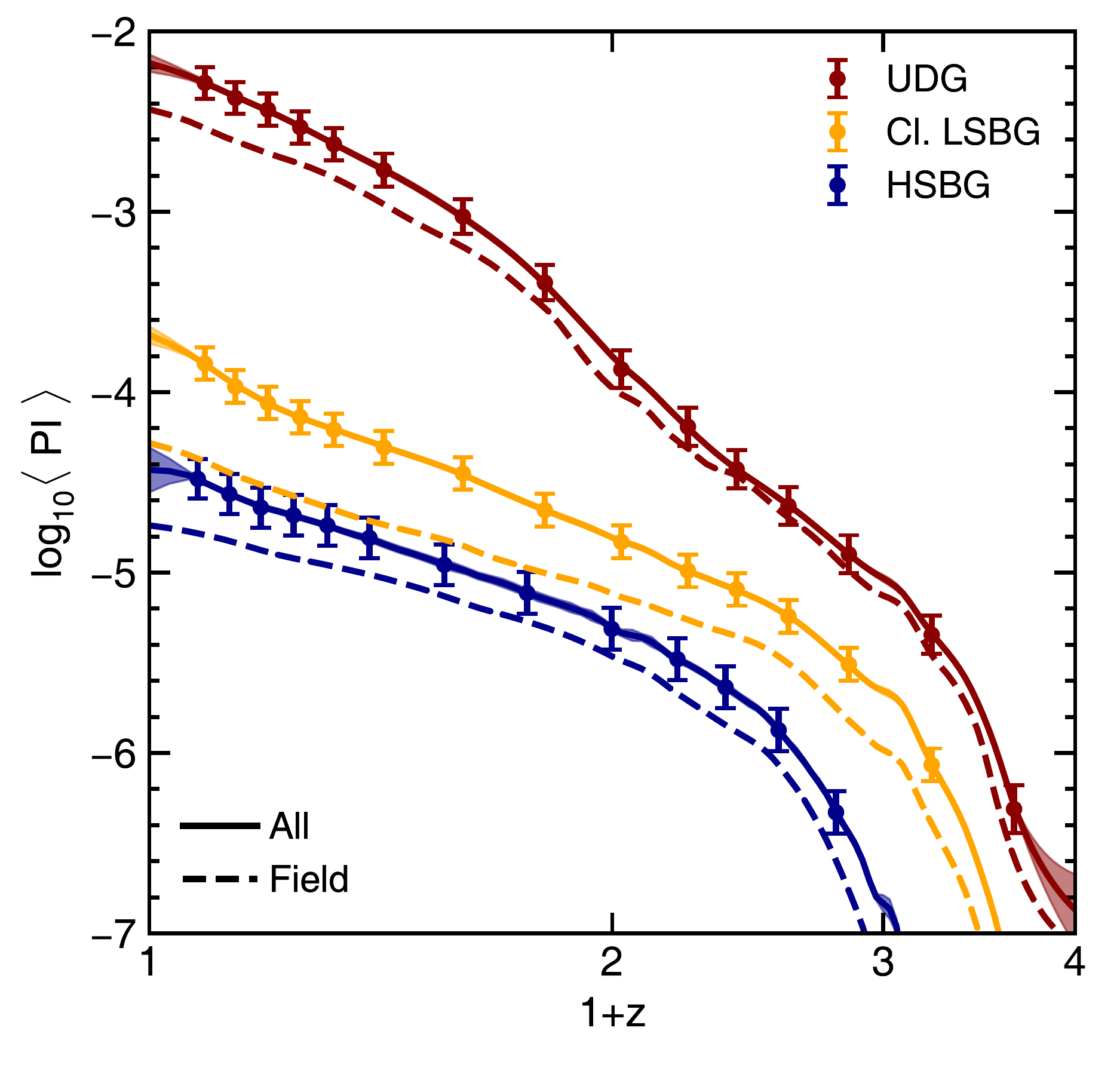}\label{fig:a}}
    \subfigure{\includegraphics[width=0.45\textwidth]{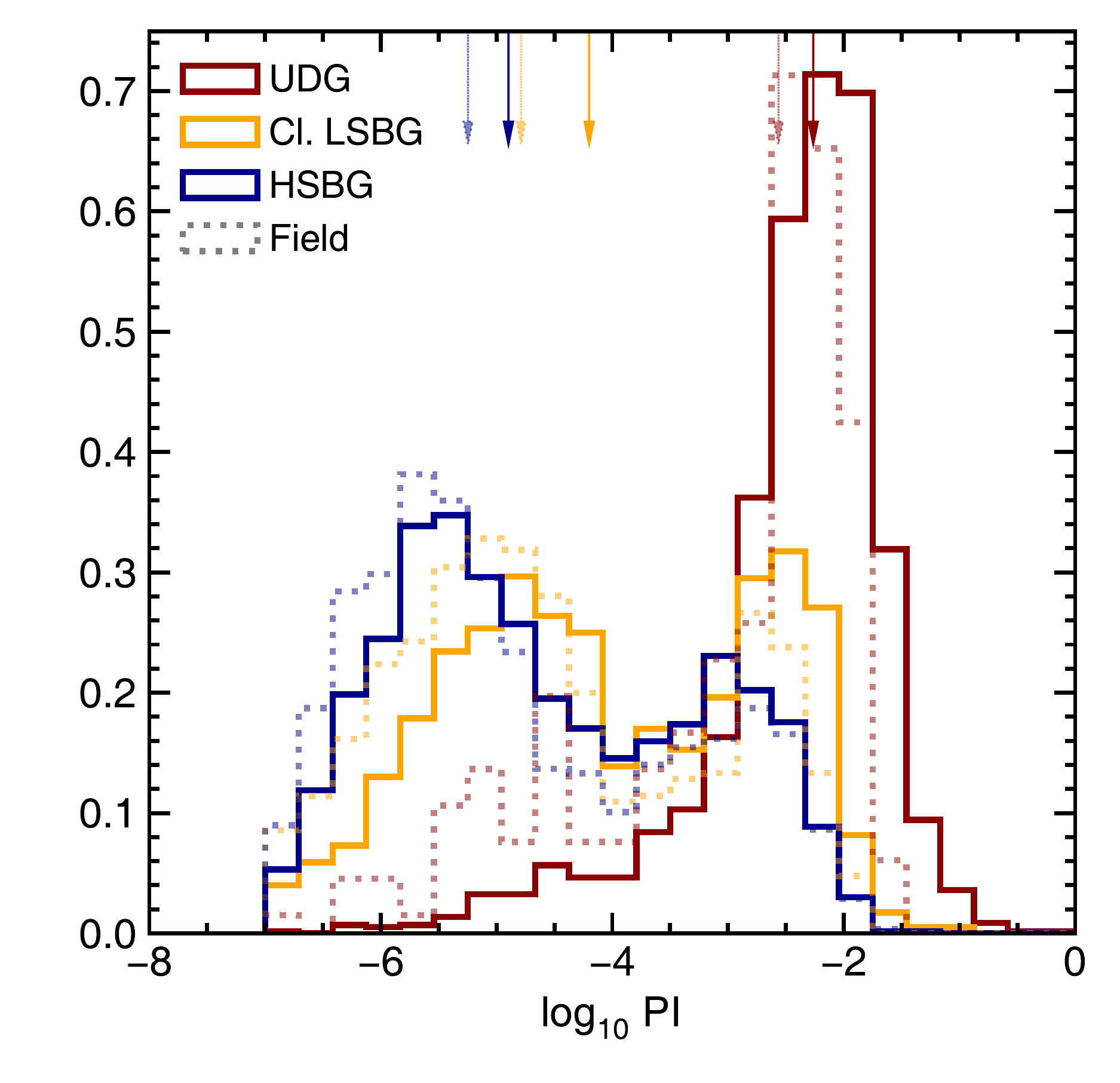}\label{fig:b}}
    \caption{\textbf{Top}: Median perturbation index (PI), as defined by \autoref{eqn:PI}, between $z=3$ and the redshift in question. Error bars indicate the errors on the median value of the PI at each redshift and solid filled regions show the $1\sigma$ confidence intervals for a Gaussian process regression to these points. Dashed lines indicate the same for galaxies in the field only. \textbf{Bottom}: Distribution of the perturbation index between $z=3$ and $z=0$. Dotted lines show the distribution for field galaxies only. Coloured arrows indicate the median value for each histogram and fainter arrows indicate the median values for field galaxies. Note that the histograms are normalised so that the field and general populations can be easily compared.}
    \label{fig:PI}
\end{figure}

In the top panel of \autoref{fig:PI} we plot the median value of the PI in each of our populations, as a function of redshift. At all redshifts galaxies that have lower surface-brightnesses exhibit consistently higher PI values. The discrepancy between the median PI values in the LSBG and HSBG populations becomes more pronounced with time. Compared with HSBGs, UDGs in all environments undergo more frequent or violent perturbations, exhibiting PI values more than 2 dex higher towards low redshift (with Cl. LSBGs reaching values around 1 dex higher). Not unexpectedly, for all populations, galaxies that inhabit the field exhibit lower PI values.  

In the bottom panel, we show the PI over the entire redshift range of the top panel ($0<z<3$), i.e. \autoref{eqn:PI} evaluated at the present day, for each of the galaxy populations. In other words, this is the cumulative impact of the tidal field experienced by the galaxy over around 90 per cent of cosmic time. The PI values for UDGs are significantly larger, with the median of the UDG distribution being around 2 orders of magnitude greater than that for the HSBGs.

We note that, if the definition of the perturbation index is changed so that it is independent of $R_{\mathrm{eff}}$ (by fixing $R_{\mathrm{eff}}$ to 1 kpc), the average perturbation index for UDGs remains significantly larger than for equivalent HSB galaxies. With such a change in definition, the median for UDGs remains 40 times higher than for HSBGs (compared to 160 times higher when radius is considered), indicating that the PI is a genuine result of stronger perturbations, rather than simply an effect of galaxy size.

It is important to note that the perturbations felt by UDGs are not a strong function of environment. As the dashed red line in the top panel and the dotted histograms in the bottom panel indicate, the majority of UDGs in field environments have still undergone very large perturbations compared with their HSB counterparts. Indeed the PI values of field UDGs are not dissimilar to that of the general UDG population (which is dominated by UDGs in groups and clusters). Finally, it is worth noting that if we only consider galaxies in low-density field environments which are not satellites, i.e. those that are truly isolated, the cumulative PI of such UDGs remains more than 10 times higher than that of field HSBGs.
Together with the fact that field UDGs have similar effective radii and star-forming gas fractions at the present day to UDGs in clusters (Figure \ref{fig:reff_fgas_evo}), this indicates that \textit{tidal interactions are likely to be the primary mechanism that drives LSBG evolution and causes these systems to both expand and lose their reservoir of star-forming gas over cosmic time.} 


\subsection{Ram pressure stripping - an additional mechanism of gas removal in cluster LSBGs}
While tidal perturbations are capable of acting on galaxies regardless of their environment, ram-pressure provides an additional process that can shape the evolution of galaxies in denser environments, particularly in clusters. The ram pressure exerted on the gas in a galaxy as it travels through a  hot intra-cluster medium (ICM) or intra-group medium (IGM) can remove gas from the galaxy and quench star formation \citep{Gunn1972}. This represents an appealing mechanism for explaining the transformation of galaxies from gas-rich, star-forming objects to quiescent systems that might resemble LSBGs at the present day. Indeed, the interaction between the ICM/IGM and the inter-stellar media of galaxies that are traversing hot, dense environments has often been used to explain the deficiency of gas and the redder colours of galaxies in clusters \citep[e.g.][]{Chamaraux1980,Lee2003,Sabatini2005,Boselli2008,Gavazzi2013,Habas2018}. This is a particularly effective mechanism in low-mass galaxies ($M_{\star}<10^{10} \mathrm{M}_{\odot}$), as gravitational potentials are typically shallow enough to allow the efficient removal of gas \citep[e.g.][]{Vollmer2001}. In this section, we explore whether ram pressure stripping may play a role in the gas exhaustion that creates our sample of LSBGs. 

\subsubsection{Ram pressure}
\label{sec:RP}
The cumulative ram pressure, $P_{\mathrm{ram}}$, felt between $z=3$ and $z$ by a galaxy moving through the local medium is given by
\begin{equation}
\label{eqn:ram-pressure}
P_{ram} \sim \int_{z=3}^{z} \rho_{\mathrm{IGM}} v_{\mathrm{gal}}^{2} \ dt \ / \ {\rm Gyr}
\end{equation}

\noindent where $v_{gal}$ is the velocity of the galaxy relative to the bulk velocity of the surrounding medium and $\rho_{\mathrm{IGM}}$ is the mean gas density of the surrounding medium within 10 times the maximum extent of the stellar distribution of the galaxy.

\begin{figure}
	\centering
    \subfigure{\includegraphics[width=0.45\textwidth]{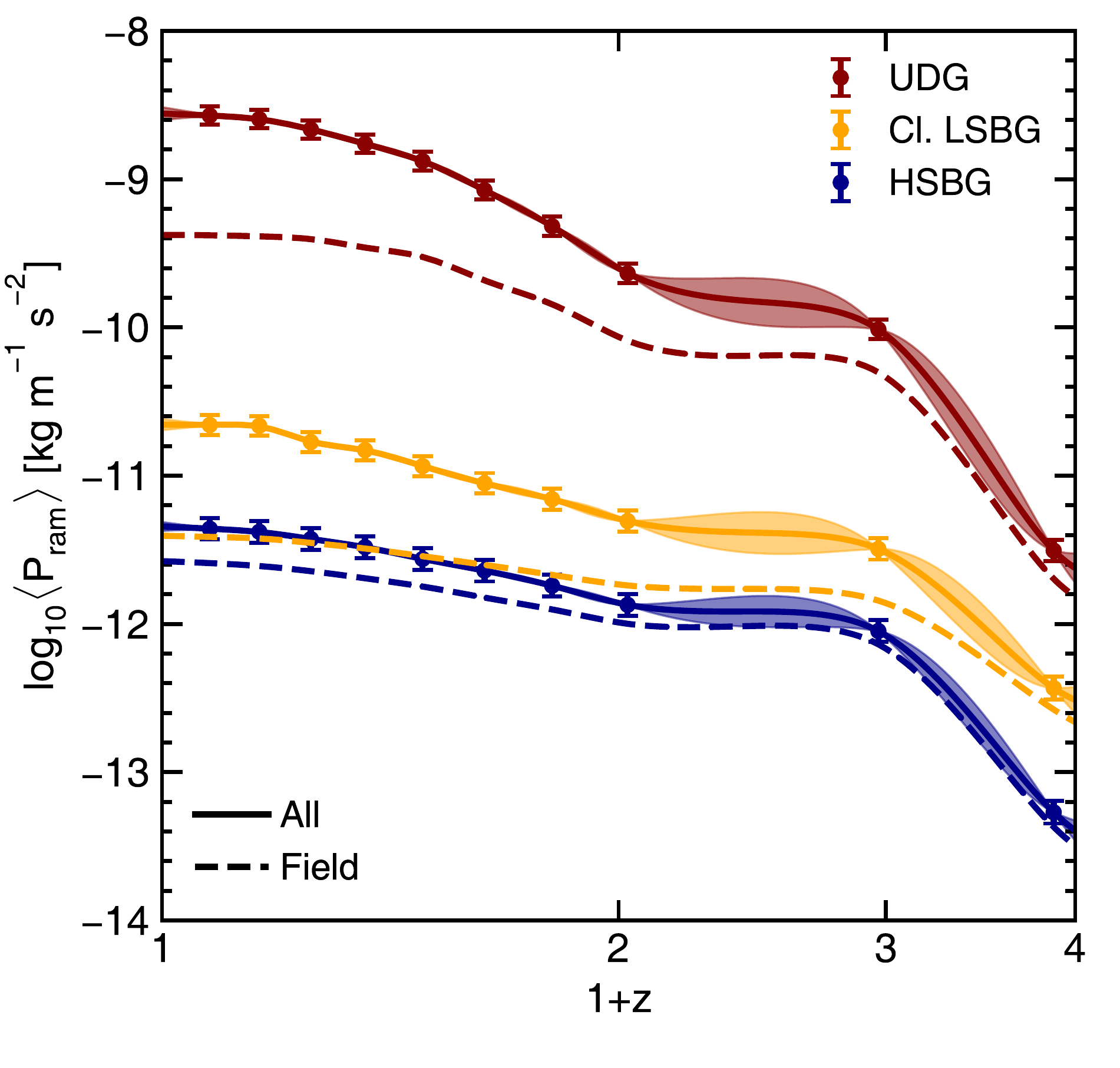}\label{fig:a}}
    \subfigure{\includegraphics[width=0.45\textwidth]{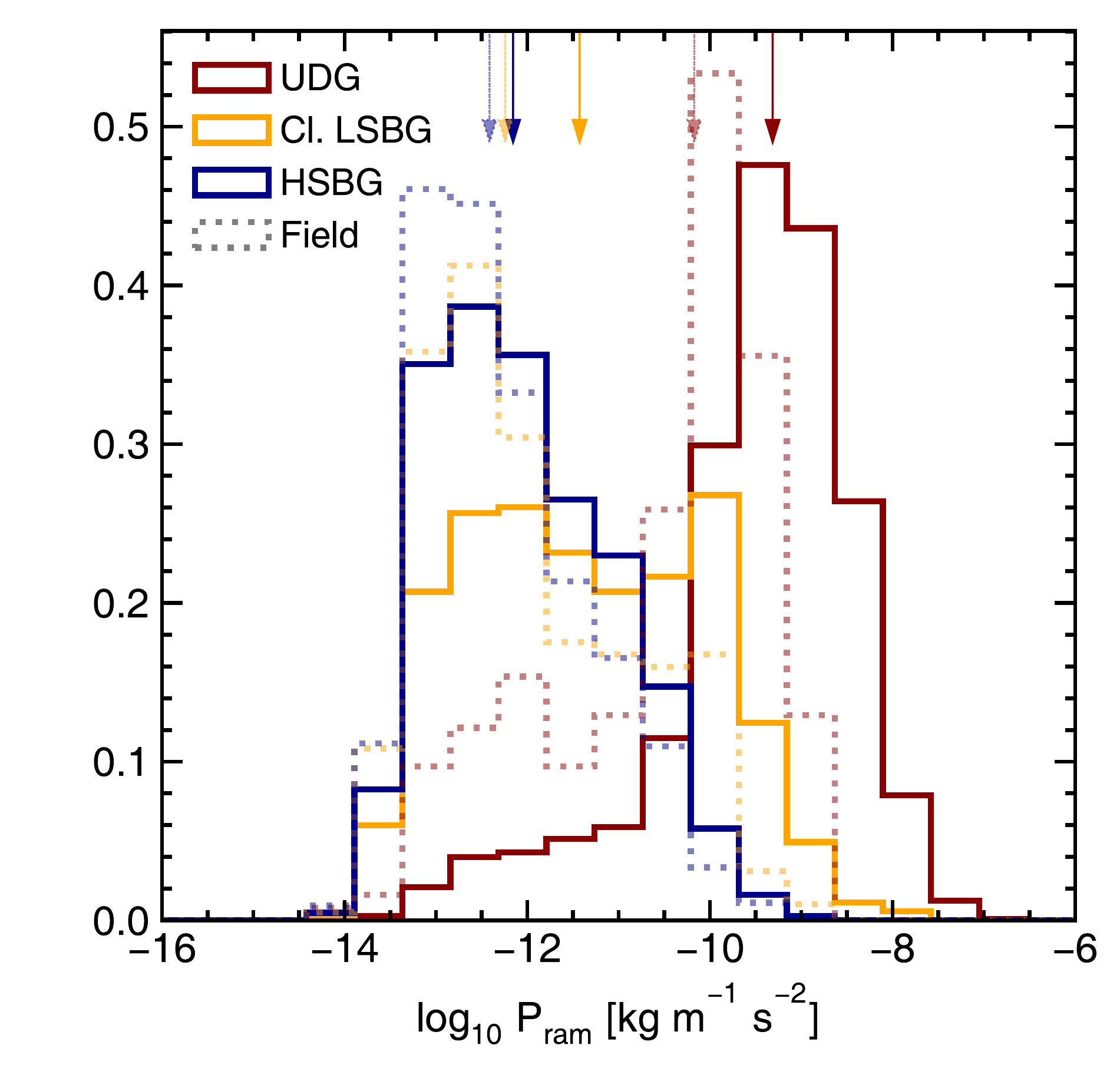}\label{fig:b}}
    \caption{\textbf{Top}: The cumulative ram pressure felt by galaxies between $z=3$ and the redshift in question, as defined by \autoref{eqn:ram-pressure}. Error bars indicate the errors on the median value of $P_{ram}$ at each redshift and solid filled regions show the $1\sigma$ confidence intervals for a Gaussian process regression to these points. Dashed lines show the cumulative ram pressure felt by field galaxies only. \textbf{Bottom}: Distribution of the cumulative ram pressure felt by galaxies between $z=3$ and $z=0$. Coloured arrows indicate the median value for each histogram and fainter arrows indicate the median values for galaxies in the field. Dotted lines indicate the total integrated ram pressure for field galaxies only. Note that the histograms are normalised so that the field and general populations can be easily compared.}
    \label{fig:ram_pressure_stripping}
\end{figure}

The top panel of \autoref{fig:ram_pressure_stripping} shows the median cumulative value of $P_{\mathrm{ram}}$ for the HSBG, Cl. LSBG and UDG populations as a function of redshift. The average ram-pressure continues to increase towards the present day for UDGs, Cl. LSBGs and HSBGs. However, the average ram-pressure felt by HSBGs and Cl. LSBGs is relatively small at all redshifts (around 2--3 orders of magnitude smaller than that of the UDG population). Ram-pressure stripping begins to have a significantly stronger impact on UDG progenitors around $z=1$. This is consistent with the typical infall epoch of galaxies into clusters \citep[e.g.][]{Tormen1998,Muldrew2015,Mistani2016,Muldrew2018}.

The cumulative ram pressure experienced by the progenitors of UDGs in the field (dashed red line) is significantly lower (by an order of magnitude) than the general population of UDG progenitors. Although the level of ram pressure in these field UDGs is high compared to that in Cl. LSBGs and HSBGs, it is low enough that significant gas stripping does not occur (as indicated by the relatively high total gas fractions retained by field UDGs at $z=0$, shown in the bottom panel of Figure \ref{fig:reff_fgas_evo}). The bottom panel of \autoref{fig:ram_pressure_stripping} shows the cumulative ram pressure experienced by HSBGs, Cl. LSBGs and UDGs between $z=0$ and $z=3$. Again, the cumulative ram pressure felt by UDGs is, on average, several orders of magnitude higher than that felt by either the Cl. LSBGs or HSBGs. 

It is worth noting here that the ram pressure experienced by UDGs in the field is higher than that experienced by Cl. LSBGs and HSBGs. This is a consequence of the fact that a larger fraction ($\sim65$ per cent) of local UDGs are satellites (i.e. their haloes are identified as sub-structures of a more massive halo) while a majority of low-mass field HSBGs at $z=0$ are not (only $\sim25$ per cent of these galaxies are satellites). UDGs are therefore typically found in regions of slightly higher gas density and experience ram pressure due to the host halo they are embedded in \citep[e.g.][]{Simpson2018}. When genuinely isolated UDGs are selected (i.e. those that are not satellites), the ram pressure felt falls significantly so that the median cumulative ram pressure felt by completely isolated UDGs, LSBGs and HSBGs agrees to within 0.2 dex.
\subsubsection{Bulk flow of gas}
\label{sec:flow}
Studying the bulk flow of gas within galaxies also allows us to quantify the degree to which ram-pressure stripping is experienced by our different galaxy populations. We explore the density weighted average angle, $\theta$, between the relative velocity between the gas and stars ($\textbf{v}_{rel}$) and the bulk motion of the stellar component in the observed frame ($\textbf{v}_{\star}$):
\begin{equation}
\label{eqn:theta}
\cos(\theta) = \frac{1}{\sum \rho_{i}}\sum_{i} \frac{\textbf{v}_{rel,i}\cdot \langle \textbf{v}_{\star} \rangle}{|\textbf{v}_{rel,i}| \cdot |\langle \textbf{v}_{\star} \rangle|}\rho_{i}
\end{equation}

\noindent where $\textbf{v}_{rel} = \textbf{v}_{gas} - \langle \textbf{v}_{\star} \rangle$ is the velocity of each gas cell relative to the average velocity of the galaxy's stellar component. In the case where the bulk motion of the gas is in the opposite direction to the stars, $\theta$ will be close to $\pi$ radians (and cos$(\theta)$ will be close to -1). When the gas and stellar components are moving together at roughly the same velocity, the angle between a given component of $\textbf{v}_{rel}$ and $\langle\textbf{v}_{\star}\rangle$ is essentially randomly distributed and therefore $\theta$ will be close to $\pi/2$ (i.e. $\cos(\theta)=0$). If the gas is either moving ahead of the stellar component of the galaxy, or being accreted in a wake behind the galaxy \citep[e.g][]{sakelliou2000}, then $\theta$ will be close to 0. When ram-pressure stripping occurs, we therefore expect $\theta$ to be close to $\pi$ radians and cos$(\theta)$ to be close to -1. Note that gas loss as a result of mechanisms other than ram pressure stripping does not produce the same signature. For example, in the case of gas loss driven by harassment or feedback processes, gas moves out of the galaxy either in a random direction or approximately isotropically, so the average value of cos$(\theta)$ will be close to 0.

\begin{figure}
	\centering
    \includegraphics[width=0.45\textwidth]{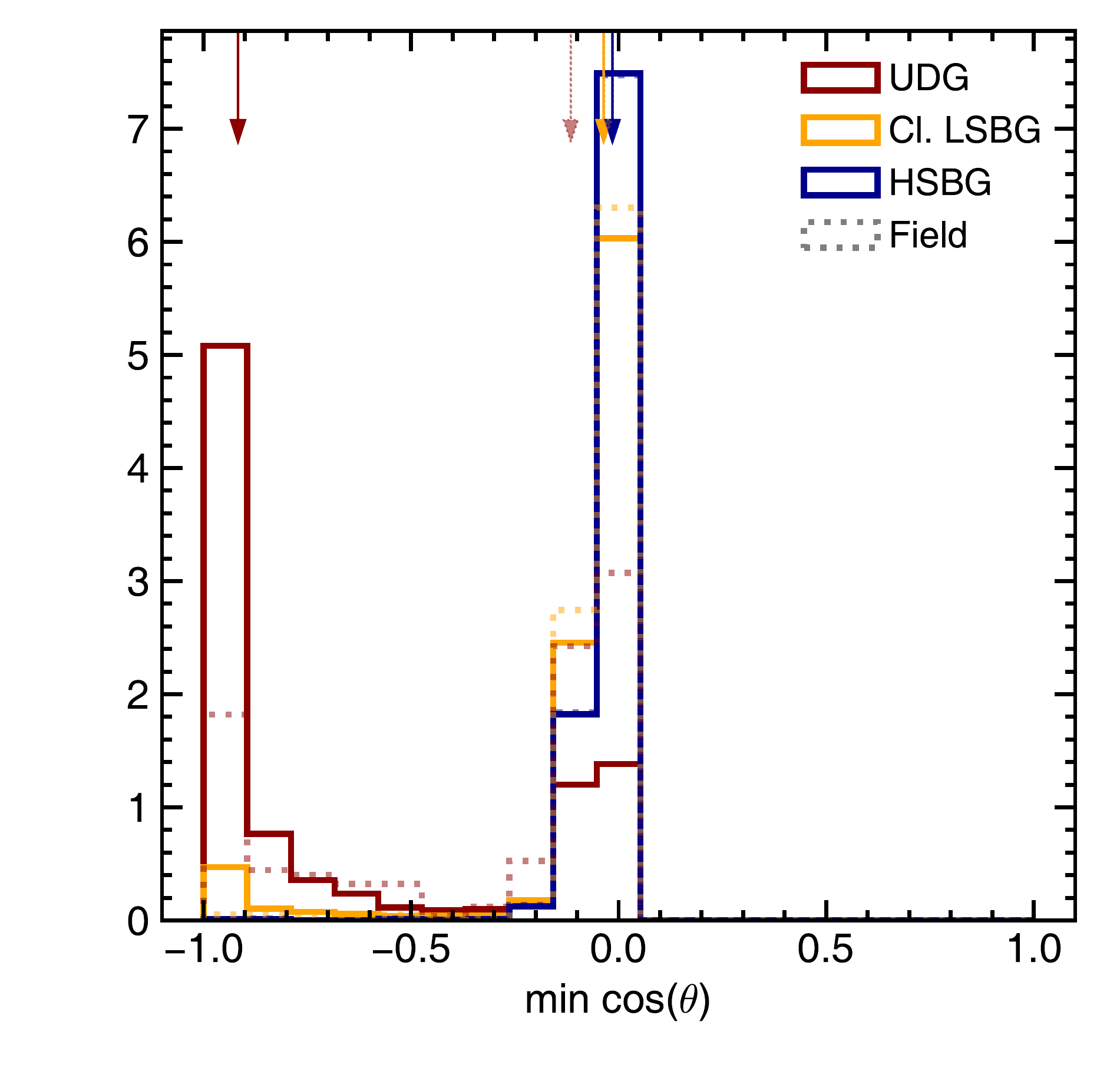}\label{fig:b}
    \caption{The minimum value of $\cos \theta$ between $z=3$ and $z=0$, where $\theta$ is the angle between the average direction of the bulk motion of gas relative to that of the stellar component in galaxies. Dotted lines indicate the largest value of $\cos \theta$ for field galaxies only. Coloured arrows indicate the median value for each histogram and fainter dotted arrows indicate the median values for field galaxies only. Note that the histograms are normalised so that the field and general populations can be easily compared.}
    \label{fig:ram_pressure_stripping2}
\end{figure}

\autoref{fig:ram_pressure_stripping2} shows the \textit{minimum} value of $\cos(\theta)$ that galaxies exhibit over cosmic time. Thus, minimum $\cos(\theta)$ values close to 0 would indicate that the ram pressure has not operated on the galaxy at any point over cosmic time. On the contrary, if we consider galaxies to have undergone some ram-pressure stripping when the minimum value of $\cos(\theta)$ is less than -0.75, then a large majority (65 per cent) of UDGs have undergone ram pressure stripping at some point in their history. The same is not true of Cl. LSBG or HSBG progenitors (or to a large extent, field UDGs). By the same definition, almost none of the HSBGs in our sample (0.3 per cent) have ever undergone significant ram-pressure stripping and a small minority of Cl. LSBGs have (6 per cent). In the field, only a modest fraction (25 per cent) of field UDGs have been ram-pressure stripped.

Taken together, \autoref{fig:ram_pressure_stripping} and \autoref{fig:ram_pressure_stripping2} indicate that ram-pressure stripping make a significant contribution to the quenching of UDG progenitors in dense environments. However, UDGs in the field are not as significantly stripped (as shown by both panels of \autoref{fig:ram_pressure_stripping}) but still have very low \textit{star-forming} gas fractions at $z=0$ (panel d of \autoref{fig:properties}). This indicates that ram-pressure stripping is not a necessary ingredient for the low star formation rates seen in today's UDGs. The high total gas fractions and low \textit{star-forming} gas fractions of field UDGs indicates that, for this subset of UDGs, their gas has been heated by other processes rather than been entirely removed from the galaxy. Thus, in cases where ram-pressure stripping is absent, other processes still act to quench UDGs by heating their gas. While ram-pressure stripping is an important mechanism for removing gas from UDGs in dense environments, UDGs (in all environments) lose their star-forming gas  through tidal perturbations, even in the absence of this process. Ram-pressure stripping is, therefore, an \textit{additional} process, to tidal perturbations, that assists in the removal of gas in LSBGs, particularly in clusters, but is not necessary for quenching their star formation. Interaction with the tidal field remains the principal driver of LSBG evolution in all environments.


\section{Summary}
\label{sec:conclusion}
In the forthcoming era of deep-wide observational surveys, the low-surface-brightness Universe represents an important new frontier in the study of galaxy evolution. While largely uncharted, due to the lack of depth of past wide-area datasets like the SDSS, low-surface-brightness galaxies (LSBGs) are essential to a complete understanding of galaxy evolution. Recent work using small deep surveys has hinted at the significant contribution that LSBGs may make to the galaxy number density of the local Universe and highlighted the need to understand the evolution of these objects across all local environments. Given the current dearth of data on LSBGs, theoretical insights, using cosmological simulations, into their demographics, the redshift evolution of their properties and the principal mechanisms that drive their formation is highly desirable. 

Here, we have used the Horizon-AGN hydrodynamical cosmological simulation to perform a comprehensive study of the formation and evolution of LSBGs. We have (1) studied the demographics and properties of local LSBGs and compared them to that of their high-surface-brightness (HSB) counterparts, (2) explored the evolution of the properties of LSBG progenitors with redshift and (3) quantified the role of key processes, in particular SN feedback, tidal perturbations and ram-pressure stripping, that lead to the formation of LSB systems. Our main conclusions are as follows:

\begin{figure}
	\centering
    \begin{tikzpicture}
    	\draw (0,0) node(start)[ellipse, minimum height=.25cm,minimum width=.25cm,draw,fill=blue!60,blue!60] {};
        \draw (0,-2.) node(2nd)[ellipse, minimum height=.5cm,minimum width=.5cm,draw,fill=blue!40,blue!40] {};
        \draw[->,black,line width=1pt] (start) -- (2nd);
        
        \draw[->,black] (0.1,0.) -- (0.35,0);
        \draw[->,black] (0.031,0.0951) -- (0.108,0.333);
        \draw[->,black] (-0.081,0.0587) -- (-0.283,0.206);
        \draw[->,black] (-0.081,-0.0587) -- (-0.283,-0.206);
        \draw[->,black] (0.031,-0.0951) -- (0.108,-0.329);
        
        \draw (3.2,-3.4) node(x)[ellipse, minimum height=3.5cm,minimum width=3.5cm,draw,fill=gray!25,gray!25] {};
        \draw (2.5,-2.7) node(cluster)[ellipse, minimum height=0.8cm,minimum width=0.8cm,draw,fill=red!20,red!20] {};
        \draw[->,black,line width=1pt] (2nd) to [out=-45, in=-225] (cluster.north west);
        \draw (3,-4.5) node(x)[ellipse, minimum height=0.25cm,minimum width=0.5cm,rotate=120,draw,fill=red!50,red!50] {};
        \draw (2.4,-4.) node(x)[ellipse, minimum height=0.25cm,minimum width=0.3cm,rotate=70,draw,fill=red!60,red!60] {};
        \draw (3.2,-3.4) node(x)[ellipse, minimum height=0.25cm,minimum width=.35cm,rotate=230,draw,fill=red!80,red!80] {};
        \draw (3.2,-1.9) node(x)[ellipse, minimum height=0.2cm,minimum width=0.4cm,rotate=290,draw,fill=red!35,red!35] {};
        \draw (4.,-4.2) node(x)[ellipse, minimum height=0.05cm,minimum width=0.3cm,rotate=15,draw,fill=red!35,red!35] {};
        \draw (3.7,-4.7) node(x)[ellipse, minimum height=0.2cm,minimum width=0.15cm,rotate=100,draw,fill=red!70,red!70] {};
        \draw[thick,blue!80!black] [domain=0:11,variable=\t,smooth,samples=75, yshift=-2.5cm,xshift=4cm]
        plot ({\t r}: {0.001*\t*\t})
        plot ({\t r}: {-0.001*\t*\t});
        \draw[thick,blue!10!black] [domain=0:9,variable=\t,smooth,samples=75, yshift=-4.7cm,xshift=2.3cm]
        plot ({\t r}: {0.0015*\t*\t})
        plot ({\t r}: {-0.0015*\t*\t});
        \draw[thick,blue!40!black] [domain=0:15,variable=\t,smooth,samples=75, yshift=-3.1cm,xshift=2.0cm]
        plot ({\t r}: {0.0008*\t*\t})
        plot ({\t r}: {-0.0008*\t*\t});
        \draw[thick,blue!55!black] [domain=0:15,variable=\t,smooth,samples=75, yshift=-3.5cm,xshift=4.3cm]
        plot ({\t r}: {0.0009*\t*\t})
        plot ({\t r}: {-0.0009*\t*\t});

        \draw (0,-4) node(field)[ellipse, minimum height=0.8cm,minimum width=0.8cm,draw,fill=red!20,red!20,rotate=120] {};
        \draw[->,black,line width=1pt] (2nd) -- (field);
        
        \draw (-0.5,-3.2) node(enc11)[ellipse, minimum height=0.,minimum width=0.,draw,fill=red!35,red!35] {};
        \draw (0.4,-5.5) node(enc10)[ellipse, minimum height=0.,minimum width=0.,draw,white] {};
        \draw[->,line width=0.7pt,gray,dashed] (enc10) to [out=45, in=-45] (enc11.south east) (enc11);
        \draw (-0.85,-3.7) node(enc21)[ellipse, minimum height=0.2cm,minimum width=0.4cm,draw,white] {};
        \draw (1.1,-3.2) node(enc20)[ellipse, minimum height=0.,minimum width=0,draw,white] {};
        \draw[-,line width=0.7pt,gray,dashed] (enc20) to [out=-60, in=-45] (enc21.south east) (enc21);
        \draw (-0.7,-4.9) node(enc31)[ellipse, minimum height=0.2cm,minimum width=0.4cm,draw,white] {};
        \draw (0.9,-3.2) node(enc30)[ellipse, minimum height=0.,minimum width=0,draw,white] {};
        \draw[-,line width=0.7pt,gray,dashed] (enc30) to [out=-90, in =-330] (enc31);
        
        \node [right, align=justify, text width=4cm] at (1.0, -0.35) {(i) rapid star formation at high redshift drives strong stellar feedback which creates flatter gas density slopes which then produce shallower stellar density slopes.};
        \node [right, align=justify,text width=3.3cm] at (-1.7, -6.4) {(iia) shallower density slopes make UDGs more susceptible
        to galaxy--galaxy interactions which heat the gas and `puff up' the stellar components, producing diffuse, gas-poor systems.};
        \node [right, align=justify,text width=3.3cm] at (2.0, -6.4) {(iib) cluster UDGs are processed further as they fall into these dense environments, undergoing ram-pressure stripping in addition to heating by the ambient tidal field.};
        
        \node [right, align=justify,color=red!70!black] at (0.45, -2.15) {40\%};
        \node [right, align=justify, color=green!50!black] at (0, -3.3) {10\%};
        \node [right, align=justify, color=blue!70!black] at (0.6, -3.3) {/ 50\%};

    \end{tikzpicture}
    \caption{A summary of the formation mechanisms of our sample of  UDGs. 40 per cent of these galaxies are found in high-density (cluster) environments at $z=0$, while 10 and 50 per cent are found low density (field) and intermediate density (group) environments, as indicated by the text next to each arrow.}
    \label{fig:cartoon}
\end{figure}
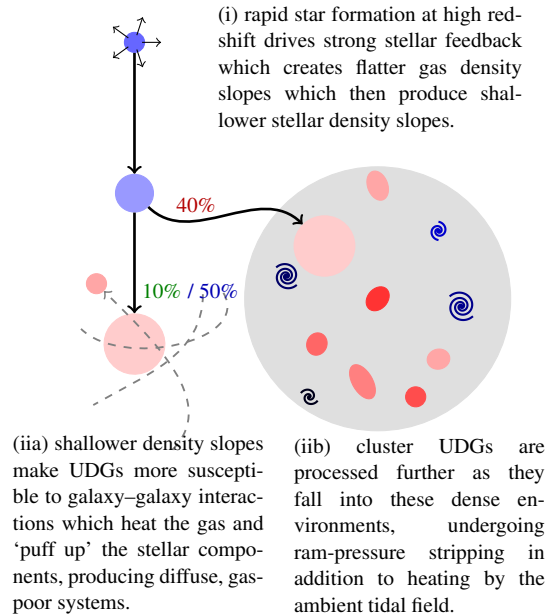

\begin{itemize}
\item \textit{LSBGs are significant contributors to the number density of galaxies in the local Universe.} For $M_{\star}$>$10^{8}~$M$_{\odot}$, LSBGs contribute 47 per cent of the local number density ($\sim$85 per cent for $M_{\star}$>$10^{7}~$M$_{\odot}$). They are, however, minority contributors to the local stellar mass and luminosity densities. For $M_{\star}$>$10^{8}~$M$_{\odot}$ ($M_{\star}$>$10^{7}~$M$_{\odot}$), the LSBGs contribute 7 (11) per cent and 6 (10) per cent to the stellar mass and luminosity densities respectively.
\\
\item \textit{Local LSBGs have similar dark matter fractions and angular momenta as their HSB counterparts but exhibit larger effective radii (2.5$\times$ for UDGs), older stellar populations (1.6$\times$ for UDGs), lower gas fractions (no star-forming gas remaining in most UDGs) and shallower density profiles}.
\\
\item \textit{LSBGs evolve from the same progenitor population as HSBGs at high redshift.} HSBGs and LSBGs originate from populations with almost identical gas fractions and effective radii at $z=3$ and evolve along the same locii in the $f_{gas}-R_{\mathrm{eff}}$ plane. However, the evolution of LSBGs (and UDGs in particular) is much more rapid, especially at $z<1$.
\\
\item \textit{UDGs experience more rapid star formation between $z=3$ and $z=1$, which triggers their creation and ultimate divergence from the HSB population.} More rapid star formation in UDG progenitors produces more concentrated SN feedback which, in turn, leads to shallower gas density profiles at high redshift ($z>1$) without quenching star formation. The star formation fuelled by this gas then produces systems which have shallow stellar density slopes (and larger effective radii). These systems are more susceptible to processes like tidal heating of both stars and gas by the ambient tidal field, and ram-pressure stripping of gas in denser environments.
\\
\item \textit{External processes (tidal perturbations and ram-pressure stripping) that drive most of the evolution of LSBGs are principally effective at low and intermediate redshifts.} At $z<1$, the total and star-forming gas fractions and effective radii of LSBGs, and UDGs in particular, change drastically after fairly gradual evolution between $z=3$ and $z=1$.
\\
\item \textit{Tidal heating (regardless of local environment) is able to produce the large sizes and low star-forming gas fractions of today's UDGs.} Flattened density profiles, produced via stronger SN feedback, are amplified by the ambient tidal field, further broadening the stellar distributions. UDGs, regardless of environment, undergo tidal perturbations of similar magnitude, with field UDGs exhibiting similar effective radii to their group/cluster counterparts at the present day. In a similar vein, tidal heating is also able to prevent gas from forming stars in UDG progenitors, regardless of their local environment. Even in field environments, where field UDGs remain star-forming down to low redshift, the tidal field is able to continually heat the gas in a large number of these systems, effectively quenching their star formation by $z=0.25$.
\\
\item \textit{In clusters, ram-pressure stripping is a significant additional mechanism that removes gas from in-falling UDG progenitors, starting around $z=1$.} Although ram-pressure stripping is very effective at stripping gas in dense environments, it acts as a secondary mechanism to tidal heating outside of these environments, for creating the low fractions of \textit{star-forming} gas found in UDGs at the present day. Our analysis shows that tidal heating would likely produce the low gas fractions found in cluster UDGs, even in the absence of ram-pressure stripping.
\\
\end{itemize}

\autoref{fig:cartoon} shows a summary of the evolutionary channels for LSBG formation described above. Our results offer insights into the formation of galaxies in the LSB regime which, given their dominance of the galaxy number densities, are essential pieces of the puzzle of galaxy evolution. Furthermore, as we have demonstrated in the analysis above, LSBGs are much more sensitive tracers of key processes that shape galaxy evolution (e.g. SN feedback, tidal perturbations and ram-pressure stripping) than their HSB counterparts. Without an understanding of the formation and evolution of LSBGs, therefore, our comprehension of galaxy evolution remains incomplete. 

The new era of deep-wide surveys like the Hyper Suprime Cam Subaru Strategic Program (HSC-SSP), and forthcoming datasets from instruments like LSST, Euclid and WFIRST will revolutionize the study of LSBGs, by yielding statistical samples of these systems, for the first time, across all environments. These datasets will enable us to perform the first statistical census of LSBG properties and their evolution with redshift, producing stringent tests of current theoretical predictions, such as those presented in this study. Together, this will create a platform for constructing a new generation of cosmological simulations, which offer a better understanding of processes (e.g. SN feedback, ram pressure stripping and tidal perturbations) to which the LSBG population is particularly sensitive, and a better reproduction of galaxies in the as-yet-unexplored LSB regime. This convergence of deep-wide surveys and cosmological hydrodynamical simulations is  likely to have a transformational impact on our understanding of galaxy evolution in the coming years. 


\section*{Acknowledgements}
We are grateful to the anonymous referee for several constructive suggestions that improved the quality of the original manuscript. We thank David Valls-Gabaud, Johan Knapen, Lee Kelvin, Cristina Martinez-Lombilla, Elias Brinks, Shaun Read and Sebastien Viaene for many interesting discussions. GM acknowledges support from the STFC [ST/N504105/1]. SK acknowledges support from the STFC [ST/N002512/1] and a Senior Research Fellowship from Worcester College Oxford. RAJ acknowledges support from the STFC [ST/R504786/1]. JD and AS acknowledge funding support from Adrian Beecroft, the Oxford Martin School and the STFC. CL is supported by a Beecroft Fellowship. CP acknowledges funding support from the ERC [670193].  This research has used the DiRAC facility, jointly funded by the STFC and the Large Facilities Capital Fund of BIS, and has been partially supported by grant Spin(e) ANR-13-BS05-0005 of the French ANR. This work was granted access to the HPC resources of CINES under the allocations 2013047012, 2014047012 and 2015047012 made by GENCI. This work is part of the Horizon-UK project.




\bibliographystyle{mnras}
\bibliography{paper_mnras} 

\appendix

\section{Extrapolating the mass and surface-brightness functions}
\label{sec:appendix_SB}
Since the Horizon-AGN mass function becomes incomplete as we approach the galaxy mass resolution limit of the simulation ($M_{\star} \sim 10^{8}$M$_{\odot}$), it is necessary to extrapolate the stellar mass and surface-brightness functions in order to obtain estimates of the contribution of LSBGs to the number, stellar mass and luminosity densities down to $M_{\star} \sim 10^{7}$M$_{\odot}$. 

\begin{figure}
	\centering
    \includegraphics[width=0.45\textwidth]{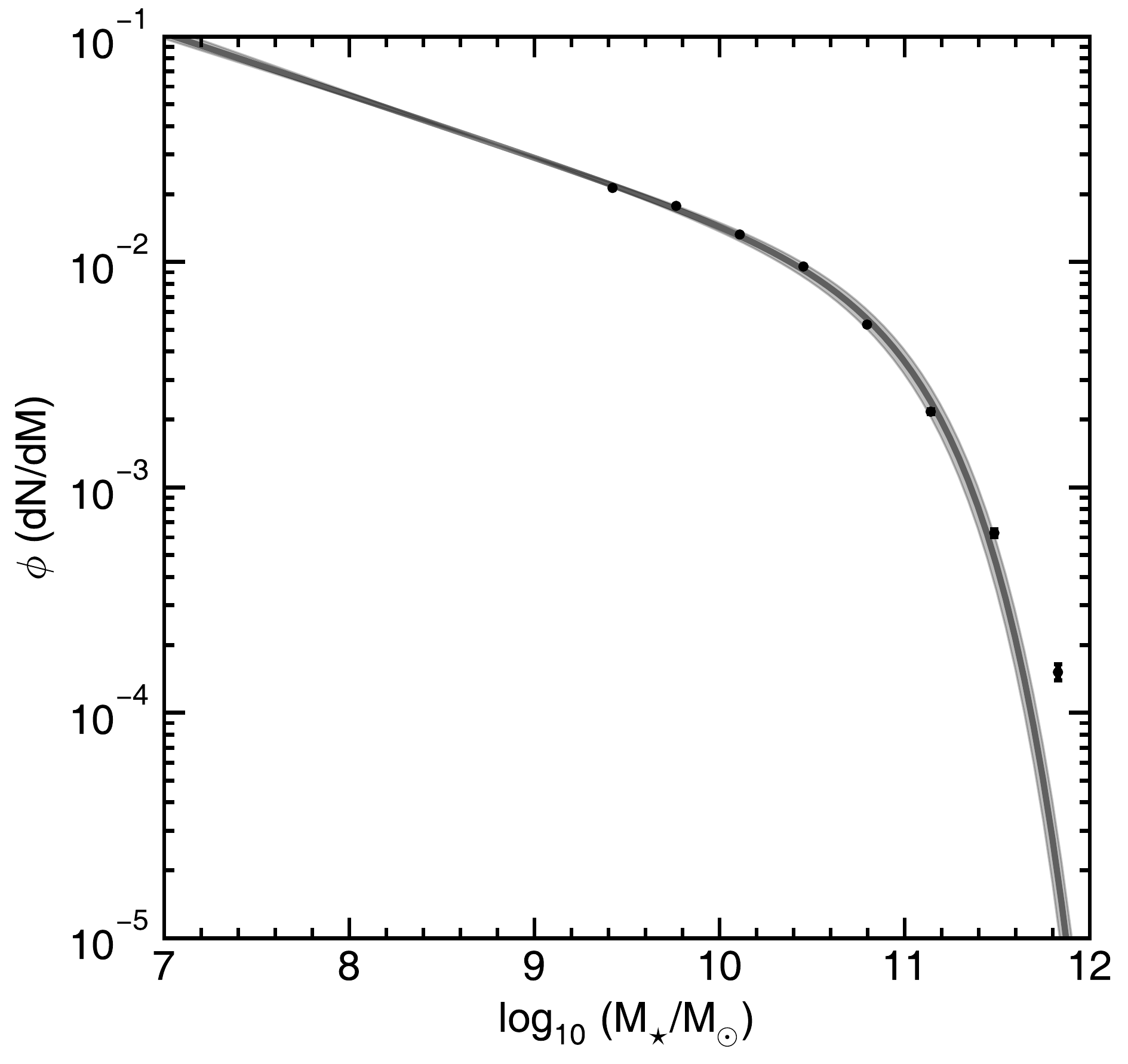}\label{fig:b}
    \caption{Schechter function fit to the Horizon-AGN mass function. Points with error bars indicate binned data with Poisson errors. The grey region indicates the 99 per cent confidence interval for the fit to the data.}
    \label{fig:m_fun}
\end{figure}

\begin{figure}
	\centering
    \includegraphics[width=0.45\textwidth]{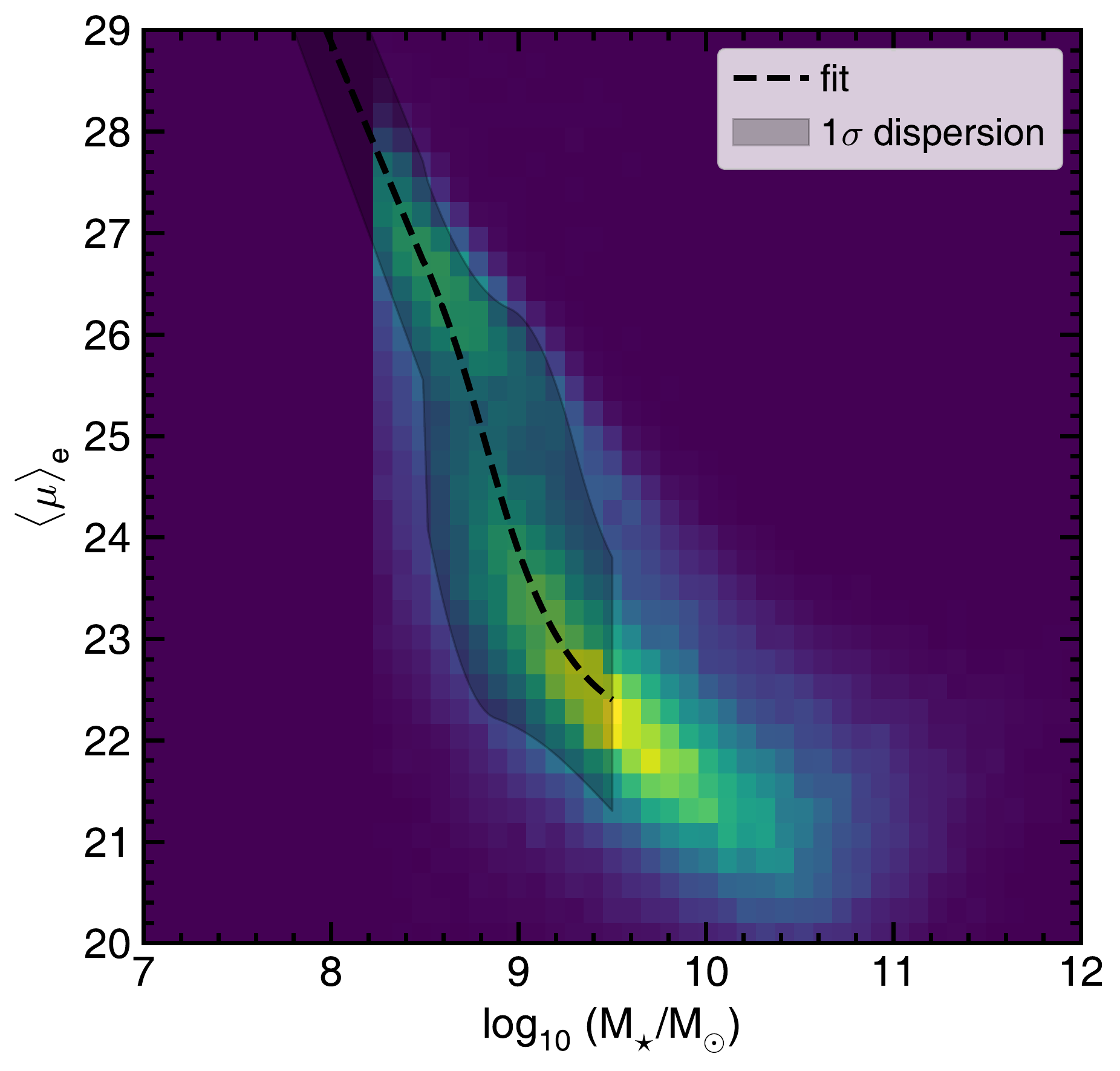}\label{fig:b}
    \caption{Density plot showing the distribution of galaxies in the stellar mass -- surface-brightness plane. The dashed black line indicates the fit to the data using a mixture of Gaussians and the filled grey region indicates the 1$\sigma$ dispersion.}
    \label{fig:m_vs_sb}
\end{figure}

To do this, we first fit a Schechter function \citep{Schechter1976} to the raw Horizon-AGN mass function yielding a slope of $1.23 \pm^{0.03}_{0.05}$ (\autoref{fig:m_fun}). We then use a Gaussian mixture model to fit the distribution of galaxies in the stellar mass -- surface-brightness plane, allowing us to estimate the shape and variance of the data. Beyond the resolution limit ($2\times10^{8}$M$_{\odot}$), we linearly extrapolate the variance down to $10^{7}$M$_{\odot}$ (\autoref{fig:m_vs_sb}).

In order to obtain the extrapolated surface-brightness function, we split the mass function into narrow mass bins and draw $N$ times from a Gaussian distribution with a variance and mean defined by our fit to the stellar mass -- surface-brightness distribution, where $N$ is the number of objects in that mass bin. Where the mass function is complete (above $10^{9}$M$_{\odot}$), we fill in the surface-brightness function using the raw data. The resultant surface-brightness is shown in \autoref{fig:sbfunction}.

\section{Effect of the spatial resolution of the simulation}
\label{sec:resolution}
In this section, we discuss the effect of the spatial resolution of the simulation (1~kpc) on the sizes (and therefore surface-brightnesses) of low-mass galaxies. Although the locus of the M$_{\star}$--R$_{\mathrm{eff}}$ distribution only barely reaches 1~kpc at our mass limit ($2\times10^{8}$M$_{\odot}$; e.g. see \autoref{fig:radii_compare}), it is still possible that the resolution may produce some spurious dynamical support. This problem would likely be compounded if the maximum resolution is not satisfied in the cells at the centres of our galaxies.

We first check the refinement level (i.e. the accuracy used by the gravity and hydrodynamics solvers; see \citet{Teyssier2002}) of the AMR grid within the central R$_{\mathrm{eff}}$. As noted in Section \ref{sec:simulation}, the AMR grid is refined according to a semi-Lagrangian criterion, where the refinement of a cell is approximately proportional to the total mass within the cell. \autoref{table:levels} shows the refinement of the AMR cells within 1 and 2 $R_{\mathrm{eff}}$ of each galaxy used for our sample of UDGs and HSBGs in Section \ref{sec:evolution}. On average almost 100 per cent of the AMR cells within $1~R_{\mathrm{eff}}$ of each galaxy are refined to the maximum level (level 17; 1~kpc) for both UDGs and HSBGs and within $2~R_{\mathrm{eff}}$. The value falls to 21 per cent for UDGs, owing to their larger effective radii (i.e. they extend much further from the centre of the total mass distribution). In both cases, all cells are refined to at least the second highest level (level 16; 2~kpc).

\begin{table}
\centering
\begin{tabular}{lll}
  \toprule UDGs & level 17 (1~kpc) & $\geqslant$ level 16 ($\geqslant$2~kpc)\\
  \midrule $R<R_{\mathrm{eff}}$ & $98 \%$ & $100 \%$ \\
  $R_{\mathrm{eff}}<R<2~R_{\mathrm{eff}}$ & $21 \%$ & $100 \%$\\
  \midrule HSBs & & \\
  \midrule $R<R_{\mathrm{eff}}$ & $100\%$ & $100\%$ \\
  $R_{\mathrm{eff}}<R<2~R_{\mathrm{eff}}$ & $88\%$ & $100\%$ \\
\bottomrule
\end{tabular}
\caption{The percentage of AMR cells within $1~R_{\mathrm{eff}}$ or $2~R_{\mathrm{eff}}$ that are refined to level 17 or at least level 16. The table is split between the UDG sample and the HSB sample.}
\label{table:levels}
\end{table}

We also check how the effective radii of equivalent galaxies in a higher resolution, 4000 Mpc$^{3}$ zoom-in of a region of Horizon-AGN (New Horizon; Dubois et al. in preparation) differ from those in the Horizon-AGN simulation. New Horizon has a spatial resolution 35~pc ($\times$30  Horizon-AGN) but uses the same underlying code \citep[\textsc{ramses}][]{Teyssier2002} and implements similar sub-grid prescriptions. The comparison is made at $z=0.7$, the lowest redshift to which the New Horizon simulation has been run. 

In order to produce a matching catalogue of galaxies, at the initial snapshot, the particle IDs of multiple (64) DM particles in the high resolution simulation were mapped onto each DM particle in Horizon-AGN. This allows us to match galaxy haloes between simulations and thereby attempt to find each galaxy's `twin' in the New Horizon simulation. We limit ourselves to haloes that share at least 75 per cent of the same DM particles, have at least 75 per cent of the mass of their matching halo and which host galaxies with stellar masses that are no more than a factor of two different from their twin. This yields a sample of 50 galaxies with masses between $2\times10^{8}~\mathrm{M}_{\odot}$ and $10^{10}~\mathrm{M}_{\odot}$.

\begin{figure}
	\centering
    \includegraphics[width=0.45\textwidth]{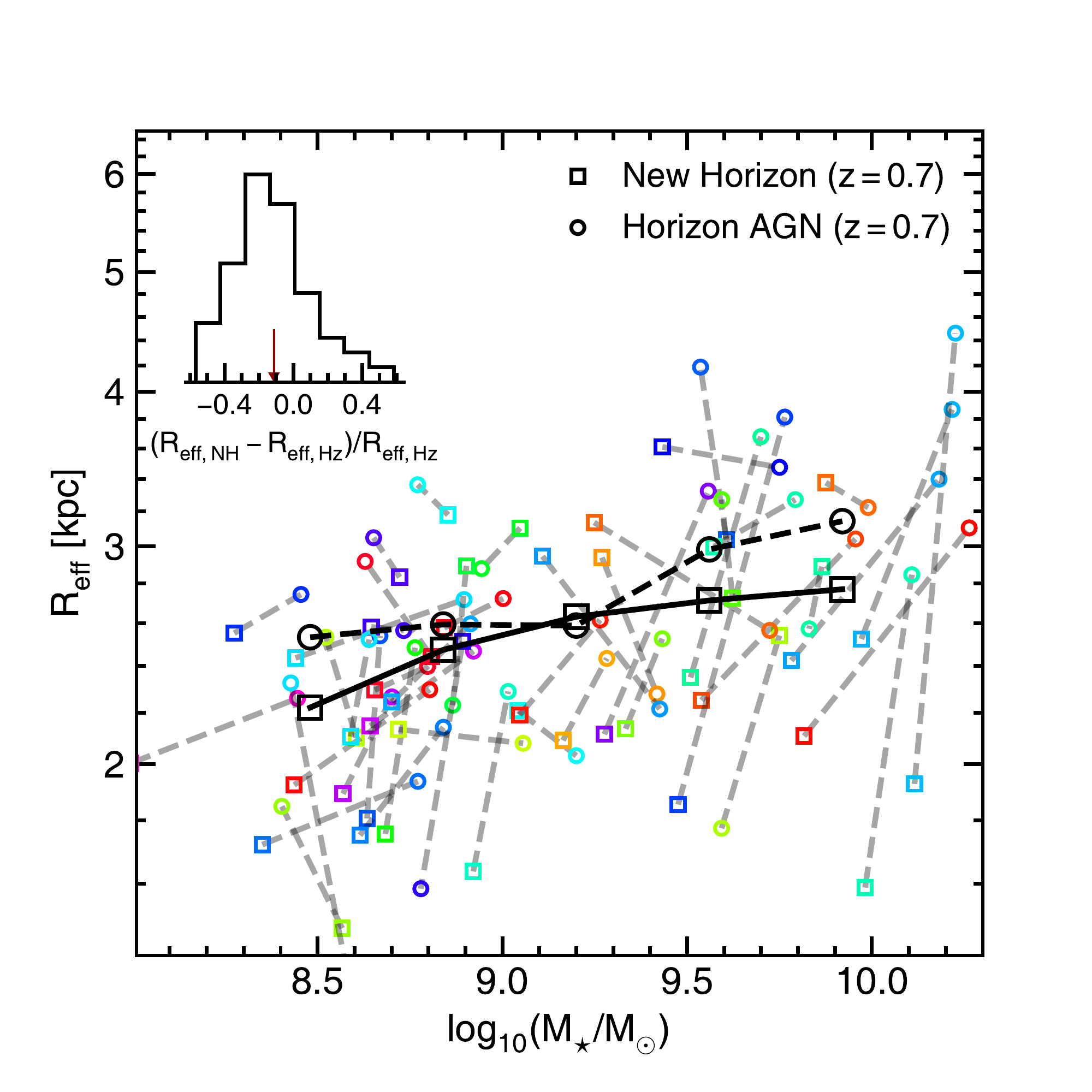}
    \caption{Comparison of the effective radius and stellar mass of galaxies with haloes matched between the Horizon-AGN simulation and higher-resolution (35~pc) New Horizon simulation. Square and circle markers of the same colour linked by a dashed line indicate the stellar mass and effective radius of a matched galaxy in the New Horizon simulation (open square) and the Horizon-AGN simulation (open circle). The black solid and dashed lines show the mean trend in $R_{\mathrm{eff}}$ vs $M_{\star}$ for New Horizon and the matching volume in Horizon-AGN respectively. Errors are not shown in the interest of legibility, but the typical error on the mean is $\pm0.1$~kpc in each bin. The inset plot shows the distribution of the fractional difference in $R_{\mathrm{eff}}$ between the two simulations and the red arrow shows the mean fractional difference.}
	\label{fig:NH_HZ_compare}
\end{figure}

\autoref{fig:NH_HZ_compare} shows the effective radii and stellar masses of galaxies that we were able to robustly match between the two simulations, with each pair of twin galaxies joined by a dashed grey line. While the much higher resolution of the New Horizon simulation produces differences in the accretion histories and star formation of haloes compared with their twin haloes in the Horizon-AGN simulation, the lower resolution of the Horizon-AGN simulation does not produce a significant systematic offset in galaxy sizes. On average, galaxies in the Horizon-AGN simulation have only marginally larger sizes. The mean of the distribution of size offsets in \autoref{fig:NH_HZ_compare} (denoted by a red arrow) is $0.1\pm0.04$, so that galaxies in Horizon-AGN are only 10 per cent larger, on average, than their twin in New Horizon. Note that the higher-resolution twin is often the larger of the two ($26$ per cent of higher-resolution galaxies are slightly larger).

The black lines in \autoref{fig:NH_HZ_compare} show the trend in $R_{\mathrm{eff}}$ vs $M_{\star}$ for the whole sample of galaxies within the same volume as New Horizon, regardless of whether they are reliably matched. Again, the average sizes of galaxies in Horizon-AGN are only around 10 per cent larger than equivalent mass galaxies in New Horizon.

We note however, that there is some degree of systematic offset between the average sizes of the simulated and observed galaxies (e.g. see blue filled points vs blue open points in Figure \ref{fig:radii_compare}). This is especially pronounced at the high stellar mass end ($\sim 10^{11.5}$M$_{\odot}$), where, compared to \citet[][open circles]{Cappellari2011}, simulated galaxy sizes are around 1.5 -- 2 times larger than observed galaxies of equivalent mass. Towards lower masses ($\sim 10^{10}$M$_{\odot}$), the typical sizes of simulated galaxies are only 1.15 times larger than those of observed galaxies. This may be an indication that the AGN feedback prescription produces artificially large galaxies at high stellar mass (where AGN feedback is most efficient), but is not such an important effect at low masses, where AGN feedback becomes relatively unimportant compared to stellar feedback. For example, \citet[][see their Figure 1]{Peirani18} show that the AGN feedback implementation used in Horizon-AGN produces galaxies that are larger than observed galaxies compared to the same simulation with no AGN feedback at masses of $M_{\star}\approx10^{11.5}$M$_{\odot}$.

\section{Relevance of this study to observed LSBG populations}
\label{sec:UDG_MF}
\begin{figure}
	\centering
    \includegraphics[width=0.45\textwidth]{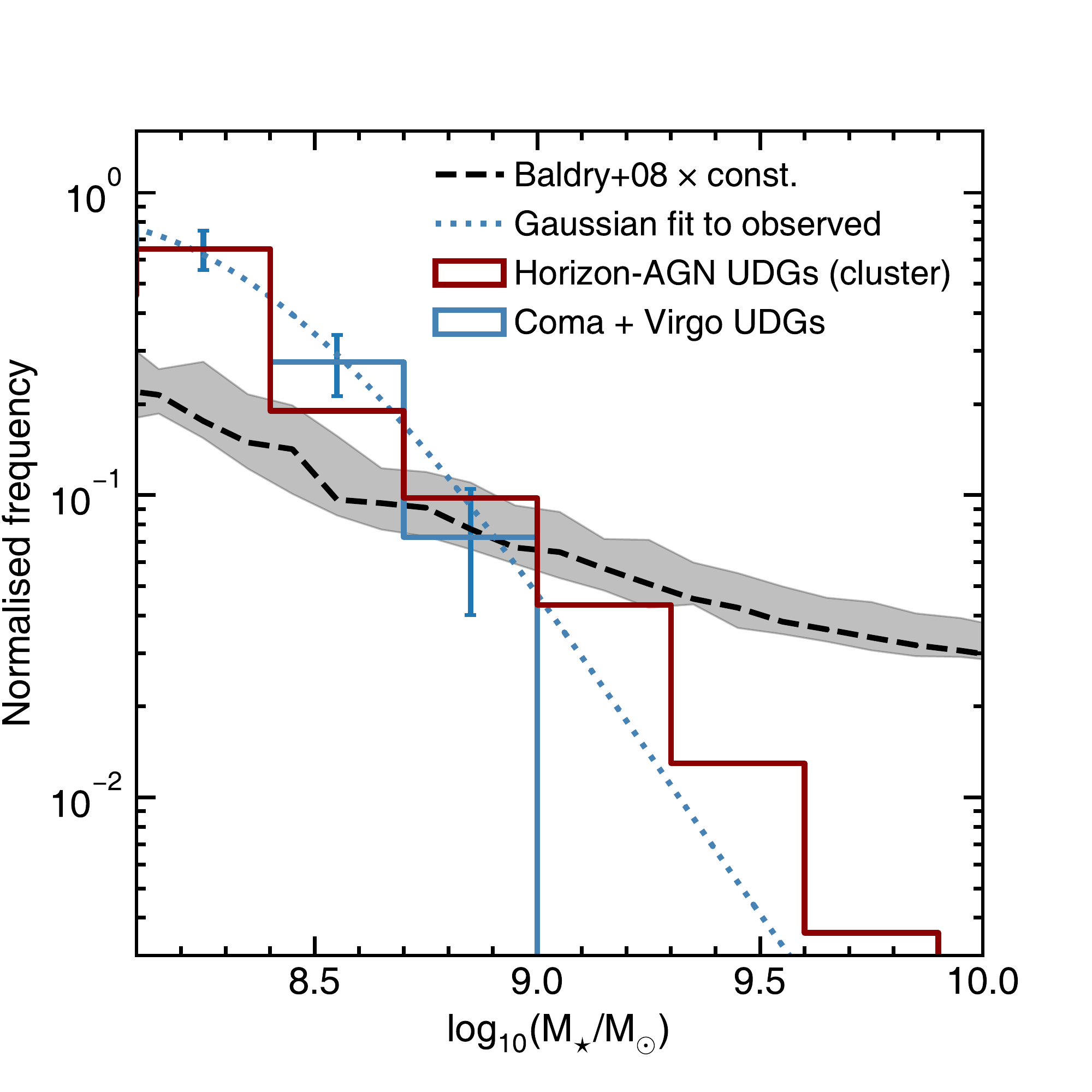}
    \caption{The blue histogram shows the normalised stellar mass distribution for UDGs in the Coma and Virgo clusters (red open squares in \autoref{fig:radii_compare}) and the blue dotted line shows a log-normal fit to the full distribution of masses. The red histogram shows the stellar mass distribution for Horizon-AGN cluster UDGs after an extrapolation of the stellar mass function has been performed as detailed in Appendix \ref{sec:appendix_SB}. The Horizon-AGN UDG and observed cluster UDG histograms are both normalised by dividing by the number of counts in the three bins where the datasets overlay (between $10^{8}$M$_{\odot}$) and $10^{9}$M$_{\odot}$. The dashed black line indicates the non-parametric galaxy stellar mass function from \citet{Baldry2008} multiplied by a constant for clarity and the shaded region indicates the error.}
	\label{fig:UDF_MF}
\end{figure}

In this section, we discuss the relevance and applicability of this study to observed UDG populations. \autoref{fig:UDF_MF} compares the mass distribution of cluster UDGs in the Horizon-AGN simulation (with a correction for incompleteness applied as detailed in Appendix \ref{sec:appendix_SB}) to that of observed UDGs in the Coma and Virgo clusters \citep{vanDokkum2015,Mihos2015,Yagi2016,Gu2018}. Within the mass range where both samples overlap, there is good agreement between the mass distribution of Horizon-AGN UDGs (red histogram) and the observed cluster UDGs (blue histogram). Assuming a log-normal distribution for the observed sample, we also find that Horizon-AGN agrees within a factor of a few with the extrapolated value for the observed sample for $M_{\star}>10^{9}$M$_{\odot}$. 

Although high mass UDGs ($M_{\star}>10^{9}$M$_{\odot}$) are largely missing from observations, the very limited volumes explored observationally to date do not preclude the existence of galaxies with significantly larger stellar masses that satisfy the same low-surface-brightness criteria as their less massive counterparts. Indeed, examples of such massive LSBGs are already known e.g. Malin 1 and UGC 1382 (see large open red squares in Figure \ref{fig:radii_compare}). Furthermore, the dashed black line in \autoref{fig:UDF_MF} indicates the galaxy stellar mass function from \citet{Baldry2008}. A decline in the UDG fraction towards higher stellar masses should be expected and is likely driven by a combination of a mass dependence in the efficiency of the physical processes (e.g. SN feedback) that drive the formation of LSBGs \citep[e.g.][]{Brook2015,vanDokkum2016,Toloba2018} and the steep decline in the galaxy stellar mass function towards higher stellar masses (which is exacerbated by the small observational volumes probed so far).

\bsp	
\label{lastpage}
\end{document}